\documentclass[a4paper,11pt]{article}
\usepackage{jheppub}
\usepackage{xcolor,float,bbm,multirow}
\usepackage{amsmath,amssymb,graphicx,slashed}
\usepackage[utf8x]{inputenc}
\usepackage[figuresright]{rotating}
\usepackage{orcidlink}
\usepackage{lineno}

\allowdisplaybreaks

\newcommand{\cyan}[1]{{\color{cyan} #1}}

\title{\bf \boldmath Matching $\nu$2HDM onto $\nu$SMEFT up to one-loop accuracy}

\author[a]{Wen-Feng Duan\,\orcidlink{0000-0001-5452-8327},}
\author[a]{Xuan-Zhen He\,\orcidlink{0009-0008-7418-7513},}
\author[a,1]{Xin-Qiang Li\,\orcidlink{0000-0002-3962-3577}\note{Corresponding author.},}
\author[a,b]{Ya-Dong Yang\,\orcidlink{0000-0002-9327-0557}}
\author[a]{and Xing-Bo Yuan\,\orcidlink{0000-0002-5647-2563}}

\affiliation[a]{Institute of Particle Physics and Key Laboratory of Quark and Lepton Physics~(MOE), Central China Normal University, Wuhan, Hubei 430079, China}
\affiliation[b]{Institute of Particle and Nuclear Physics, Henan Normal University, Xinxiang 453007, China}
\emailAdd{dufewe@mails.ccnu.edu.cn}
\emailAdd{hexuanzhen@mails.ccnu.edu.cn}
\emailAdd{xqli@mail.ccnu.edu.cn}
\emailAdd{yangyd@mail.ccnu.edu.cn}
\emailAdd{y@mail.ccnu.edu.cn}

\preprint{\today} 

\abstract{The neutrinophilic two-Higgs-doublet model ($\nu$2HDM) represents one of the most economical extensions of the Standard Model (SM) while still giving rise to a rich phenomenology. In this work, we carry out a full tree-level and one-loop matching of this ultraviolet completion onto the SM Effective Field Theory extended with right-handed neutrinos ($\nu$SMEFT), including all operators up to mass dimension six. This is achieved by integrating out the heavy scalar doublet at a scale well above the electroweak scale. For the matching procedure, we employ the functional method in combination with the naive dimensional regularization and modified minimal subtraction schemes. After converting from the Green's basis to the minimal, non-redundant Warsaw-extended basis, the effects of high-scale new physics are encoded in the Wilson coefficients of both physical and evanescent operators in the $\nu$SMEFT. These coefficients are given in terms of the $\nu$2HDM parameters prior to spontaneous symmetry breaking. Our final expressions at the high-energy matching scale are provided as ancillary files on \href{https://github.com/dufewe/v2HDM-to-vSMEFT}{GitHub}. This setup forms the foundation for forthcoming systematic and global analyses of the $\nu$2HDM parameter space using effective field theory techniques.}

\arxivnumber{2611.xxxxx} 

\begin{document}
\maketitle
\flushbottom

\section{Introduction}
\label{sec:intro}

The observation of the Brout-Englert-Higgs boson~\cite{Higgs:1964pj,Higgs:1964ia,Englert:1964et,Guralnik:1964eu,Kibble:1967sv} by the ATLAS~\cite{ATLAS:2012yve} and CMS~\cite{CMS:2012qbp} collaborations in 2012 completed the experimental verification of the particle content of the Standard Model (SM). While the measured properties of this $125$-$\mathrm{GeV}$ scalar are largely in agreement with the SM predictions~\cite{ATLAS:2022vkf,CMS:2022dwd}, the existence of additional scalar bosons is still allowed, provided their contribution to spontaneous symmetry breaking (SSB) of the electroweak gauge symmetry remains small. Crucially, even if the current indications of possible new physics (NP) at the Large Hadron Collider (LHC) eventually prove to be merely statistical fluctuations, there remains strong motivation to explore the phenomena of extensions of the SM scalar sector~\cite{Gildener:1976ih}. Among the simplest such extensions, the two-Higgs-doublet model (2HDM)~\cite{Lee:1973iz} exhibits a variety of appealing features and rich phenomena, including, among others, new sources of CP violation, the possibility of a strong first-order electroweak phase transition, and viable dark matter candidates~\cite{Gunion:1989we,Branco:2011iw,Ivanov:2017dad}. 

Within the SM particle spectrum, neutrinos are by far the most elusive and, due to their puzzling properties~--~namely, possessing finite but extremely small masses and exhibiting substantial mixing among their different flavors~\cite{Navas:2024jxa}~--~they provide a promising window to NP beyond the SM. A wide range of theoretical frameworks have been proposed to account for neutrino masses and the observed patterns of neutrino mixings (for reviews, see e.g. refs.~\cite{King:2013eh,King:2014nza,Xing:2020ijf,Chauhan:2023faf} and references therein). One way to obtain small neutrino masses is to mimic the Higgs mechanism by introducing a second Higgs doublet with a much smaller vacuum expectation value (vev). In a 2HDM realization, one scalar doublet is responsible for giving mass to the charged fermions, while the second develops a tiny vev and generates neutrino masses with Yukawa couplings of $\mathcal{O}(1)$. This construction is known as the neutrinophilic 2HDM ($\nu$2HDM)~\cite{Ma:2000cc,Wang:2006jy,Gabriel:2006ns,Davidson:2009ha,Haba:2011nb}, whose phenomena have been extensively investigated in the context of flavor physics~\cite{Gabriel:2006ns,Davidson:2009ha,Haba:2011nb,Haba:2010zi,Davidson:2010sf,Machado:2015sha,Bertuzzo:2015ada,Tang:2017owe,Li:2018rax,Dey:2018yht,Duan:2021whx,Cherchiglia:2023utd,Quintard:2026zub} and cosmology~\cite{Haba:2011ra,Sher:2011mx,Zhou:2011rc,Chao:2012pt,Mohanty:2018iop,Camargo:2019ukv,Liu:2020mxj,Sarma:2021icl,Li:2021tlv,Li:2022yna,Liu:2022rst}. Another attractive feature of this model is its robustness against radiative corrections~\cite{Haba:2011fn,Morozumi:2011zu}. 

In this work, we follow the framework proposed in refs.~\cite{Davidson:2009ha,Davidson:2010spd}, in which the SM field content is enlarged by adding a second scalar doublet $\Phi_2$ and three right-handed neutrinos $\nu_{R_i}$. The new scalar $\Phi_2$ carries the same gauge quantum numbers as the SM scalar doublet $\Phi_1$, while the $\nu_{R_i}$ are gauge singlets that combine with the three left-handed SM neutrinos to form Dirac fermions. To forbid Majorana mass terms for $\nu_{R_i}$ and, at the same time, enforce a Yukawa sector where only $\Phi_2$ couples to $\nu_{R_i}$, we introduce a global $U(1)$ symmetry under which $\Phi_2$ and $\nu_{R_i}$ have charge $+1$, whereas all SM fields remain uncharged. By explicitly breaking this $U(1)$ via a soft term $m_{12}^{\prime 2} \Phi_1^\dagger \Phi_2 +\text{h.c.}$ in the scalar potential,\footnote{In principle, the quartic Higgs and Yukawa interactions may also induce soft $U(1)$ breaking, but they lie outside the scope of the present work.} a tiny vev for $\Phi_2$ is induced, leading to appropriately small Dirac neutrino masses for sufficiently small $m_{12}^{\prime 2}$. A notable feature of this setup is the absence of an ultralight scalar state and the preservation of the conventional lepton number. The smallness of $m_{12}^{\prime 2}$, which constitutes the source of explicit global $U(1)$ breaking in the model, may be justified by physics at higher scales~--~for instance, through spontaneous $U(1)$ breaking in a hidden sector that is transmitted to the scalar sector via heavy messenger fields or higher-loop effects~\cite{Davidson:2009ha}. We will not delve into the details of such a UV realization.

The lack of direct evidence for NP~--~in particular, the absence of heavy Higgs-like resonances~--~in current experiments suggests a substantial mass gap between the electroweak and NP scales. This motivates us, in this work, to regard the heavy scalar doublet of the $\nu$2HDM as a single {\it fundamental} particle endowed with four degrees of freedom (d.o.f.). When these heavy d.o.f. cannot be probed directly, their impact on low-energy observables can be described in a model-independent way using effective field theory (EFT) techniques. Since the right-handed neutrinos $\nu_{R_i}$ of $\nu$2HDM belong to the spectrum of experimentally accessible states, they should not be integrated out together with the high-energy modes. In this situation, the appropriate EFT is obtained by enlarging the SM field content to explicitly include the right-handed neutrinos as dynamical d.o.f. This setup is commonly known as the Standard Model Effective Field Theory (SMEFT)~\cite{Buchmuller:1985jz,Grzadkowski:2010es,Brivio:2017vri,Isidori:2023pyp,Aebischer:2025qhh} extended by right-handed neutrinos, or, more briefly, the $\nu$SMEFT~\cite{delAguila:2008ir,Aparici:2009fh,Bhattacharya:2015vja,Liao:2016qyd,Li:2021tsq}.\footnote{In the literature it is also referred to as SMNEFT, NSMEFT, NRSMEFT, or $\nu_R$SMEFT.} The renormalization group (RG) equations of the $\nu$SMEFT at one- and two-loop orders are available in refs.~\cite{Datta:2020ocb,Datta:2021akg,Chala:2020pbn,Fuyuto:2024oii,Ardu:2024tzb,Zhang:2025ywe}, and have been implemented in the python package {\tt wilson}~\cite{Aebischer:2024csk}. Furthermore, the matching to the low-energy formulation of the EFT~--~usually called $\nu$LEFT~--~incorporating the $SU(3)_C\otimes U(1)_{em}$-invariant operators with right-handed neutrinos has been worked out at tree level~\cite{Chala:2020vqp,Li:2020lba,Bischer:2019ttk}. Since its inception~\cite{delAguila:2008ir}, the $\nu$SMEFT has evolved into a robust and systematic framework to analyze the low-energy manifestations of ultraviolet (UV) dynamics in scenarios with right-handed neutrinos~\cite{Duarte:2015iba,Duarte:2016miz,Duarte:2016caz,Caputo:2017pit,Bischer:2019ttk,Alcaide:2019pnf,Butterworth:2019iff,Jones-Perez:2019plk,Han:2020pff,Li:2020lba,Dekens:2020ttz,Biekotter:2020tbd,Li:2020wxi,DeVries:2020jbs,Barducci:2020icf,Cirigliano:2021peb,Cottin:2021lzz,Beltran:2021hpq,Zhou:2021ylt,Zapata:2022qwo,Barducci:2022hll,Datta:2022czw,Delgado:2022fea,Han:2022uho,Mitra:2022nri,Felkl:2023ayn,Duarte:2023tdw,Beltran:2023nli,Fernandez-Martinez:2023phj,Zapata:2023wsz,deVries:2024mla,Colangelo:2024sbf,Mitra:2024ebr,Biswas:2024gtr,Beltran:2025ilg,Bolton:2025tqw,Li:2025slp,Alonso:2025gzl,Duarte:2025zrg,Mitra:2026kxt,Abada:2026dlb,Duarte:2026ssh}. 

To connect the phenomenology of the $\nu$2HDM to the experimentally accessible low-energy domain and to identify which regions of its parameter space remain compatible with current experimental data, we need to match it onto the $\nu$SMEFT at the NP scale. After performing this matching, the heavy d.o.f. of the $\nu$2HDM are integrated out, and the resulting Wilson coefficients (WCs) of the higher-dimensional operators in the $\nu$SMEFT are expressed in terms of the {\it high-scale} NP parameters.\footnote{Throughout this work, we use the terms {\it high-scale} and {\it low-scale} to indicate that the heavy fields are integrated out {\it before} and {\it after} SSB, respectively.} This procedure induces specific correlations among the WCs across different sectors, thereby substantially reducing the number of independent parameters in the $\nu$SMEFT compared to a completely model-independent analysis within the same EFT framework.

The matching of a UV theory such as the $\nu$2HDM onto an EFT framework like the $\nu$SMEFT is typically performed using two main strategies:\footnote{The on-shell amplitude approach has also been proposed in the literature; see, e.g., refs.~\cite{DeAngelis:2023bmd,Li:2023edf,Chala:2024llp,Dong:2025wvf}.} the diagrammatic and the functional method (see, e.g., refs.~\cite{Henning:2014wua,Henning:2016lyp,Ellis:2016enq,Fuentes-Martin:2016uol,Zhang:2016pja,Ellis:2017jns,Cohen:2020fcu,Dittmaier:2021fls,Fuentes-Martin:2023ljp,Kramer:2019fwz,Ellis:2020ivx,Angelescu:2020yzf,Li:2024ciy,Fuentes-Martin:2024agf} and references therein). In the diagrammatic approach, one computes one-light-particle-irreducible Green's functions both in the UV theory and in the EFT for arbitrary off-shell kinematics, and then extracts the WCs of the local operators in the EFT in terms of the NP parameters. The functional method instead relies on the evaluation of effective actions, constructed using covariant derivatives and NP interactions, and heavy physics is integrated out directly in the path integral, leaving a non-local effective action that can be power expanded in terms of local operators. Compared with the diagrammatic method, the functional approach has the benefit of retaining background fields (the sources of the generating functionals) throughout the intermediate stages, which enables a very direct derivation of the EFT action. It also guaranties manifest gauge invariance at every step and removes the need to explicitly handle permutations of external legs~\cite{Cohen:2020fcu,Fuentes-Martin:2024agf}. In both approaches, however, the resulting operators are in general not in a physical basis (usually denoted as the Green's basis~\cite{Jiang:2018pbd,Gherardi:2020det}), and they have to be reduced to
the physical basis by means of field redefinitions~\cite{Deans:1978wn,Politzer:1980me,Arzt:1993gz,Passarino:2016saj,Criado:2018sdb} (equivalent to the application of the equations of motion (EOMs) at the linear level). Since both approaches involve algebraically lengthy computations, automation is essential. Several public tools are now available to perform matching and running: {\tt MatchMaker}~\cite{delAguila:2016zcb}, {\tt MatchingTools}~\cite{Criado:2017khh}, {\tt DsixTools}~\cite{Celis:2017hod,Fuentes-Martin:2020zaz}, {\tt wilson}~\cite{Aebischer:2018bkb,Aebischer:2024csk}, {\tt CoDEx}~\cite{DasBakshi:2018vni}, {\tt STrEAM}~\cite{Cohen:2020qvb}, {\tt SuperTracer}~\cite{Fuentes-Martin:2020udw}, {\tt Matchmakereft}~\cite{Carmona:2021xtq}, {\tt Matchete}~\cite{Fuentes-Martin:2022jrf}, and {\tt SOLD}~\cite{Guedes:2023azv,Guedes:2024vuf}.

In this work, we carry out the complete tree-level and one-loop matching of the $\nu$2HDM onto the $\nu$SMEFT up to operators of mass dimension six by integrating out the heavy scalar doublet, employing the functional method together with the packages {\tt Matchmakereft}, {\tt Matchete}, {\tt STrEAM} and {\tt SuperTracer}. Unlike previous analyses of the matching between 2HDM and SMEFT~\cite{Dawson:2022cmu,Dawson:2023ebe,DasBakshi:2024krs}, which focus on the {\it before-SSB} formulation but present their matching results in terms of low-energy physical quantities defined {\it after SSB}, we consistently regard the heavy scalar doublet as a {\it fundamental} field and integrate it out to derive the WCs of higher-dimensional operators directly in terms of the $\nu$2HDM parameters, always maintaining the {\it before-SSB} perspective~\cite{DasBakshi:2024krs,Kraml:2025fpv}. After mapping the Green's basis to a minimal, non-redundant basis, the high-scale NP effects are encoded in the WCs of both physical and evanescent operators. Their final expressions at the matching scale are provided as ancillary files on \href{https://github.com/dufewe/v2HDM-to-vSMEFT}{GitHub}, and can be cross-checked by other matching techniques. A detailed phenomenological analysis within this setup, along with a comparison to the results obtained in the full $\nu$2HDM, will be presented in a forthcoming publication.

This paper is structured as follows. Section~\ref{sec:v2HDM&vSMEFT} recapitulates the $\nu$2HDM and $\nu$SMEFT structures. In sections~\ref{sec:tree} and \ref{sec:one-loop}, we perform the tree-level and one-loop matching of the $\nu$2HDM onto the $\nu$SMEFT, respectively. Our conclusions are made in section~\ref{sec:conc}. For reference, all dimension-six operators in the {\it Warsaw-extended} basis are compiled in appendix~\ref{appA:vSMEFT_basis}, while the Green's operators and their transformations to the former are listed in appendix~\ref{appB:Green_basis}. The relevant interaction matrices are provided in appendix~\ref{appC:Inter_Matrix}. Finally, the one-loop Fierz rearrangements for four four-fermion operators are collected in appendix~\ref{appD:1-loop_Fierz}.

\section{\texorpdfstring{$\boldsymbol{\nu}$2HDM}{v2HDM} and \texorpdfstring{$\boldsymbol{\nu}$SMEFT}{vSMEFT}}
\label{sec:v2HDM&vSMEFT}

To clarify and shed light on the matching procedure from $\nu$2HDM to $\nu$SMEFT, we first present the structure of the underlying high-scale UV theory and the corresponding low-energy EFT, before carrying out the explicit calculations.

\subsection[\texorpdfstring{$\nu$2HDM}{v2HDM}]{\texorpdfstring{$\boldsymbol{\nu}$2HDM}{v2HDM}}

\subsubsection{Higgs basis}

The $\nu$2HDM enlarges the SM field content by introducing a second $SU(2)_L$ scalar doublet $\Phi_2$, which carries the same weak hypercharge $Y = 1/2$ as the SM doublet $\Phi_1$, along with three right-handed gauge-singlet fields $\nu_{R_i}$. Imposing that the vacuum configuration preserves the $U(1)_{em}$ symmetry, and exploiting the freedom in choosing the basis for the two scalar doublets, we can, without loss of generality, parametrize them as~\cite{Branco:2011iw}
\begin{equation}
	\Phi_{a}=\mathrm{e}^{i \theta_{a}}\left[\begin{array}{c}
		\phi_{a}^{+} \\[0.1cm]
		\frac{1}{\sqrt{2}}\left(v_{a}+\rho_{a}+i \eta_{a}\right)
	\end{array}\right]\,, \qquad (a=1,2)\,,
\end{equation}
where the vevs may in general be complex, $\langle 0\vert \Phi_a^T(x)\vert 0 \rangle=(0,v_a e^{i \theta_a}/\sqrt{2})$, with $v_a \geq 0$. Because only the relative phase $\theta \equiv \theta_2-\theta_1$ is physically meaningful, we can choose $\theta_1 = 0$ via an appropriate $U(1)_{Y}$ transformation.

To clearly distinguish the physical scalar fields from the Goldstone modes, it is convenient to perform a suitable global $SU(2)$ transformation that rotates the original scalar doublets $(\Phi_1,\Phi_2)$ into the so-called Higgs basis $(H_1,H_2)$~\cite{Branco:2011iw,Donoghue:1978cj,Georgi:1978ri,Botella:1994cs,Lavoura:1994fv,Davidson:2005cw},  
\begin{equation}
	\left( \begin{array}{c} H_1 \\[0.1cm] -H_2 \end{array} \right) \equiv
	\left( \begin{array}{cc} \cos\beta & \sin\beta \\[0.1cm] \sin\beta & -\cos\beta \end{array} \right)\,
	\left( \begin{array}{c} \Phi_1 \\[0.1cm]  \mathrm{e}^{-i\theta}\Phi_2 \end{array} \right)\,,
\end{equation}
so that the vev is entirely in the first doublet $H_1$, $\langle 0\vert H_1^T(x)\vert 0 \rangle=(0,v/\sqrt{2})$, with $v=\sqrt{v_{1}^{2}+v_{2}^{2}}=246~\mathrm{GeV}$, while the second doublet $H_2$ has zero vev. The mixing angle $\beta$ is defined through $\tan\beta \equiv e^{-i\theta} \langle \Phi_{2}^{0} \rangle / \langle \Phi_{1}^{0} \rangle = v_2/v_1$ and, by convention, is restricted to the first quadrant. In this Higgs basis, the scalar doublets can be parametrized as~\cite{Branco:2011iw,Pich:2009sp}
\begin{equation} \label{eq:Higgsbasis}
	H_1=\left[ \begin{array}{c} G^+ \\[0.1cm] \frac{1}{\sqrt{2}}\, (v+S_1+iG^0) \end{array} \right] \,,
	\qquad
	H_2 = \left[ \begin{array}{c} H^+ \\[0.1cm] \frac{1}{\sqrt{2}}\, (S_2+iS_3) \end{array} \right] \,.
\end{equation}
One can observe that the Goldstone bosons $G^{\pm}$ and $G^{0}$ now reside exclusively in the first doublet, mirroring the situation in the SM. The physical scalar spectrum consists of a charged scalar pair $H^{\pm}$ and three neutral states $h$, $H$, and $A$, which are obtained as linear combinations of the neutral fields $S_{1,2,3}$ via an orthogonal transformation, $\varphi^0_i =\{h, H, A\}=\mathcal{R}_{ij}S_j$. The orthogonal matrix $\mathcal{R}$ specifies the mass eigenstates of the three neutral scalars, and its explicit form is determined by the structure of the scalar potential~\cite{Branco:2011iw,Pich:2009sp}.

\subsubsection{Scalar potential}

In the Higgs basis, the most general renormalizable scalar potential of the model consistent with the SM gauge symmetry can be written as~\cite{Davidson:2005cw,Wu:1994ja}
\begin{align}
V_{\nu{\rm 2HDM}}^{\rm Higgs} =&\; m_{11}^2\, H_1^\dagger H_1 + m_{22}^2\, H_2^\dagger H_2 - \left(m_{12}^2\, H_1^\dagger H_2 + \text{h.c.}\right) + \frac{\lambda_1}{2} \left(H_1^\dagger H_1\right)^2 + \frac{\lambda_2}{2} \left(H_2^\dagger H_2\right)^2 \notag\\[1.0mm]
&+ \lambda_3 \left(H_1^\dagger H_1\right)\left(H_2^\dagger H_2\right) + \lambda_4 \left(H_1^\dagger H_2\right)\left(H_2^\dagger H_1\right) + \bigg[\frac{\lambda_5}{2} \left(H_1^\dagger H_2\right)^2 \notag\\[0.8mm]
&+ \lambda_6 \left(H_1^\dagger H_1\right)\left(H_1^\dagger H_2\right) + \lambda_7 \left(H_2^\dagger H_2\right)\left(H_1^\dagger H_2\right) + \text{h.c.}\bigg], \label{eq:v2HDM_Higgs}
\end{align}
where ``h.c.'' stands for the hermitian conjugate. The parameters $m_{11,22}^2$ and $\lambda_{1,2,3,4}$ are real by hermiticity, while $m_{12}^2$ and $\lambda_{5,6,7}$ are, in general, complex. These couplings are constrained by requirements of vacuum stability (the potential is bounded from below (BFB)) and perturbative unitarity (ensuring a reliable perturbative expansion)~\cite{Branco:2011iw,Ivanov:2017dad}.

Within the {\it high-scale} perspective, each scalar doublet of the model, despite containing four d.o.f., can be regarded as both {\it physical} and {\it fundamental}. Focusing on the case prior to SSB, we take the dimensionful parameter $m_{22}^2$ to be large, so that the heavier doublet $H_2$ can be integrated out in the matching procedure. After eliminating $H_2$, the resulting theory is described by an effective Lagrangian expressed as an expansion in inverse powers of $m_{22}^2$. Consequently, this parameter is identified with the mass scale squared of the $\nu$SMEFT, $\Lambda^2=m_{22}^2$. Under these circumstances, it is both natural and essential to work in the decoupling~\cite{Gunion:2002zf,Haber:2013mia,Dawson:2023ebe} and alignment~\cite{Craig:2013hca,Bernon:2015qea,Bernon:2015wef} limits of the model, where $m_{22}^2$ is interpreted as the mass of $H_2$, and the mixing parameter $m_{12}^2$ is taken to be {\it sufficiently small}.\footnote{The smallness of $m_{12}^2$ can be enforced by imposing a softly broken symmetry $U(1)$ between the two scalar doublets, such that the mass parameters satisfy the hierarchy $\left|m_{12}\right|^2 \ll m_{11}^2 \ll m_{22}^2$.} If ignoring the loop-order contributions, we also require $H_1$ to obtain the usual SM vev through SSB with $m_{11}^2 < 0$ to minimize the potential. In the matching computation below, we demonstrate that there is no need to diagonalize the $(H_1,H_2)$ mass matrix before integrating out $H_2$: it can instead be incorporated as an effective scalar interaction coupling. Once $H_2$ is decoupled, the remaining effective scalar potential becomes SM-like, featuring an SM-structured scalar doublet $H_1$, so that the vacuum admits a (meta-)stable configuration and SSB emerges automatically. Operationally, this setup leads to an EFT expansion naturally organized in powers of $1/m_{22}$, i.e., an $m_{22}$-based EFT power counting.

\subsubsection{Yukawa coupling}

For the Yukawa sector, the most general Lagrangian of the model can be written as~\cite{Davidson:2009ha,Davidson:2010spd}
\begin{align}
\mathcal{L}_{\nu{\rm 2HDM}}^{\rm Yuk} = & - \overline{Q}_L \left(Y_U^1 \widetilde{H}_1 + Y_U^2 \widetilde{H}_2\right) U_R - \overline{Q}_L \left(Y_D^1 H_1 + Y_D^2 H_2\right) D_R \notag\\[0.8mm]
&- \overline{L}_L \left(Y_\nu^1 \widetilde{H}_1 + Y_\nu^2 \widetilde{H}_2\right) \nu_R - \overline{L}_L \left(Y_E^1 H_1 + Y_E^2 H_2\right) E_R + \text{h.c.}, \label{eq:v2HDM_Yukawa}
\end{align}
where $Q_L$ and $L_L$ denote the left-handed quark and lepton $SU(2)_L$ doublets, while the right-handed $SU(2)_L$ singlets are represented by $U_R$ (up-type quark), $D_R$ (down-type quark), $\nu_R$ (neutrino), and $E_R$ (charged lepton), respectively. For brevity, the generation indices of these fermionic fields are omitted. The charge-conjugated scalar fields $\widetilde{H}_i=i\tau_2 H_i^\ast$, with $\tau_2$ the second Pauli matrix, carry weak hypercharge $Y=-1/2$.

The Yukawa matrices $Y_i^1$ in eq.~\eqref{eq:v2HDM_Yukawa} are, in general, unrelated to $Y_i^2$ for $i=U,D,\nu,E$. As a result, they cannot be diagonalized simultaneously in flavor space once we move to the fermion mass-eigenstate basis. This misalignment typically generates tree-level flavor-changing neutral currents (FCNCs), which can be phenomenologically dangerous. To comply with the strong experimental limits on FCNCs, we can impose additional flavor symmetries~\cite{Branco:2011iw}, such as {\it minimal flavor violation} (MFV)~\cite{Chivukula:1987py,Buras:2000dm,DAmbrosio:2002vsn,Cirigliano:2005ck,Buras:2010mh} or {\it natural flavor conservation} (NFC)~\cite{Glashow:1976nt,Paschos:1976ay,Buras:2010mh}. In MFV, the magnitude of FCNCs is controlled by a global flavor symmetry acting on the Yukawa sector, while NFC eliminates tree-level FCNCs by enforcing suitable discrete $Z_2$ symmetries. The {\it Yukawa alignment} scenario~\cite{Pich:2009sp,Tuzon:2010vt,Cree:2011uy}, which can be viewed as a particular realization of MFV and a generalized form of NFC, also provides a mechanism to evade stringent experimental constraints from FCNCs, by assuming that the Yukawa coupling matrices $Y_i^1$ and $Y_i^2$ are proportional to each other in flavor space.

To prevent $\Phi_1$ from coupling to the right-handed neutrinos $\nu_{R_i}$ and simultaneously forbid Majorana mass terms for $\nu_{R_i}$, we introduce a global $U(1)$ symmetry. Under this symmetry, $\Phi_2$ and $\nu_{R_i}$ carry charge $+1$, whereas all other fields are neutral~\cite{Davidson:2009ha,Davidson:2010spd}. The symmetry also removes the $\lambda_{5,6,7}$ interactions from the scalar potential in eq.~\eqref{eq:v2HDM_Higgs}. Nevertheless, we may explicitly break the global $U(1)$ by retaining the soft term $m_{12}^2 H_1^\dagger H_2 + \text{h.c.}$ in the scalar potential. Doing so allows $\Phi_2$ to acquire a very small vev, which in turn yields appropriately tiny Dirac neutrino masses for sufficiently small $m_{12}^2$. This setup is additionally characterized by heavy extra scalar states and by the conventional lepton number being preserved as an accidental symmetry of the theory~\cite{Davidson:2009ha,Davidson:2010spd}. As in the SM, the naturalness or hierarchy problem of this framework can be addressed by embedding it into a supersymmetric theory or a strong-dynamics scenario at the TeV scale~\cite{Craig:2022eqo}.

\subsubsection{Full Lagrangian}

Together with the scalar potential and Yukawa Lagrangian given in eqs.~\eqref{eq:v2HDM_Higgs} and \eqref{eq:v2HDM_Yukawa}, the complete $\nu$2HDM Lagrangian, compatible with the SM gauge symmetry, can be written as
\begin{equation}
\mathcal{L}_{\nu{\rm 2HDM}} = \mathcal{L}_{\nu{\rm 2HDM}}^{\rm kin} + \mathcal{L}_{\nu{\rm 2HDM}}^{\rm Yuk} - V_{\nu{\rm 2HDM}}^{\rm Higgs} + \mathcal{L}_\xi + \mathcal{L}_\eta + \mathcal{L}_\theta, \label{eq:v2HDM}
\end{equation}
where the kinetic sector reads
\begin{align}
\mathcal{L}_{\nu{\rm 2HDM}}^{\rm kin} =&\; \overline{Q}_L i \slashed{D} Q_L + \overline{U}_R i\slashed{D} U_R + \overline{D}_R i\slashed{D} D_R + \overline{L}_L i\slashed{D} L_L + \overline{\nu}_R i\slashed{D}\nu_R + \overline{E}_R i\slashed{D} E_R \notag \\[0.8mm]
&- H_1^\dagger D^2 H_1 - H_2^\dagger D^2 H_2 - \frac{1}{4} B_{\mu\nu} B_{\mu\nu} - \frac{1}{4} W^I_{\mu\nu} W^I_{\mu\nu} - \frac{1}{4} G^A_{\mu\nu} G^A_{\mu\nu}, 
\end{align}
and the gauge-fixing and Faddeev–Popov ghost terms are given, respectively, by
\begin{align}
\mathcal{L}_\xi &= - \frac{1}{2\xi_B} \left(\partial_\mu B_\mu\right)^2 - \frac{1}{2\xi_W} \left(D_\mu W_\mu^I\right)^2 - \frac{1}{2\xi_G} \left(D_\mu G_\mu^A\right)^2, \\[1.0mm]
\mathcal{L}_\eta &= - \overline{\eta}_W^I \partial^\mu D_\mu \eta_W^I - \overline{\eta}_G^A \partial^\mu D_\mu \eta_G^A.
\end{align}
Here, $\xi_{B,W,G}$ denote the gauge-fixing parameters, and the covariant derivative can be written in a universal form as
\begin{equation}
D_\mu \equiv \partial_\mu - ig_1 B_\mu Y_\mathcal{F} - ig_2 W_\mu^I S^I - ig_s G_\mu^A T^A,
\end{equation}
where $g_1$, $g_2$, and $g_s$ are the gauge coupling constants of $U(1)_Y$, $SU(2)_L$, and $SU(3)_C$, respectively; $Y_\mathcal{F}$ is the weak hypercharge of the field $\mathcal{F}$, and $S^I \equiv \tau^I/2$ and $T^A \equiv \lambda^A/2$, with $\tau^I$ and $\lambda^A$ being the Pauli and Gell-Mann matrices, are the generators of the $SU(2)_L$ and $SU(3)_C$ groups, respectively. The electroweak field strength tensors are specified as
\begin{align}
    B_{\mu \nu} = \partial_\mu B_\nu-\partial_\nu B_\mu, \qquad
    W_{\mu \nu}^I = \partial_\mu W_\nu^I-\partial_\nu W_\mu^I+g_2 \epsilon^{I J K} W_\mu^J W_\nu^K.
\end{align}
The term $\mathcal{L}_\theta$ stands for the renormalizable CP-odd Lagrangian. Note that $\mathcal{L}_\xi$ only affects the structure of the gauge-boson propagators, while $\mathcal{L}_\eta$ and $\mathcal{L}_\theta$ do not enter the one-loop matching calculation. Hence, we can effectively drop these terms by absorbing the $\xi$-dependent pieces into a redefinition of the gauge propagators.

We emphasize that all fermions and gauge bosons appearing in eq.~\eqref{eq:v2HDM} remain light and massless d.o.f. before SSB, and that the small but non-vanishing $m_{12}^2$ term in eq.~\eqref{eq:v2HDM_Higgs} may arise from the soft breaking of a global $U(1)$ symmetry. Once this global $U(1)$ symmetry is softly broken, pseudo–Nambu-Goldstone bosons do not appear, and lepton-number violation is forbidden, which in turn rules out the Weinberg operator~\cite{Weinberg:1979sa} and the usual type-I seesaw mechanism~\cite{Minkowski:1977sc,Yanagida:1979as,Gell-Mann:1979vob,Mohapatra:1979ia} for neutrino mass generation~\cite{Davidson:2009ha,Davidson:2010spd}. However, the original Dirac seesaw mechanism introduced in ref.~\cite{Gu:2006dc} continues to operate in this setup to generate neutrino masses~\cite{Davidson:2009ha,Davidson:2010spd}.

\subsection[\texorpdfstring{$\nu$SMEFT}{vSMEFT}]{\texorpdfstring{$\boldsymbol{\nu}$SMEFT}{vSMEFT}}

After integrating out the heavy doublet $H_2$, all information about the high-scale dynamics of the $\nu$2HDM is encoded in the WCs of higher-dimensional operators in the $\nu$SMEFT. We therefore briefly outline the construction of the $\nu$SMEFT Lagrangian~\cite{delAguila:2008ir,Aparici:2009fh,Bhattacharya:2015vja,Liao:2016qyd,Li:2021tsq}. Schematically, it can be written as
\begin{equation}
\mathcal{L}_{\nu{\rm SMEFT}} = \mathcal{L}_{\nu{\rm SM}} + \sum_{n,i} \left[C_i^{\mathcal{Q},(n)} \mathcal{Q}_i^{(n)} + C_i^{E,(n)} E_i^{(n)}\right], \label{eq:vSMEFT}
\end{equation}
where $C_i^{\mathcal{Q},(n)}$ and $C_i^{E,(n)}$ denote the WCs of the physical and evanescent~\cite{Buras:1989xd,Dugan:1990df,Herrlich:1994kh} operators of mass dimension $n$. In our setup, $n = 2,4,5,6,\cdots$, where the operators of dimensions two and four correspond to the effective quadratic and quartic interactions of the light doublet $H_1$ that arise after integrating out $H_2$. The explicit expressions for the dimension-five and dimension-six physical and evanescent operators are given in appendix~\ref{appA:vSMEFT_basis}, in the so-called Warsaw-extended basis. These evanescent operators are defined in dimensional regularization (DR) with $d = 4 - 2\epsilon$ spacetime dimensions~\cite{Aebischer:2024xnf,Aebischer:2025hsx} and included here for completeness; they do not affect physical observables at tree level and induce only negligible corrections to the physical operators through one-loop RG evolution~\cite{Aebischer:2024xnf,Aebischer:2025hsx}.

The first term in eq.~\eqref{eq:vSMEFT} corresponds to the SM Lagrangian extended by right-handed neutrinos, which takes the form
\begin{equation}
\mathcal{L}_{\nu{\rm SM}} = \mathcal{L}_{\nu{\rm SM}}^{\rm kin} + \mathcal{L}_{\nu{\rm SM}}^{\rm Yuk} - V_{\nu{\rm SM}}^{\rm Higgs} + \mathcal{L}_\xi + \mathcal{L}_\eta + \mathcal{L}_\theta,
\end{equation}
where
\begin{align}
\mathcal{L}_{\nu{\rm SM}}^{\rm kin} =&\; \overline{Q}_L i\slashed{D} Q_L + \overline{U}_R i\slashed{D} U_R + \overline{D}_R i\slashed{D} D_R + \overline{L}_L i\slashed{D} L_L + \overline{\nu}_R i\slashed{D} \nu_R + \overline{E}_R i\slashed{D} E_R \notag\\[0.8mm]
&- H_1^\dagger D^2 H_1 - \frac{1}{4} B_{\mu\nu} B_{\mu\nu} - \frac{1}{4} W^I_{\mu\nu} W^I_{\mu\nu} - \frac{1}{4} G^A_{\mu\nu} G^A_{\mu\nu}, 
\\[1.2mm]
\mathcal{L}_{\nu{\rm SM}}^{\rm Yuk} =&\; - \overline{Q}_L Y_U^1 \widetilde{H}_1 U_R - \overline{Q}_L Y_D^1 H_1 D_R - \overline{L}_L Y_\nu^1 \widetilde{H}_1 \nu_R - \overline{L}_L Y_E^1 H_1 E_R + \text{h.c.}, \\[1.2mm]
V_{\nu{\rm SM}}^{\rm Higgs} =&\; m_{11}^2 H_1^\dagger H_1 + \frac{\lambda_1}{2} \left(H_1^\dagger H_1\right)^2, \label{eq:nuSM_Higgs_potential}
\end{align}
and $\mathcal{L}_{\xi,\eta,\theta}$ are carried over from the $\nu$2HDM Lagrangian. More precisely, the effective non-renormalizable part of $\mathcal{L}_{\nu{\rm SMEFT}}$ in eq.~\eqref{eq:vSMEFT} arises from integrating out the heavy d.o.f. of the $\nu$2HDM, whereas the renormalizable part of $\mathcal{L}_{\nu{\rm SM}}$ corresponds to the sector of the UV theory at high scale that involves only the light d.o.f.

Starting from the complete Lagrangian in eq.~\eqref{eq:vSMEFT}, the corresponding EOMs without any suppression by the high scale are obtained as
\begin{align}
i\slashed{D} Q_L &= Y_U^1 U_R \widetilde{H}_1 + Y_D^1 D_R H_1,& i\slashed{D} U_R &= \widetilde{H}^\dagger_1 Y_U^{1\dagger} Q_L,& i\slashed{D} D_R &= H_1^\dagger Y_D^{1\dagger} Q_L, \label{eq:EOM_fermion_1} \\[1.0mm]
i\slashed{D} L_L &= Y_\nu^1 \nu_R \widetilde{H}_1 + Y_E^1 E_R H_1,& i\slashed{D} \nu_R &= \widetilde{H}^\dagger_1 Y_\nu^{1\dagger} L_L,& i\slashed{D} E_R &= H_1^\dagger Y_E^{1\dagger} L_L, \label{eq:EOM_fermion_2}
\end{align}
for the fermionic fields, and
\begin{align}
D_\mu G_{\mu\nu}^A = &- g_s \left(\overline{Q}_L \gamma_\nu T^A Q_L + \overline{U}_R \gamma_\nu T^A U_R + \overline{D}_R \gamma_\nu T^A D_R\right), \label{eq:EOM_boson_1} \\[2.5mm]
D_\mu W_{\mu\nu}^I = &- \frac{g_2}{2} \left(\overline{Q}_L \tau^I \gamma_\nu Q_L + \overline{L}_L \tau^I \gamma_\nu L_L + H_1^\dagger i \overleftrightarrow{D}_\nu^I H_1\right), \label{eq:EOM_boson_2} \\[1.0mm]
\partial_\mu B_{\mu\nu} = &- g_1 \bigg(Y_{Q_L} \overline{Q}_L \gamma_\nu Q_L + Y_{U_R} \overline{U}_R \gamma_\nu U_R + Y_{D_R} \overline{D}_R \gamma_\nu D_R \notag\\[0.5mm]
&+ Y_{L_L} \overline{L}_L \gamma_\nu L_L + Y_{E_R} \overline{E}_R \gamma_\nu E_R + Y_{\nu_R} \overline{\nu}_R \gamma_\nu \nu_R + Y_H H_1^\dagger i\overleftrightarrow{D}_\nu H_1\bigg), \label{eq:EOM_boson_3} \\[1.0mm]
D^2 H_1 = &- m_{11}^2 H_1 - \lambda_1 \left(H_1^\dagger H_1\right) H_1 - \overline{U}_R^\ast Y_U^{1T} \widetilde{Q}_L - \overline{D}_R Y_D^{1\dagger} Q_L - \overline{\nu}_R^\ast Y_\nu^{1T} \widetilde{L}_L - \overline{E}_R Y_E^{1\dagger} L_L, \label{eq:EOM_boson_4}
\end{align}
for the bosonic fields, where we define $\widetilde{Q}_L \equiv i\tau^2 Q_L^\ast$, $\widetilde{L}_L \equiv i\tau^2 L_L^\ast$, and 
\begin{align}
H_1^\dagger i\overleftrightarrow{D}_\mu^I H_1 &\equiv i \left(H_1^\dagger \tau^I D_\mu H_1 - H_1^\dagger \overleftarrow{D}_\mu \tau^I H_1\right), \\[0.8mm]
H_1^\dagger i\overleftrightarrow{D}_\mu H_1 &\equiv i \left(H_1^\dagger D_\mu H_1 - H_1^\dagger \overleftarrow{D}_\mu H_1\right).
\end{align}
Here, the covariant derivatives acting on the gauge fields are taken in the adjoint representation, whereas all other covariant derivatives are in the fundamental representation. As will be demonstrated later, these EOMs, together with integration-by-parts (IBP) relations, Fierz transformations, and other relations, allow us to eliminate linearly dependent or unphysical operators in the Green's basis, yielding the non-redundant and independent operator basis summarized in appendix~\ref{appA:vSMEFT_basis}.

\section{Tree-level matching}
\label{sec:tree}

Having set up the required model structures, we now present the explicit matching computation, retaining terms up to $\mathcal{O}(1/m_{22}^2)$ and operators up to mass dimension six. To demonstrate the tree-level and one-loop matching procedure between $\nu$2HDM and $\nu$SMEFT within the functional framework, we start by introducing the particle multiplets $\varphi_i$ and their conjugates $\overline{\varphi}_i$ for all d.o.f. in the $\nu$2HDM:
\begin{align}
\varphi_i &\in \{\varphi_{Q_L},\; \varphi_{L_L},\; \varphi_{U_R},\; \varphi_{D_R},\; \varphi_{\nu_R},\; \varphi_{E_R},\; \varphi_{H_1},\; \varphi_{H_2},\; \varphi_{B},\; \varphi_{W},\; \varphi_{G},\; \varphi_{\eta_W},\; \varphi_{\eta_G}\}, \\[1.0mm]
\overline{\varphi}_i &\in \{\overline{\varphi}_{Q_L},\; \overline{\varphi}_{L_L},\; \overline{\varphi}_{U_R},\; \overline{\varphi}_{D_R},\; \overline{\varphi}_{\nu_R},\; \overline{\varphi}_{E_R},\; \overline{\varphi}_{H_1},\; \overline{\varphi}_{H_2},\; \overline{\varphi}_{B},\; \overline{\varphi}_{W},\; \overline{\varphi}_{G},\; \overline{\varphi}_{\eta_W},\; \overline{\varphi}_{\eta_G}\},
\end{align}
where
\begin{align}
\varphi_f &\equiv \begin{pmatrix} f \\ f^c \end{pmatrix},& \varphi_H &\equiv \begin{pmatrix} H \\ H^\ast \end{pmatrix},& \varphi_V &\equiv V,& \varphi_\eta &\equiv \begin{pmatrix} \eta \\ \eta^c \end{pmatrix}, \\[0.8mm]
\overline{\varphi}_f &\equiv \left(\overline{f},\overline{f^c}\right),& \overline{\varphi}_H &\equiv  \left(H^\dagger,H^T\right),& \overline{\varphi}_V &\equiv V,& \overline{\varphi}_\eta &\equiv \left(\overline{\eta},\overline{\eta^c}\right).
\end{align}
Using these field multiplets, the full $\nu$2HDM Lagrangian in eq.~\eqref{eq:v2HDM} can be written in a compact form as~\cite{Cohen:2020fcu}
\begin{equation} \label{eq:Lnu2HDM_compact}
\mathcal{L}_{\nu{\rm 2HDM}} = \frac{1}{2} \sum_{i=f}^{H,V,\eta} \overline{\varphi}_i K_i \varphi_i - \frac{1}{2} \sum_{i,j=f}^{H,V,\eta} \overline{\varphi}_i X_{ij} \varphi_j,
\end{equation}
where the sum runs over fermions $f\in \{Q_L,L_L,U_R,D_R,\nu_R,E_R\}$, scalars $H\in \{H_1,H_2\}$, vectors $V\in \{B,W,G\}$, and ghosts $\eta\in \{\eta_W,\eta_G\}$. The corresponding inverse propagators $K_i$ are given by\footnote{We neglect the effect of $m_{11}^2$, even though $m_{11} \ll m_{22}$ may correspond to the mass of the light scalar doublet $H_1$ in the case of $m_{11}^2>0$.}
\begin{align}
    K_i = \left\{\begin{array}{ll}
          - D^2 & \quad \text{for $H_1$}, \\[0.8mm]
          - D^2 - m_{22}^2 & \quad \text{for $H_2$}, \\[0.8mm]
          i\slashed{D} & \quad \text{for fermions}, \\[0.8mm]
          g_{\mu\nu} D^2 - \left(1 - 1/\xi_V\right) D_\mu D_\nu & \quad \text{for vectors}, \\[0.8mm]
          - D^2 & \quad \text{for ghosts}, 
         \end{array} 
         \right.
\end{align}
where $g_{\mu\nu}=\text{diag}(1,-1,-1,-1)$ denotes the Minkowski metric tensor. The matrices $X_{ij}$ in eq.~\eqref{eq:Lnu2HDM_compact} encode the interactions between the fields $\varphi_{i}$ and $\varphi_{j}$, and can be obtained by taking the second derivative of the three- and higher-point interactions in the UV Lagrangian in eq.~\eqref{eq:v2HDM} and setting the heavy scalar field $H_2$ to its EOM solution, $H_2=H_2^c$, as given by eq.~\eqref{eq:H2c-solution} below; for convenience, their explicit expressions are collected in appendix~\ref{appC:Inter_Matrix}.

We can then determine the classical trajectory of the heavy field $\varphi_{H_2}$, or equivalently $H_2$, by solving the Euler–Lagrange equation, obtaining
\begin{equation} \label{eq:H2c-solution}
H_2^c \equiv H_2^{c[1]} + H_2^{c[3]} + H_2^{c[5]},
\end{equation}
valid up to $\mathcal{O}(1/m_{22}^2)$, where
\begin{align}
H_2^{c[1]} &= \frac{m_{12}^{\ast2}}{m_{22}^2} H_1, \notag\\[0.8mm]
H_2^{c[3]} &= \frac{1}{m_{22}^2} B_{\rm Yuk}^{[3]} - \frac{1}{m_{22}^2} \biggl[\left(D^2 + \lambda_3 H_1^\dagger H_1 + \lambda_4 H_1 H_1^\dagger\right) H_2^{c[1]} + \lambda_5^\ast \left(H_2^{c[1]\dagger} H_1\right) H_1\biggr], \notag\\[1.5mm]
H_2^{c[5]} &= - \frac{1}{m_{22}^2} \biggl[\left(D^2 + \lambda_3 H_1^\dagger H_1 + \lambda_4 H_1 H_1^\dagger\right) H_2^{c[3]} + \lambda_5^\ast \left(H_2^{c[3]\dagger} H_1\right) H_1\biggr], 
\end{align}
and
\begin{equation}
B_{\rm Yuk}^{[3]} \equiv - \overline{U}_R^\ast Y_U^{2T} \widetilde{Q}_L - \overline{D}_R Y_D^{2\dagger} Q_L - \overline{\nu}_R^\ast Y_\nu^{2T} \widetilde{L}_L - \overline{E}_R Y_E^{2\dagger} L_L - \lambda_6^\ast \left(H_1^\dagger H_1\right) H_1.
\end{equation}
Substituting the classical solution $H_2^c$ back into the full $\nu$2HDM Lagrangian in eq.~\eqref{eq:v2HDM}, we obtain the tree-level matching onto operators up to mass dimension six~\cite{Cohen:2020fcu,Fuentes-Martin:2024agf}, with the resulting tree-level $\nu$SMEFT Lagrangian given by
\begin{align}
\mathcal{L}_{\nu{\rm SMEFT}}^{\rm tree} =&\; \frac{\left|m_{12}\right|^4}{m_{22}^2} H_1^\dagger H_1 - \bigg[\frac{m_{12}^2 \lambda_6^\ast}{m_{22}^2} \mathcal{Q}_{H_1}^{(4)} + \frac{m_{12}^2}{m_{22}^2} \left(Y_U^2\right)_{pr} \mathcal{Q}_{\underset{pr}{qu}}^{(4)} + \frac{m_{12}^{\ast2}}{m_{22}^2} \left(Y_D^2\right)_{pr} \mathcal{Q}_{\underset{pr}{qd}}^{(4)} \notag\\[0.8mm]
&+ \frac{m_{12}^2}{m_{22}^2} \left(Y_\nu^2\right)_{pr} \mathcal{Q}_{\underset{pr}{\ell\nu}}^{(4)} + \frac{m_{12}^{\ast2}}{m_{22}^2} \left(Y_E^2\right)_{pr} \mathcal{Q}_{\underset{pr}{\ell e}}^{(4)} + \text{h.c.}\bigg] + \frac{1}{m_{22}^2} \bigg[\left(Y_E^2\right)_{pr} \left(Y_D^{2\dagger}\right)_{st} \mathcal{Q}_{\underset{prst}{\ell edq}}^{(6)} \notag\\[0.8mm]
&+ \left(Y_\nu^2\right)_{pr} \left(Y_E^{2}\right)_{st} \mathcal{Q}_{\underset{prst}{\ell\nu\ell e}}^{(6)} + \left(Y_\nu^2\right)_{pr} \left(Y_U^{2\dagger}\right)_{st} \mathcal{Q}_{\underset{prst}{\ell\nu uq}}^{(6)} - \left(Y_E^2\right)_{pr} \left(Y_U^{2}\right)_{st} \mathcal{Q}_{\underset{prst}{\ell equ}}^{(1),(6)} \notag\\[0.8mm]
&+ \left(Y_\nu^2\right)_{pr} \left(Y_D^{2}\right)_{st} \mathcal{Q}_{\underset{prst}{\ell\nu qd}}^{(1),(6)} + \left(Y_U^2\right)_{pr} \left(Y_D^{2}\right)_{st} \mathcal{Q}_{\underset{prst}{quqd}}^{(1),(6)} + \lambda_6 \left(Y_U^2\right)_{pr} \mathcal{Q}_{\underset{pr}{uH_1}}^{(6)} \notag\\[0.8mm]
&+ \lambda_6^\ast \left(Y_D^2\right)_{pr} \mathcal{Q}_{\underset{pr}{dH_1}}^{(6)} + \lambda_6 \left(Y_\nu^2\right)_{pr} \mathcal{Q}_{\underset{pr}{\nu H_1}}^{(6)} + \lambda_6^\ast \left(Y_E^2\right)_{pr} \mathcal{Q}_{\underset{pr}{eH_1}}^{(6)} + \text{h.c.}\bigg] \notag\\[0.8mm]
&+ \frac{1}{m_{22}^2} \bigg[\left|\lambda_6\right|^2 \mathcal{Q}_{H_1}^{(6)} + \left(Y_E^2\right)_{pr} \left(Y_E^{2\dagger}\right)_{st} \mathcal{O}_{\underset{prst}{\ell ee\ell}}^{S,(6)} + \left(Y_\nu^2\right)_{pr} \left(Y_\nu^{2\dagger}\right)_{st} \mathcal{O}_{\underset{prst}{\ell\nu\nu\ell}}^{S,(6)} \notag\\[0.8mm]
&+ \left(Y_D^2\right)_{pr} \left(Y_D^{2\dagger}\right)_{st} \mathcal{O}_{\underset{prst}{qddq}}^{S,(6)} + \left(Y_U^2\right)_{pr} \left(Y_U^{2\dagger}\right)_{st} \mathcal{O}_{\underset{prst}{quuq}}^{S,(6)}\bigg]. \label{eq:2HDM_SMEFT_tree}
\end{align}
The dimension-four operators $\mathcal{Q}_{i}^{(4)}$ present above are defined, respectively, as
\begin{align}
\mathcal{Q}_{\underset{pr}{qu}}^{(4)} &\equiv \overline{Q}_{Lp} \widetilde{H}_1 U_{Rr},& \mathcal{Q}_{\underset{pr}{qd}}^{(4)} &\equiv \overline{Q}_{Lp} H_1 D_{Rr}, \\[0.8mm]
\mathcal{Q}_{\underset{pr}{\ell\nu}}^{(4)} &\equiv \overline{L}_{Lp} \widetilde{H}_1 \nu_{Rr},& \mathcal{Q}_{\underset{pr}{\ell e}}^{(4)} &\equiv \overline{L}_{Lp} H_1 E_{Rr},& \mathcal{Q}_{H_1}^{(4)} &\equiv \left(H_1^\dagger H_1\right)^2.
\end{align}
The physical dimension-six operators $\mathcal{Q}_i^{(6)}$ and the redundant (or evanescent) operators $\mathcal{O}_i^{S,(6)}$ are listed explicitly in appendices~\ref{appA:vSMEFT_basis} and \ref{appB:Green_basis}, respectively. In deriving eq.~\eqref{eq:2HDM_SMEFT_tree}, we work entirely at the operator level, without invoking any algebraic simplifications, such as EOMs, IBP relations, Fierz transformations, or Greek identities.

From the above results, it is clear that the parameter $m_{12}$ plays a multifaceted role: it affects not only the tree-level WCs but also alters both the scalar potential and the Yukawa interactions of the standard SM Lagrangian. Crucially, the tiny neutrino masses arise via two separate mechanisms: (i) the usual $Y_\nu^1$ term in $\mathcal{L}_{\nu{\rm SM}}$, and (ii) a $1/m_{22}^2$-suppressed contribution proportional to the $Y_\nu^2$ coupling. Because the dimensionful parameters satisfy the strong hierarchy $\left|m_{12}\right|^2 \ll m_{11}^2 \ll m_{22}^2$, the $m_{12}^2$-induced effects in the quark and charged-lepton sectors are extremely suppressed and therefore phenomenologically irrelevant. In contrast, the same $m_{12}^2$ contributions~--~although small~--~provide a phenomenologically viable way to generate neutrino masses at the eV scale, even for $Y_\nu^2$ of order unity. This setup closely parallels the Dirac-type seesaw framework introduced in ref.~\cite{Gu:2006dc} and the symmetry-suppressed neutrino mass generation mechanism discussed in ref.~\cite{Davidson:2009ha}.

\section{One-loop matching}
\label{sec:one-loop}

\subsection{Generic results}

As explicitly shown in refs.~\cite{Cohen:2020fcu,Fuentes-Martin:2024agf}, the one-loop EFT Lagrangian can be derived directly from the UV Lagrangian in eq.~\eqref{eq:v2HDM} using the functional approach, without the need to first construct an EFT operator basis or to compute the low-energy amplitudes. The one-loop matching relation can be written schematically as~\cite{Cohen:2020fcu,Fuentes-Martin:2024agf}
\begin{align} \label{eq:matching-master}
\int d^dx \cdot \mathcal{L}_{\nu{\rm SMEFT}}^{\rm 1-loop} =&\; \frac{i}{2}\,{\rm STr} \log\left(P^2 - m_{22}^2\right) \notag \\[0.8mm]
& - \frac{i}{2} \sum_{i_1,\cdots,i_n} \frac{1}{r}\,{\rm STr} \left[\frac{1}{K_{i_1}} X_{i_1i_2} \frac{1}{K_{i_2}} X_{i_2i_3} \cdots \frac{1}{K_{i_n}} X_{i_ni_1}\right]_{\rm hard},
\end{align}
where $P_\mu \equiv iD_\mu$ denotes the hermitian form of the covariant derivative appearing in the \textit{log-type} supertrace. The factor $1/r$ is a symmetry factor associated with a possible $Z_r$ symmetry of the terms in the \textit{power-type} supertrace, and the subscript ``hard'' indicates that only the hard-region contribution of the loop integrals is retained.\footnote{The method of expansion by regions~\cite{Beneke:1997zp,Jantzen:2011nz} provides a framework for decomposing loop integrals in DR into distinct momentum regions, which are determined by the singularity structure of the internal propagators. As demonstrated in refs.~\cite{Fuentes-Martin:2016uol,Zhang:2016pja,Fuentes-Martin:2024agf}, the loop contributions in the EFT cancel against the contributions from the soft region of the loops in the full UV theory. This cancellation allows for a particularly straightforward extraction of the effective action, which can be obtained directly from the hard-region part of the UV quantum effective action.} Consequently, the one-loop matching computation reduces to evaluating the two kinds of functional supertraces in eq.~\eqref{eq:matching-master}, and the resulting one-loop $\nu$SMEFT Lagrangian can be written as
\begin{equation}
    \mathcal{L}_{\nu{\rm SMEFT}}^{\mathrm{1-loop}} = \mathcal{L}_{\nu{\rm SMEFT}}^{\mathrm{1-loop, log-type}} + \mathcal{L}_{\nu{\rm SMEFT}}^{\mathrm{1-loop, power-type}}.
\end{equation}
In this work, we adopt the naive dimensional regularization (NDR) scheme with an anti-commuting $\gamma_5$ (see, e.g., ref.~\cite{Belusca-Maito:2023wah} for a recent review) and the Dirac trace convention
\begin{equation}
{\rm tr}\left(\gamma_\mu \gamma_\nu \gamma_\rho \gamma_\sigma \gamma_5\right) = - 4i \epsilon_{\mu\nu\rho\sigma},
\end{equation}
for algebraic manipulations in $d$-dimensional spacetime, with $\epsilon_{0123}=-1$. The UV divergences arising from the loop integrals always appear in the combination $\tfrac{1}{\widehat{\epsilon}} \equiv \tfrac{1}{\epsilon}-\gamma_E+\ln 4\pi$ and will be removed using the modified minimal subtraction ($\overline{\rm MS}$) scheme. In addition, we make use of the background field method~\cite{Abbott:1980hw,Abbott:1981ke,Thomsen:2024abg} and adopt the 't Hooft–Feynman gauge with $\xi_V = 1$ for the gauge vector fields when performing the functional derivations.

Within the NDR scheme, the log-type supertrace in eq.~\eqref{eq:matching-master} is evaluated as~\cite{Cohen:2020fcu}
\begin{align}
  \mathcal{L}_{\nu{\rm SMEFT}}^{\mathrm{1-loop, log-type}} = &\; \frac{1}{16\pi^2} \Bigg\{- \left(\frac{1}{\widehat{\epsilon}} + \log\frac{\mu^2}{m_{22}^2}\right) \frac{1}{24} \left(4g_1^2 Y_H^2 B_{\mu\nu} B_{\mu\nu} + g_2^2 W_{\mu\nu}^I W_{\mu\nu}^I\right) \notag\\[0.8mm]
  &- \frac{1}{120m_{22}^2} \left(4g_1^2 Y_H^2 \mathcal{O}_{BD}^{(6)} + g_2^2 \mathcal{O}_{WD}^{(6)}\right) + \frac{1}{360m_{22}^2} g_2^3 \mathcal{Q}_{W}^{(6)} \Bigg\}, \label{eq:1-loop_log}
\end{align}
where the Green's operators $\mathcal{O}_{BD,WD}^{(6)}$ are defined in appendix~\ref{appB:Green_basis}, and the Warsaw-extended operator $\mathcal{Q}_{W}^{(6)}$ in table~\ref{tab:Warsaw_1}. The UV divergence proportional to $1/\widehat{\epsilon}$ is canceled out by the $\overline{\rm MS}$ counterterms.

For the power-type supertraces in eq.~\eqref{eq:matching-master}, their explicit analytic forms can be obtained through the following two steps~\cite{Cohen:2020fcu,Fuentes-Martin:2024agf}:
\begin{itemize}
   \item Step~1: we use the Mathematica packages {\sf STrEAM}~\cite{Cohen:2020qvb} and {\sf SuperTracer}~\cite{Fuentes-Martin:2020udw} to compute these power-type supertraces in a generic way, i.e., without fixing the explicit structure of the interaction matrices $X_{ij}$.\footnote{We have partially cross-checked the power-type supertraces by hand against the results from these Mathematica packages, employing the covariant derivative expansion methods~\cite{Cheyette:1985ue,Gaillard:1985uh,Chan:1986jq,Henning:2014wua,Cohen:2019btp}. In doing so, we use the substitutions
   \begin{equation}
      {\rm STr} \to \int\frac{d^dq}{\left(2\pi\right)^d} {\rm tr}, \qquad P^\mu \to \widetilde{P}^\mu \equiv - q^\mu + \widetilde{G}^{\mu\nu} \frac{\partial}{\partial q^\nu}, \qquad 
      X \to \widetilde{X},
   \end{equation}
   where
   \begin{align}
      \widetilde{G}^{\mu\nu} &\equiv \sum_{n=0}^\infty \frac{n+1}{\left(n+2\right)!}\left[P^{\alpha_1}, \cdots \left[P^{\alpha_n},\left[D^\mu,D^\nu\right]\right]\right] \frac{\partial^n}{\partial q^{\alpha_1} \cdots \partial q^{\alpha_n}}, \\[0.8mm]
      \widetilde{X} &\equiv \sum_{n=0}^\infty \frac{1}{n!} \left[P^{\alpha_1}, \cdots \left[P^{\alpha_n},X\right]\right] \frac{\partial^n}{\partial q^{\alpha_1} \cdots \partial q^{\alpha_n}}.
   \end{align}}

\item Step~2: we insert the specific $\nu$2HDM interaction matrices $X_{ij}$, listed in appendix~\ref{appC:Inter_Matrix}, into the general power-type supertrace formulae obtained in Step 1, and then perform the remaining traces.
\end{itemize}
Proceeding in this way yields the full set of (generally redundant) operators that appear in the one-loop EFT Lagrangian, together with their WCs. This one-loop EFT Lagrangian, however, features many redundant operators as well as finite corrections to the kinetic terms. To bring the kinetic terms to canonical form, we perform the following field redefinitions for both fermions and gauge fields:
\begin{align}
Q_{Lp} &\to Q_{Lp} - \frac{1}{128\pi^2} \left(Y_U^2 Y_U^{2\dagger} + Y_D^2 Y_D^{2\dagger}\right)_{pr} Q_{Lr}, \\[0.8mm]
U_{Rp} &\to U_{Rp} - \frac{1}{64\pi^2} \left(Y_U^{2\dagger} Y_U^2\right)_{pr} U_{Rr}, \qquad 
D_{Rp} \to D_{Rp} - \frac{1}{64\pi^2} \left(Y_D^{2\dagger} Y_D^2\right)_{pr} D_{Rr}, \\[0.8mm]
L_{Lp} &\to L_{Lp} - \frac{1}{128\pi^2} \left(Y_\nu^2 Y_\nu^{2\dagger} + Y_E^2 Y_E^{2\dagger}\right)_{pr} L_{Lr}, \\[0.8mm]
\nu_{Rp} &\to \nu_{Rp} - \frac{1}{64\pi^2} \left(Y_\nu^{2\dagger} Y_\nu^2\right)_{pr} \nu_{Rr}, \,\qquad\, 
E_{Rp} \to E_{Rp} - \frac{1}{64\pi^2} \left(Y_E^{2\dagger} Y_E^2\right)_{pr} E_{Rr}, 
\end{align}
and
\begin{equation}
B_\mu \to \left(1 - \frac{1}{2} \delta Z_B\right) B_\mu, \qquad 
W_\mu^I \to \left(1 - \frac{1}{2} \delta Z_W\right) W_\mu^I,
\end{equation}
with
\begin{align}
\delta Z_B \equiv \frac{4g_1^2Y_H^2}{96\pi^2} \log \frac{\mu^2}{m_{22}^2}, \qquad 
\delta Z_W \equiv \frac{g_2^2}{96\pi^2} \log \frac{\mu^2}{m_{22}^2},
\end{align}
accompanied by the corresponding gauge-coupling rescaling, $g_{1/2} \to \left(1 + \tfrac{1}{2}\,\delta Z_{B/W}\right) g_{1/2}$. After these manipulations and retaining only operators up to mass dimension six, the resulting one-loop EFT Lagrangian is expressed in an IBP-independent Green's basis as
\begin{align}
\mathcal{L}_{\nu{\rm SMEFT}}^{\mathrm{1-loop, power-type}} =&\; C_{H_1}^{(2)\prime} \mathcal{Q}_{H_1}^{(2)} + C_{H_1}^{(4)\prime} \mathcal{Q}_{H_1}^{(4)} + \sum_{i = qu,qd}^{\ell\nu,\ell e} \left(C_{i}^{\mathcal{Q},(4)\prime} \mathcal{Q}_{i}^{(4)} + \text{h.c.}\right) \notag\\[0.8mm]
&+ \sum_{i} \left(C_i^{\mathcal{Q},(6)\prime} \mathcal{Q}_i^{(6)} + C_i^{\mathcal{O},(6)} \mathcal{O}_i^{(6)}\right),
\end{align}
where $\mathcal{Q}_{H_1}^{(2)} = H_1^\dagger H_1$, and the UV-finite WCs in the $\overline{\rm MS}$ scheme are evaluated at the matching scale $\mu = m_{22}$. Explicitly, we have
\begin{align}
C_{H_1}^{(2)\prime} &= \frac{m_{22}^2}{16\pi^2} \left(2\lambda_3 + \lambda_4\right), \\[0.8mm]
C_{H_1}^{(4)\prime} &= \frac{3}{16\pi^2} \left(2\left|\lambda_6\right|^2 - \lambda_6 \lambda_7^\ast - \lambda_6^\ast \lambda_7\right), \\[0.8mm]
C_{\underset{pr}{\ell e}}^{\mathcal{Q},(4)\prime} &= - \frac{3\lambda_7^\ast}{16\pi^2} \left(Y_E^2\right)_{pr} + \frac{1}{128\pi^2} \left(2Y_E^1 Y_E^{2\dagger} Y_E^2 + Y_E^2 Y_E^{2\dagger} Y_E^1 + Y_\nu^2 Y_\nu^{2\dagger} Y_E^1 - 8 Y_\nu^2 Y_\nu^{1\dagger} Y_E^2\right)_{pr}, \\[0.8mm]
C_{\underset{pr}{\ell\nu}}^{\mathcal{Q},(4)\prime} &= - \frac{3\lambda_7}{16\pi^2} \left(Y_\nu^2\right)_{pr} + \frac{1}{128\pi^2} \left(2Y_\nu^1 Y_\nu^{2\dagger} Y_\nu^2 + Y_\nu^2 Y_\nu^{2\dagger} Y_\nu^1 + Y_E^2 Y_E^{2\dagger} Y_\nu^1 - 8 Y_E^2 Y_E^{1\dagger} Y_\nu^2\right)_{pr}, \\[0.8mm]
C_{\underset{pr}{qd}}^{\mathcal{Q},(4)\prime} &= - \frac{3\lambda_7^\ast}{16\pi^2} \left(Y_D^2\right)_{pr} + \frac{1}{128\pi^2} \left(2Y_D^1 Y_D^{2\dagger} Y_D^2 + Y_D^2 Y_D^{2\dagger} Y_D^1 + Y_U^2 Y_U^{2\dagger} Y_D^1 - 8 Y_U^2 Y_U^{1\dagger} Y_D^2\right)_{pr}, \\[0.8mm]
C_{\underset{pr}{qu}}^{\mathcal{Q},(4)\prime} &= - \frac{3\lambda_7}{16\pi^2} \left(Y_U^2\right)_{pr} + \frac{1}{128\pi^2} \left(2Y_U^1 Y_U^{2\dagger} Y_U^2 + Y_U^2 Y_U^{2\dagger} Y_U^1 + Y_D^2 Y_D^{2\dagger} Y_U^1 - 8 Y_D^2 Y_D^{1\dagger} Y_U^2\right)_{pr},
\end{align}
and the remaining WCs are available on \href{https://github.com/dufewe/v2HDM-to-vSMEFT}{GitHub} as ancillary files.

To map the above results onto the $\nu$SMEFT Lagrangian in eq.~\eqref{eq:vSMEFT}, we make use of the operator-basis transformation relations summarized in appendix~\ref{appB:Green_basis}. The final analytic forms of the corresponding WCs in the Warsaw-extended basis are provided as ancillary files on \href{https://github.com/dufewe/v2HDM-to-vSMEFT}{GitHub}. These files contain the full UV-finite contributions obtained after integrating out the heavy scalar doublet $H_2$ at both tree and one-loop level. We have validated our results by comparing them with those produced by the automated tools {\sf Matchmakereft}~\cite{Carmona:2021xtq} and {\sf Matchete}~\cite{Fuentes-Martin:2022jrf} and find great agreement, including the treatment of evanescent reductions. Strictly speaking, the results depend on the chosen evanescent scheme, since one can always redefine evanescent operators by adding $\mathcal{O}(\epsilon)$ terms, thereby moving to a different evanescent basis. Nonetheless, this freedom in redefining the evanescent basis cannot change any physical observables at either the high or low scales. The only distinction between two such schemes lies in their RG evolution from high to low scales. Even a small $\mathcal{O}(\epsilon)$ shift between two evanescent bases can induce sizable changes in the mixing into physical operators. In this work, we adopt the evanescent basis defined in eqs.~\eqref{eq:evan_beg}--\eqref{eq:evan_end}, and the RG running effects will be detailed when performing the thorough phenomenological analysis in the forthcoming publication.

\subsection{Specific examples}

As two specific examples, we show here parts of the matching results that are related to the BFB conditions~\cite{Branco:2011iw,Maniatis:2006fs} and the evanescent operators~\cite{Dugan:1990df,Herrlich:1994kh,Buras:1989xd,Fuentes-Martin:2022vvu,Aebischer:2022tvz,Aebischer:2022aze}.

\subsubsection{BFB conditions}

The requirement that the scalar potential in eq.~\eqref{eq:v2HDM_Higgs} be stable implies that it must be BFB, i.e., there must be no direction in field space along which the potential tends to negative infinity. This, in turn, guaranties the existence of a non-trivial minimum of the potential, in the sense that the field $H_1$ develops a non-zero vev. The BFB conditions constrain the parameters of the scalar potential by demanding that the part $V_n$ of the scalar potential, where $n$ is the highest mass dimension of the scalar operators, remain positive as the scalar fields become arbitrarily large. For a renormalizable theory, stability is usually imposed in a marginal sense~\cite{Branco:2011iw,Maniatis:2006fs}, i.e., $V_4 \ge 0$ for all $H_i \to \infty$ and $V_2 \ge 0$ whenever $V_4 = 0$. Under this marginal-stability requirement, the quartic couplings $\lambda_i$ in eq.~\eqref{eq:v2HDM_Higgs} are strongly restricted. As a simple example, the scalar potential in eq.~\eqref{eq:nuSM_Higgs_potential} satisfies the BFB condition through the trivial requirement $\lambda_1 > 0$. In contrast, the BFB conditions for the $\nu$2HDM scalar potential are more involved. As shown in refs.~\cite{Maniatis:2006fs,Tooby-Smith:2026kzj}, the necessary and sufficient conditions for the quartic potential $V_4$ to remain positive along all field directions are
$\lambda_1 \ge 0$, $\lambda_2 \ge 0$, $\lambda_3 \ge - \sqrt{\lambda_1 \lambda_2}$, and $\lambda_3 + \lambda_4 - \left|\lambda_5\right| \ge - \sqrt{\lambda_1 \lambda_2}$, with $\lambda_{6,7} = 0$. These constraints can also be obtained via algebraic techniques, for example, by using copositivity of the scalar mixing matrices~\cite{Kannike:2012pe}.

However, for a {\it high-scale} $\nu$2HDM matched onto the $\nu$SMEFT, the BFB conditions do not directly restrict the UV-complete theory, since the $\nu$2HDM is defined above the heavy scale $m_{22}$ and does not trigger any non-trivial vacuum via SSB. By contrast, at scales $\mu \in \left[\Lambda_{\rm EW}, m_{22}\right]$, where $\Lambda_{\rm EW}$ denotes the electroweak scale, the stability requirements must be enforced on the $\nu$SMEFT potential. Up to mass dimension six, this leads to $V_6 \ge 0$ for $H_1 \to \infty$ and $V_4 \ge 0$ when $V_6 = 0$, where $V_4$ and $V_6$ denote, respectively, the quartic and sextic terms of the $\nu$SMEFT Higgs potential, which at $\mu = m_{22}$ is given by
\begin{equation}
V_{\nu{\rm SMEFT}}^{\rm Higgs} = - \mu^2 H_1^\dagger H_1 + \frac{\lambda}{2} \left(H_1^\dagger H_1\right)^2 - C_{H_1}^{(6)} \mathcal{Q}_{H_1}^{(6)},
\end{equation}
with
\begin{align}
\mu^2 &= - m_{11}^2 + \frac{m_{22}^2}{16\pi^2} \left(2\lambda_3 + \lambda_4\right) + \frac{\left|m_{12}\right|^4}{m_{22}^2}, \label{eq:vSMEFT_mu2} \\[1.5mm]
\lambda &= \lambda_1 - \frac{3}{8\pi^2} \left(2\left|\lambda_6\right|^2 - \lambda_6 \lambda_7^\ast - \lambda_6^\ast \lambda_7\right) + \frac{2m_{12}^2 \lambda_6^\ast + 2m_{12}^{\ast2} \lambda_6}{m_{22}^2}, \\[1.2mm]
C_{H_1}^{(6)} &= \frac{\left|\lambda_6\right|^2}{m_{22}^2} + \frac{1}{16\pi^2 m_{22}^2} \bigg[\lambda_5 \left(-7\lambda_6^{\ast2} + 3 \lambda_6^\ast \lambda_7^\ast\right) + 3 \left(-\lambda_1 + \lambda_3 + \lambda_4\right) \lambda_6 \lambda_7^\ast + h.c.\bigg] \notag\\[0.8mm]
&- \frac{g_2^4 \lambda_1}{1920\pi^2 m_{22}^2} + \frac{1}{96\pi^2 m_{22}^2} \bigg[\lambda_1 \left(\lambda_4^2 + \left|\lambda_5\right|^2 + 102 \left|\lambda_6\right|^2\right) - 2\lambda_3^3 - 3\lambda_3^2 \lambda_4 \notag\\[1.0mm]
&- 3\lambda_3 \lambda_4^2 - \lambda_4^3 - 3 \left(\lambda_3 + \lambda_4\right) \left|\lambda_5\right|^2 + 6 \left(3\lambda_2 - 12 \lambda_3 - 13 \lambda_4\right) \left|\lambda_6\right|^2\bigg].
\end{align}
Consequently, the BFB conditions demand $C_{H_1}^{(6)} \le 0$ for $\mu \in \left[\Lambda_{\rm EW}, m_{22}\right]$. One can immediately see that this inequality is generically violated at tree level unless we work in the special limit $\lambda_6 = 0$. Setting $\lambda_6 = 0$ and neglecting one-loop corrections reduces the BFB conditions to $\lambda = \lambda_1 \ge 0$, coinciding with the SM requirement.

\subsubsection{Evanescent operators}

Matching calculations beyond tree level are most commonly performed in DR due to its simplicity and convenience. Nevertheless, loop-level computations in this scheme must account for the presence of evanescent operators~\cite{Fuentes-Martin:2022vvu,Aebischer:2022tvz,Aebischer:2022aze}, because the four-dimensional operator bases for the EFT do not span the operator space in $d$ dimensions. These evanescent operators are formally of $\mathcal{O}(\epsilon)$ and therefore do not contribute to tree-level amplitudes. However, when inserted into divergent loop amplitudes, they yield additional finite contributions that have to be incorporated beyond leading-order computations.

As illustrated in eq.~\eqref{eq:vSMEFT}, we present our final matching results in terms of the WCs of the evanescent operators introduced in appendix~\ref{appA:vSMEFT_basis}. These evanescent operators are defined as the differences between the redundant Green's operators and their four-dimensional Fierz rearrangements or their Greek reductions. Consequently, they do not affect the $S$-matrix elements of physical observables in the limit $d\to4$. At one-loop and higher orders, however, they induce finite, non-vanishing contributions, which have been examined in detail in refs.~\cite{Fuentes-Martin:2022vvu,Aebischer:2022tvz,Aebischer:2022aze}. In this work, all the non-zero evanescent WCs are listed below.
\begin{itemize}
\item for $\left(\overline{L}R\right) \left(\overline{R}L\right)$ operators:
\begin{align}
C_{\underset{prst}{\ell\nu\nu\ell}}^{E,S,(6)} &= \frac{\left(Y_\nu^2\right)_{pr} \left(Y_\nu^{2\dagger}\right)_{st}}{m_{22}^2} + \frac{4g_1^2\left(Y_H + 2Y_{L_L}\right)^2 + 3g_2^2}{128\pi^2m_{22}^2} \left(1+2a_{\rm ev}\right) \left(Y_\nu^2\right)_{pr} \left(Y_\nu^{2\dagger}\right)_{st} \notag\\[0.8mm]
&+ \frac{3}{16\pi^2 m_{22}^2} \Bigg[\lambda_2 \left(Y_\nu^2\right)_{pr} \left(Y_\nu^{2\dagger}\right)_{st} -
\lambda_7^\ast \left(Y_\nu^1\right)_{pr} \left(Y_\nu^{2\dagger}\right)_{st} - \lambda_7 \left(Y_\nu^2\right)_{pr} \left(Y_\nu^{1\dagger}\right)_{st}\Bigg] \notag\\[0.8mm]
&+ \frac{3}{64\pi^2 m_{22}^2} \Bigg[\left(Y_E^2 Y_E^{1\dagger} Y_\nu^2\right)_{pr} \left(Y_\nu^{1\dagger}\right)_{st} + \left(Y_\nu^1\right)_{pr} \left(Y_\nu^{2\dagger} Y_E^1 Y_E^{2\dagger}\right)_{st}\Bigg] \notag\\[0.8mm]
&+ \frac{1}{128\pi^2 m_{22}^2} \Bigg[7 \left(Y_E^2 Y_E^{2\dagger} Y_\nu^2\right)_{pr} \left(Y_\nu^{2\dagger}\right)_{st} + 7 \left(Y_\nu^2\right)_{pr} \left(Y_\nu^{2\dagger} Y_E^2 Y_E^{2\dagger}\right)_{st} \notag\\[0.8mm]
&- 3 \left(Y_\nu^2 Y_\nu^{2\dagger} Y_\nu^2\right)_{pr} \left(Y_\nu^{2\dagger}\right)_{st} - 3 \left(Y_\nu^2\right)_{pr} \left(Y_\nu^{2\dagger} Y_\nu^2 Y_\nu^{2\dagger}\right)_{st} \Bigg], \\[0.8mm]
C_{\underset{prst}{\ell ee\ell}}^{E,S,(6)} &= \frac{\left(Y_E^2\right)_{pr} \left(Y_E^{2\dagger}\right)_{st}}{m_{22}^2} + \frac{4g_1^2\left(Y_H - 2Y_{L_L}\right)^2 + 3g_2^2}{128\pi^2m_{22}^2} \left(1+2a_{\rm ev}\right) \left(Y_E^2\right)_{pr} \left(Y_E^{2\dagger}\right)_{st} \notag\\[0.8mm]
&+ \frac{3}{16\pi^2 m_{22}^2} \Bigg[\lambda_2 \left(Y_E^2\right)_{pr} \left(Y_E^{2\dagger}\right)_{st} - \lambda_7 \left(Y_E^1\right)_{pr} \left(Y_E^{2\dagger}\right)_{st} - \lambda_7^\ast \left(Y_E^2\right)_{pr} \left(Y_E^{1\dagger}\right)_{st} \Bigg] \notag\\[0.8mm]
&+ \frac{3}{64\pi^2 m_{22}^2} \Bigg[\left(Y_\nu^2 Y_\nu^{1\dagger} Y_E^2\right)_{pr} \left(Y_E^{1\dagger}\right)_{st} + \left(Y_E^1\right)_{pr} \left(Y_E^{2\dagger} Y_\nu^1 Y_\nu^{2\dagger}\right)_{st} \Bigg] \notag\\[0.8mm]
&+ \frac{1}{128\pi^2 m_{22}^2} \Bigg[7 \left(Y_\nu^2 Y_\nu^{2\dagger} Y_E^2\right)_{pr} \left(Y_E^{2\dagger}\right)_{st} + 7 \left(Y_E^2\right)_{pr} \left(Y_E^{2\dagger} Y_\nu^2 Y_\nu^{2\dagger}\right)_{st} \notag\\[0.8mm]
&- 3 \left(Y_E^2 Y_E^{2\dagger} Y_E^2\right)_{pr} \left(Y_E^{2\dagger}\right)_{st} - 3 \left(Y_E^2\right)_{pr} \left(Y_E^{2\dagger} Y_E^2 Y_E^{2\dagger}\right)_{st} \Bigg], \\[0.8mm]
C_{\underset{prst}{quuq}}^{E,S,(6)} &= \frac{\left(Y_U^2\right)_{pr} \left(Y_U^{2\dagger}\right)_{st}}{m_{22}^2} + \frac{12g_1^2 \left(Y_H + 2 Y_{Q_L}\right)^2 + 9g_2^2 - 8g_s^2}{384\pi^2m_{22}^2} \left(1+2a_{\rm ev}\right) \left(Y_U^2\right)_{pr} \left(Y_U^{2\dagger}\right)_{st} \notag\\[0.8mm]
&+ \frac{3}{16\pi^2 m_{22}^2} \Bigg[\lambda_2 \left(Y_U^2\right)_{pr} \left(Y_U^{2\dagger}\right)_{st} - \lambda_7 \left(Y_U^2\right)_{pr} \left(Y_U^{1\dagger}\right)_{st} - \lambda_7^\ast \left(Y_U^1\right)_{pr} \left(Y_U^{2\dagger}\right)_{st} \Bigg] \notag\\[0.8mm]
&+ \frac{3}{64\pi^2 m_{22}^2} \Bigg[\left(Y_D^2 Y_D^{1\dagger} Y_U^2\right)_{pr} \left(Y_U^{1\dagger}\right)_{st} + \left(Y_U^1\right)_{pr} \left(Y_U^{2\dagger} Y_D^1 Y_D^{2\dagger}\right)_{st}\Bigg] \notag\\[0.8mm]
&+ \frac{1}{128\pi^2 m_{22}^2} \Bigg[7 \left(Y_D^2 Y_D^{2\dagger} Y_U^2\right)_{pr} \left(Y_U^{2\dagger}\right)_{st} + 7 \left(Y_U^2\right)_{pr} \left(Y_U^{2\dagger} Y_D^2 Y_D^{2\dagger}\right)_{st} \notag\\[0.8mm]
&- 3 \left(Y_U^2 Y_U^{2\dagger} Y_U^2\right)_{pr} \left(Y_U^{2\dagger}\right)_{st} - 3 \left(Y_U^2\right)_{pr} \left(Y_U^{2\dagger} Y_U^2 Y_U^{2\dagger}\right)_{st} \Bigg], \\[0.8mm]
C_{\underset{prst}{qddq}}^{E,S,(6)} &= \frac{\left(Y_D^2\right)_{pr} \left(Y_D^{2\dagger}\right)_{st}}{m_{22}^2} + \frac{12g_1^2 \left(Y_H - 2 Y_{Q_L}\right)^2 + 9g_2^2 - 8g_s^2}{384\pi^2 m_{22}^2} \left(1+2a_{\rm ev}\right) \left(Y_D^2\right)_{pr} \left(Y_D^{2\dagger}\right)_{st} \notag\\[0.8mm]
&+ \frac{3}{16\pi^2 m_{22}^2} \Bigg[\lambda_2 \left(Y_D^2\right)_{pr} \left(Y_D^{2\dagger}\right)_{st} - \lambda_7 \left(Y_D^1\right)_{pr} \left(Y_D^{2\dagger}\right)_{st} - \lambda_7^\ast \left(Y_D^2\right)_{pr} \left(Y_D^{1\dagger}\right)_{st}\Bigg] \notag\\[0.8mm]
&+ \frac{3}{64\pi^2 m_{22}^2} \Bigg[\left(Y_U^2 Y_U^{1\dagger} Y_D^2\right)_{pr} \left(Y_D^{1\dagger}\right)_{st} + \left(Y_D^1\right)_{pr} \left(Y_D^{2\dagger} Y_U^1 Y_U^{2\dagger}\right)_{st}\Bigg] \notag\\[0.8mm]
&+ \frac{1}{128\pi^2 m_{22}^2} \Bigg[7 \left(Y_U^2 Y_U^{2\dagger} Y_D^2\right)_{pr} \left(Y_D^{2\dagger}\right)_{st} + 7 \left(Y_D^2\right)_{pr} \left(Y_D^{2\dagger} Y_U^2 Y_U^{2\dagger}\right)_{st} \notag\\[0.8mm]
&- 3 \left(Y_D^2 Y_D^{2\dagger} Y_D^2\right)_{pr} \left(Y_D^{2\dagger}\right)_{st} - 3 \left(Y_D^2\right)_{pr} \left(Y_D^{2\dagger} Y_D^2 Y_D^{2\dagger}\right)_{st} \Bigg], \\[0.8mm]
C_{\underset{prst}{\ell\nu\nu\ell}}^{E,T,(6)} &= \frac{3\left(4g_1^2 \left(Y_{L_L} + Y_{\nu_R}\right)^2 + 3g_2^2\right)}{512\pi^2 m_{22}^2} \left(Y_{\nu}^2\right)_{pr} \left(Y_{\nu}^{2\dagger}\right)_{st}, \\[0.8mm]
C_{\underset{prst}{\ell ee\ell}}^{E,T,(6)} &= \frac{3\left(4g_1^2 \left(Y_{L_L} + Y_{E_R}\right)^2 + 3g_2^2\right)}{512\pi^2 m_{22}^2} \left(Y_E^2\right)_{pr} \left(Y_E^{2\dagger}\right)_{st}, \\[0.8mm]
C_{\underset{prst}{quuq}}^{E,T,(6)} &= \frac{12g_1^2 \left(Y_{Q_L} + Y_{U_R}\right)^2 + 9g_2^2 - 8g_s^2}{512\pi^2 m_{22}^2} \left(Y_U^2\right)_{pr} \left(Y_U^{2\dagger}\right)_{st}, \\[0.8mm]
C_{\underset{prst}{qddq}}^{E,T,(6)} &= \frac{12g_1^2 \left(Y_{Q_L} + Y_{D_R}\right)^2 + 9g_2^2 - 8g_s^2}{512\pi^2 m_{22}^2} \left(Y_D^2\right)_{pr} \left(Y_D^{2\dagger}\right)_{st}, \\[0.8mm]
C_{\underset{prst}{\ell\nu uq}}^{E,T,(6)} &= \frac{3\left(4g_1^2 \left(Y_{L_L} + Y_{\nu_R}\right) \left(Y_{Q_L} + Y_{U_R}\right) + 3g_2^2\right)}{512\pi^2 m_{22}^2} \left(Y_{\nu}^2\right)_{pr} \left(Y_U^{2\dagger}\right)_{st}, \\[0.8mm]
C_{\underset{prst}{\ell edq}}^{E,T,(6)} &= \frac{3\left(4g_1^2 \left(Y_{L_L} + Y_{E_R}\right) \left(Y_{Q_L} + Y_{D_R}\right) + 3g_2^2\right)}{512\pi^2 m_{22}^2} \left(Y_E^2\right)_{pr} \left(Y_D^{2\dagger}\right)_{st},
\end{align}
and
\begin{align}
C_{\underset{prst}{quuq}}^{E,T,(8),(6)} &= \frac{3 g_s^2 \left(Y_U^2\right)_{pr} \left(Y_U^{2\dagger}\right)_{st}}{64\pi^2 m_{22}^2},& C_{\underset{prst}{quuq}}^{E,(8),(6)} &= \frac{(1 + 2 a_{\rm ev}) g_s^2}{16\pi^2 m_{22}^2} \left(Y_U^2\right)_{pr} \left(Y_U^{2\dagger}\right)_{st}, \\[0.8mm]
C_{\underset{prst}{qddq}}^{E,T,(8),(6)} &= \frac{3 g_s^2 \left(Y_D^2\right)_{pr} \left(Y_D^{2\dagger}\right)_{st}}{64\pi^2 m_{22}^2},& C_{\underset{prst}{qddq}}^{E,(8),(6)} &= \frac{(1 + 2 a_{\rm ev}) g_s^2}{16\pi^2 m_{22}^2} \left(Y_D^2\right)_{pr} \left(Y_D^{2\dagger}\right)_{st}, 
\end{align}
\item for $\left(\overline{L}L\right) \left(\overline{L}L\right)$ operators:
\begin{align}
C_{\underset{prst}{\ell\ell}}^{E,(3),(6)} &= \frac{3}{128\pi^2 m_{22}^2} \left(Y_{\nu}^1 Y_{\nu}^{2\dagger} + Y_E^2 Y_E^{1\dagger}\right)_{pr} \left(Y_{\nu}^2 Y_{\nu}^{1\dagger} + Y_E^1 Y_E^{2\dagger}\right)_{st} - \frac{g_2^4 \delta_{pr} \delta_{st}}{3840\pi^2 m_{22}^2} \notag\\[0.8mm]
&- \frac{1}{1152\pi^2 m_{22}^2} \left[9\left(Y_{\nu}^2 Y_{\nu}^{2\dagger} + Y_E^2 Y_E^{2\dagger}\right)_{pr} - 2g_2^2 \delta_{pr}\right] \left(Y_{\nu}^2 Y_{\nu}^{2\dagger} + Y_E^2 Y_E^{2\dagger}\right)_{st}, \\[0.8mm]
C_{\underset{prst}{qq}}^{E,(8),(6)} &= -\frac{g_s^2 \delta_{pr}}{72\pi^2 m_{22}^2} \left(Y_U^2 Y_U^{2\dagger} + Y_D^2 Y_D^{2\dagger}\right)_{st},
\end{align}
\item for $\left(\overline{R}R\right) \left(\overline{R}R\right)$ operators:
\begin{align}
C_{\underset{prst}{ud}}^{E,(6)} &= \frac{3}{64\pi^2 m_{22}^2} \left(Y_U^{1\dagger} Y_D^2 + Y_U^{2\dagger} Y_D^1\right)_{pr} \left(Y_D^{1\dagger} Y_U^2 + Y_D^{2\dagger} Y_U^1\right)_{st} \notag\\[0.8mm]
&- \frac{1}{16\pi^2 m_{22}^2} \left(Y_U^{2\dagger} Y_D^2\right)_{pr} \left(Y_D^{2\dagger} Y_U^2\right)_{st}, \\[0.8mm]
C_{\underset{prst}{\nu e}}^{E,(6)} &= \frac{3}{64\pi^2 m_{22}^2} \left(Y_\nu^{1\dagger} Y_E^2 + Y_\nu^{2\dagger} Y_E^1\right)_{pr} \left(Y_E^{1\dagger} Y_\nu^2 + Y_E^{2\dagger} Y_\nu^1\right)_{st} \notag\\[0.8mm]
&- \frac{1}{16\pi^2 m_{22}^2} \left(Y_\nu^{2\dagger} Y_E^2\right)_{pr} \left(Y_E^{2\dagger} Y_\nu^2\right)_{st},
\end{align}
and
\begin{align}
C_{\underset{prst}{uu}}^{E,(8),(6)} = - \frac{g_s^2 \delta_{pr}}{36\pi^2 m_{22}^2} \left(Y_U^{2\dagger} Y_U^2\right)_{st}, \qquad C_{\underset{prst}{dd}}^{E,(8),(6)} = - \frac{g_s^2 \delta_{pr}}{36\pi^2 m_{22}^2} \left(Y_D^{2\dagger} Y_D^2\right)_{st},
\end{align}
\item for $\left(\overline{L}R\right) \left(\overline{L}R\right)$ operators:
\begin{align}
C_{\underset{prst}{quqd}}^{E,(1),(6)} &= \frac{12g_1^2 \left(Y_{Q_L} + Y_{U_R}\right)\left(Y_{Q_L} + Y_{D_R}\right) - 9g_2^2 - 8g_s^2}{384\pi^2 m_{22}^2} \left(Y_U^2\right)_{pr} \left(Y_D^2\right)_{st} \notag\\[0.8mm]
&- \frac{g_s^2}{32\pi^2 m_{22}^2} \left(Y_D^2\right)_{pt} \left(Y_U^2\right)_{sr}, \\[0.8mm]
C_{\underset{prst}{quqd}}^{E,(8),(6)} &= - \frac{12g_1^2 \left(Y_{Q_L} + Y_{U_R}\right)\left(Y_{Q_L} + Y_{D_R}\right) - 9g_2^2 - 8g_s^2}{768\pi^2 m_{22}^2} \left(Y_D^2\right)_{pt} \left(Y_U^2\right)_{sr} \notag\\[0.8mm]
&+ \frac{g_s^2}{16\pi^2m_{22}^2} \left(Y_U^2\right)_{pr} \left(Y_D^2\right)_{st}, \\[0.8mm]
C_{\underset{prst}{\ell\nu\ell e}}^{E,(6)} &= \frac{4g_1^2 \left(Y_{L_L} + Y_{\nu_R}\right)\left(Y_{L_L} + Y_{E_R}\right) - 3g_2^2}{256\pi^2 m_{22}^2} \left(2\left(Y_\nu^2\right)_{pr} \left(Y_E^2\right)_{st} - \left(Y_E^2\right)_{pt} \left(Y_\nu^2\right)_{sr}\right).
\end{align}
\end{itemize}
From the above expressions, we observe that all evanescent WCs, except for $C_{\ell\nu\nu\ell}^{E,S,(6)}$, $C_{\ell ee\ell}^{E,S,(6)}$, $C_{quuq}^{E,S,(6)}$, and $C_{qddq}^{E,S,(6)}$, are generated only at the one-loop level. In practice, the effects of evanescent operators can be equivalently recast as loop-induced contributions to physical operators. Once this is done, the matching relations involve solely physical operators with appropriately shifted WCs~\cite{Fuentes-Martin:2022vvu}. As a result, the RG evolution in this setup is different from that in the original scheme with evanescent operators. Nevertheless, the physical predictions remain unchanged, even though the RG descriptions differ. As concrete examples, the one-loop substitutions for evanescent operators~--~equivalently, the one-loop Fierz transformations~--~are given in ref.~\cite{Fuentes-Martin:2022vvu}. We further generalize these results to the set $\left\{C_{\ell\nu\nu\ell}^{E,S,(6)}, C_{\ell ee\ell}^{E,S,(6)}, C_{quuq}^{E,S,(6)}, C_{qddq}^{E,S,(6)}\right\}$ in appendix~\ref{appD:1-loop_Fierz}, now including the right-handed Dirac neutrinos.

\section{Conclusion}
\label{sec:conc}

In this work, we have analyzed the so-called {\it high-scale} $\nu$2HDM, in which the SM is augmented by three right-handed Dirac neutrinos and a second heavy scalar doublet $H_2$. Making use of the modern functional matching techniques, we have carried out a full tree-level and one-loop matching of this UV completion onto the $\nu$SMEFT up to operators of mass dimension six, by integrating out the heavy scalar doublet prior to SSB. The matching is performed within the NDR and $\overline{\rm MS}$ schemes, which necessitate a careful treatment of the evanescent operators. To translate the matching results, initially obtained in an IBP-independent Green's basis, into a Warsaw-extended basis, we have employed the EOMs of the light d.o.f., and the evanescent-relevant Fierz transformations and Greek reductions.

The resulting WCs for the full $\nu$SMEFT Lagrangian in eq.~\eqref{eq:vSMEFT} are expressed in terms of the original $\nu$2HDM parameters and are provided as ancillary files on \href{https://github.com/dufewe/v2HDM-to-vSMEFT}{GitHub}. This lays the groundwork for systematic and global investigations of the $\nu$2HDM parameter space using EFT techniques. In particular, since these WCs are defined at the matching scale $\mu = m_{22}$, one must use the RG evolutions of the $\nu$SMEFT and $\nu$LEFT operators to explore the low-energy phenomena of the $\nu$2HDM. A comprehensive phenomenological study in such a setup, together with a comparison to the results obtained in the full $\nu$2HDM, will be presented in a forthcoming publication.

\acknowledgments
The authors are grateful to Javier Fuentes-Mart{\'\i}n, Matthias K{\"o}nig, Julie Pag{\`e}s, Anders Eller Thomsen, and Felix Wilsch for the inspiring correspondences about evanescent operators. This work is supported by the National Natural Science Foundation of China under Grant Nos.~12635005, 12475094, 12135006, and 12575099, the Fundamental Research Funds for the Central Universities under Grand No.~XJ2026000201, as well as the Science and Technology Innovation Leading Talent Support Program of Henan Province under Grant No.~254000510039.

\appendix

\section{Warsaw-extended basis of the \texorpdfstring{$\boldsymbol{\nu}$SMEFT}{vSMEFT}}
\label{appA:vSMEFT_basis}

\begin{table}[t] 
\renewcommand\arraystretch{1.5}
\setlength{\tabcolsep}{1pt}
\centering
\resizebox{\textwidth}{!}{
\begin{tabular}{||c|c||c|c||c|c||}
\hline \hline
\multicolumn{2}{||c||}{$X^3$} & \multicolumn{2}{|c||}{$H_1^6$ ~\&~ $H_1^4 D^2$} & \multicolumn{2}{|c||}{$\psi^2H_1^3$} \\
\hline
$\mathcal{Q}_G^{(6)}$ & $f^{ABC} G_\mu^{A\nu} G_\nu^{B\rho} G_\rho^{C\mu}$ & $\mathcal{Q}_{H_1}^{(6)}$ & $\left(H_1^\dagger H_1\right)^3$ & $Q_{eH_1}^{(6)}$ & $\left(H_1^\dagger H_1\right) \left(\overline{L}_{Lp} E_{Rr} H_1\right)$ \\
$\mathcal{Q}_{\widetilde{G}}^{(6)}$ & $f^{ABC} \widetilde{G}_\mu^{A\nu} G_\nu^{B\rho} G_\rho^{C\mu} $ &
$\mathcal{Q}_{H_1\Box}^{(6)}$ & $\left(H_1^\dagger H_1\right) \Box \left(H_1^\dagger H_1\right)$ & $\mathcal{Q}_{uH_1}^{(6)}$ & $\left(H_1^\dagger H_1\right) \left(\overline{Q}_{Lp} U_{Rr} \widetilde{H}_1\right)$ \\
$\mathcal{Q}_W^{(6)}$ & $\epsilon^{IJK} W_\mu^{I\nu} W_\nu^{J\rho} W_\rho^{K\mu}$ &
$\mathcal{Q}_{H_1 D}^{(6)}$ & $\left(H_1^\dagger D_\mu H_1\right)^\ast \left(H_1^\dagger D_\mu H_1\right)$ &
$\mathcal{Q}_{dH_1}^{(6)}$ & $\left(H_1^\dagger H_1\right) \left(\overline{Q}_{Lp} D_{Rr} H_1\right)$ \\
$\mathcal{Q}_{\widetilde{W}}^{(6)}$ & $\epsilon^{IJK} \widetilde{W}_\mu^{I\nu} W_\nu^{J\rho} W_\rho^{K\mu}$ &&&& \\
\hline \hline
\multicolumn{2}{||c||}{$X^2H_1^2$} & \multicolumn{2}{|c||}{$\psi^2 XH_1$} & \multicolumn{2}{|c||}{$\psi^2H_1^2D$}\\ 
\hline
$\mathcal{Q}_{H_1G}^{(6)}$ & $H_1^\dagger H_1 G^A_{\mu\nu} G^A_{\mu\nu}$ & $\mathcal{Q}_{eW}^{(6)}$ & $\left(\overline{L}_{Lp} \sigma_{\mu\nu} E_{Rr}\right) \tau^I H_1 W_{\mu\nu}^I$ & $\mathcal{Q}_{H_1 \ell}^{(1),(6)}$ & $\left(H_1^\dagger i\overleftrightarrow{D}_\mu H_1\right) \left(\overline{L}_{Lp} \gamma_\mu L_{Lr}\right)$ \\
$\mathcal{Q}_{H_1 \widetilde{G}}^{(6)}$ & $H_1^\dagger H_1 \widetilde{G}^A_{\mu\nu} G^A_{\mu\nu}$ &
$\mathcal{Q}_{eB}^{(6)}$ & $\left(\overline{L}_{Lp} \sigma_{\mu\nu} E_{Rr}\right) H_1 B_{\mu\nu}$ & $\mathcal{Q}_{H_1 \ell}^{(3),(6)}$ & $\left(H_1^\dagger i\overleftrightarrow{D}^I_\mu H_1\right) \left(\overline{L}_{Lp} \tau^I \gamma_\mu L_{Lr}\right)$ \\
$\mathcal{Q}_{H_1W}^{(6)}$ & $H_1^\dagger H_1 W^I_{\mu\nu} W^I_{\mu\nu}$ & $\mathcal{Q}_{uG}^{(6)}$ & $\left(\overline{Q}_{Lp} \sigma_{\mu\nu} T^A U_{Rr}\right) \widetilde{H}_1 G_{\mu\nu}^A$ & $\mathcal{Q}_{H_1 e}^{(6)}$ & $\left(H_1^\dagger i\overleftrightarrow{D}_\mu H_1\right) \left(\overline{E}_{Rp} \gamma_\mu E_{Rr}\right)$ \\
$\mathcal{Q}_{H_1\widetilde{W}}^{(6)}$ & $H_1^\dagger H_1 \widetilde{W}^I_{\mu\nu} W^I_{\mu\nu}$ & $\mathcal{Q}_{uW}^{(6)}$ & $\left(\overline{Q}_{Lp} \sigma_{\mu\nu} U_{Rr}\right) \tau^I \widetilde{H}_1 W_{\mu\nu}^I$ & $\mathcal{Q}_{H_1q}^{(1),(6)}$ & $\left(H_1^\dagger i\overleftrightarrow{D}_\mu H_1\right) \left(\overline{Q}_{Lp} \gamma^\mu Q_{Lr}\right)$ \\
$\mathcal{Q}_{H_1B}^{(6)}$ & $H_1^\dagger H_1 B_{\mu\nu} B_{\mu\nu}$ & $\mathcal{Q}_{uB}^{(6)}$ & $\left(\overline{Q}_{Lp} \sigma_{\mu\nu} U_{Rr}\right) \widetilde{H}_1 B_{\mu\nu}$ & $\mathcal{Q}_{H_1q}^{(3),(6)}$ & $\left(H_1^\dagger i\overleftrightarrow{D}_\mu^I H_1\right) \left(\overline{Q}_{Lp} \tau^I \gamma_\mu Q_{Lr}\right)$ \\
$\mathcal{Q}_{H_1\widetilde{B}}^{(6)}$ & $H_1^\dagger H_1 \widetilde{B}_{\mu\nu} B_{\mu\nu}$ & $\mathcal{Q}_{dG}^{(6)}$ & $\left(\overline{Q}_{Lp} \sigma_{\mu\nu} T^A D_{Rr}\right) H_1 G_{\mu\nu}^A$ & $\mathcal{Q}_{H_1u}^{(6)}$ & $\left(H_1^\dagger i\overleftrightarrow{D}_\mu H_1\right) \left(\overline{U}_{Rp} \gamma_\mu U_{Rr}\right)$ \\
$\mathcal{Q}_{H_1WB}^{(6)}$ & $H_1^\dagger \tau^I H_1 W^I_{\mu\nu} B_{\mu\nu}$ & $\mathcal{Q}_{dW}^{(6)}$ & $\left(\overline{Q}_{Lp} \sigma_{\mu\nu} D_{Rr}\right) \tau^I H_1 W_{\mu\nu}^I$ & $\mathcal{Q}_{H_1d}^{(6)}$ & $\left(H_1^\dagger i\overleftrightarrow{D}_\mu H_1\right) \left(\overline{D}_{Rp} \gamma_\mu D_{Rr}\right)$ \\
$\mathcal{Q}_{H_1\widetilde{W}B}^{(6)}$ & $H_1^\dagger \tau^I H_1 \widetilde{W}^I_{\mu\nu} B_{\mu\nu}$ & $\mathcal{Q}_{dB}^{(6)}$ & $\left(\overline{Q}_{Lp} \sigma_{\mu\nu} D_{Rr}\right) H_1 B_{\mu\nu}$ & $\mathcal{Q}_{H_1ud}^{(6)}$ & $i\left(\widetilde{H}_1^\dagger \overleftrightarrow{D}_\mu H_1\right) \left(\overline{U}_{Rp} \gamma_\mu D_{Rr}\right)$ \\
\hline \hline
\end{tabular}
}
\caption{Dimension-six bosonic and two-fermion operators in Warsaw basis~\cite{Grzadkowski:2010es}. \label{tab:Warsaw_1}}
\end{table}

\begin{sidewaystable}[htbp]
\renewcommand\arraystretch{1.7}
\setlength{\tabcolsep}{6pt}
\centering
\resizebox{\textwidth}{!}{
\begin{tabular}{||c|c||c|c||c|c||}
\hline\hline
\multicolumn{2}{||c||}{$\left(\overline{L}L\right) \left(\overline{L}L\right)$} & 
\multicolumn{2}{|c||}{$\left(\overline{R}R\right) \left(\overline{R}R\right)$} &
\multicolumn{2}{|c||}{$\left(\overline{L}L\right) \left(\overline{R}R\right)$} \\
\hline
$\mathcal{Q}_{\ell\ell}^{(6)}$ & $\left(\overline{L}_{Lp} \gamma_\mu L_{Lr}\right) \left(\overline{L}_{Ls} \gamma_\mu L_{Lt}\right)$ & $\mathcal{Q}_{ee}^{(6)}$ & $\left(\overline{E}_{Rp} \gamma_\mu E_{Rr}\right) \left(\overline{E}_{Rs} \gamma_\mu E_{Rt}\right)$ & $\mathcal{Q}_{\ell e}^{(6)}$ & $\left(\overline{L}_{Lp} \gamma_\mu L_{Lr}\right) \left(\overline{E}_{Rs} \gamma_\mu E_{Rt}\right)$ \\
$\mathcal{Q}_{qq}^{(1),(6)}$ & $\left(\overline{Q}_{Lp} \gamma_\mu Q_{Lr}\right) \left(\overline{Q}_{Ls} \gamma_\mu Q_{Lt}\right)$ & $\mathcal{Q}_{uu}^{(6)}$ & $\left(\overline{U}_{Rp} \gamma_\mu U_{Rr}\right) \left(\overline{U}_{Rs} \gamma_\mu U_{Rt}\right)$ & $\mathcal{Q}_{\ell u}^{(6)}$ & $\left(\overline{L}_{Lp} \gamma_\mu L_{Lr}\right) \left(\overline{U}_{Rs} \gamma_\mu U_{Rt}\right)$ \\
$\mathcal{Q}_{qq}^{(3),(6)}$ & $\left(\overline{Q}_{Lp} \gamma_\mu \tau^I Q_{Lr}\right) \left(\overline{Q}_{Ls} \gamma_\mu \tau^I Q_{Lt}\right)$ & $\mathcal{Q}_{dd}^{(6)}$ & $\left(\overline{D}_{Rp} \gamma_\mu D_{Rr}\right) \left(\overline{D}_{Rs} \gamma_\mu D_{Rt}\right)$ & $\mathcal{Q}_{\ell d}^{(6)}$ & $\left(\overline{L}_{Lp} \gamma_\mu L_{Lr}\right) \left(\overline{D}_{Rs} \gamma_\mu D_{Rt}\right)$ \\
$\mathcal{Q}_{\ell q}^{(1),(6)}$ & $\left(\overline{L}_{Lp} \gamma_\mu L_{Lr}\right) \left(\overline{Q}_{Ls} \gamma_\mu Q_{Lt}\right)$ & $\mathcal{Q}_{eu}^{(6)}$ & $\left(\overline{E}_{Rp} \gamma_\mu E_{Rr}\right) \left(\overline{U}_{Rs} \gamma_\mu U_{Rt}\right)$ & $\mathcal{Q}_{qe}^{(6)}$ & $\left(\overline{Q}_{Lp} \gamma_\mu Q_{Lr}\right) \left(\overline{E}_{Rs} \gamma_\mu E_{Rt}\right)$ \\
$\mathcal{Q}_{\ell q}^{(3),(6)}$ & $\left(\overline{L}_{Lp} \gamma_\mu \tau^I L_{Lr}\right) \left(\overline{Q}_{Ls} \gamma_\mu \tau^I Q_{Lt}\right)$ & $\mathcal{Q}_{ed}^{(6)}$ & $\left(\overline{E}_{Rp} \gamma_\mu E_{Rr}\right) \left(\overline{D}_{Rs} \gamma_\mu D_{Rt}\right)$ & $\mathcal{Q}_{qu}^{(1),(6)}$ & $\left(\overline{Q}_{Lp} \gamma_\mu Q_{Lr}\right) \left(\overline{U}_{Rs} \gamma_\mu U_{Rt}\right)$ \\ 
&& $\mathcal{Q}_{ud}^{(1),(6)}$ & $\left(\overline{U}_{Rp} \gamma_\mu U_{Rr}\right) \left(\overline{D}_{Rs} \gamma_\mu D_{Rt}\right)$ & $\mathcal{Q}_{qu}^{(8),(6)}$ & $\left(\overline{Q}_{Lp} \gamma_\mu T^A Q_{Lr}\right) \left(\overline{U}_{Rs} \gamma_\mu T^A U_{Rt}\right)$ \\ 
&& $\mathcal{Q}_{ud}^{(8),(6)}$ & $\left(\overline{U}_{Rp} \gamma_\mu T^A U_{Rr}\right) \left(\overline{D}_{Rs} \gamma_\mu T^A D_{Rt}\right)$ & $\mathcal{Q}_{qd}^{(1),(6)}$ & $\left(\overline{Q}_{Lp} \gamma_\mu Q_{Lr}\right) \left(\overline{D}_{Rs} \gamma_\mu D_{Rt}\right)$ \\
&&&& $\mathcal{Q}_{qd}^{(8),(6)}$ & $\left(\overline{Q}_{Lp} \gamma_\mu T^A Q_{Lr}\right) \left(\overline{D}_{Rs} \gamma_\mu T^A D_{Rt}\right)$ \\
\hline\hline
\multicolumn{2}{||c||}{$\left(\overline{L}R\right)\left(\overline{R}L\right)$ ~\&~ $\left(\overline{L}R\right)\left(\overline{L}R\right)$} &
\multicolumn{4}{|c||}{baryon/lepton-number-violating}\\\hline
$\mathcal{Q}_{\ell edq}^{(6)}$ & $\left(\overline{L}_{Lp}^j E_{Rr}\right)\left(\overline{D}_{Rs} Q_{Lt}^j\right)$ & $\mathcal{Q}_{duq\ell}^{(6)}$ & \multicolumn{3}{|c||}{$\epsilon_{\alpha\beta\gamma} \epsilon_{jk} \left(D_{R\alpha p}^T C U_{R\beta r}\right) \left(Q^{jT}_{L\gamma s} C L^k_{Lt}\right)$} \\
$\mathcal{Q}_{quqd}^{(1),(6)}$ & $\left(\overline{Q}_{Lp}^j U_{Rr}\right) \epsilon_{jk} \left(\overline{Q}_{Ls}^k D_{Rt}\right)$ & $\mathcal{Q}_{qque}^{(6)}$ & \multicolumn{3}{|c||}{$\epsilon_{\alpha\beta\gamma} \epsilon_{jk} \left(Q_{L\alpha p}^{jT} C Q_{L\beta r}^k\right) \left(U^{T}_{R\gamma s} C E_{Rt}\right)$} \\
$\mathcal{Q}_{quqd}^{(8),(6)}$ & $\left(\overline{Q}_{Lp}^j T^A U_{Rr}\right) \epsilon_{jk} \left(\overline{Q}_{Ls}^k T^A D_{Rt}\right)$ & $\mathcal{Q}_{qqq\ell}^{(6)}$ & \multicolumn{3}{|c||}{$\epsilon_{\alpha\beta\gamma} \epsilon_{jk} \epsilon_{mn} \left(Q_{L\alpha p}^{jT} C Q_{L\beta r}^m\right) \left(Q^{nT}_{L\gamma s} C L^k_{Lt}\right)$} \\
$\mathcal{Q}_{\ell equ}^{(1),(6)}$ & $\left(\overline{L}_{Lp}^j E_{Rr}\right) \epsilon_{jk} \left(\overline{Q}_{Ls}^k U_{Rt}\right)$ & $\mathcal{Q}_{duue}^{(6)}$ & \multicolumn{3}{|c||}{$\epsilon_{\alpha\beta\gamma} \left(D_{R\alpha p}^T C U_{R\beta r}\right) \left(U^{T}_{R\gamma s} C E_{Rt}\right)$} \\
$\mathcal{Q}_{\ell equ}^{(3),(6)}$ & $\left(\overline{L}_{Lp}^j \sigma_{\mu\nu} E_{Rr}\right) \epsilon_{jk} \left(\overline{Q}_{Ls}^k \sigma_{\mu\nu} U_{Rt}\right)$ && \multicolumn{3}{|c||}{} \\
\hline\hline
\end{tabular}
}
\caption{Dimension-six four-fermion operators in Warsaw basis~\cite{Grzadkowski:2010es}. \label{tab:Warsaw_2}}
\end{sidewaystable}

In this appendix, we augment the SMEFT Warsaw basis~\cite{Grzadkowski:2010es}, summarized in tables~\ref{tab:Warsaw_1} and \ref{tab:Warsaw_2}, by including operators involving the right-handed neutrinos $\nu_{R_i}$~\cite{Aparici:2009fh,delAguila:2008ir,Bhattacharya:2015vja,Liao:2016qyd,Li:2021tsq} as well as the non-trivial one-loop evanescent operators, up to mass dimension six.\footnote{Here and in the following, the hermitian conjugate (h.c.) of the relevant operators is implied when needed.} The extension contains two dimension-five operators~\cite{Aparici:2009fh}
\begin{align}
\mathcal{Q}_{\underset{pr}{\nu H_1}}^{(5)} &\equiv \left(H_1^\dagger H_1\right) \left(\nu_{Rp}^T C \nu_{Rr}\right), 
& \mathcal{Q}_{\underset{pr}{\nu B}}^{(5)} &\equiv \left(\nu_{Rp}^T C \sigma_{\mu\nu} \nu_{Rr}\right) B_{\mu\nu},
\end{align}
together with sixteen dimension-six operators that conserve baryon and lepton numbers:
\begin{itemize}
\item $\psi^2H_1^3$ class:
\begin{align}
\mathcal{Q}_{\underset{pr}{\nu H_1}}^{(6)} \equiv \left(H_1^\dagger H_1\right) \left(\overline{L}_{Lp} \nu_{Rr} \widetilde{H}_1\right),
\end{align}
\item $\psi^2XH_1$ class:
\begin{align}
\mathcal{Q}_{\underset{pr}{\nu W}}^{(6)} &\equiv \left(\overline{L}_{Lp} \sigma_{\mu\nu} \nu_{Rr}\right) \tau^I \widetilde{H}_1 W_{\mu\nu}^I, 
& \mathcal{Q}_{\underset{pr}{\nu B}}^{(6)} &\equiv \left(\overline{L}_{Lp} \sigma_{\mu\nu} \nu_{Rr}\right) \widetilde{H}_1 B_{\mu\nu},
\end{align}
\item $\psi^2H_1^2D$ class:
\begin{align}
\mathcal{Q}_{\underset{pr}{H_1\nu}}^{(6)} &\equiv \left(H_1^\dagger i\overleftrightarrow{D}_\mu H_1\right) \left(\overline{\nu}_{Rp} \gamma_\mu \nu_{Rr}\right), 
& \mathcal{Q}_{\underset{pr}{H_1\nu e}}^{(6)} &\equiv i \left(\widetilde{H}_1^\dagger \overleftrightarrow{D}_\mu H_1\right) \left(\overline{\nu}_{Rp} \gamma_\mu E_{Rr}\right),
\end{align}
\item $\left(\overline{R}R\right)\left(\overline{R}R\right)$ class:
\begin{align}
\mathcal{Q}_{\underset{prst}{\nu\nu}}^{(6)} &\equiv \left(\overline{\nu}_{Rp} \gamma_\mu \nu_{Rr}\right) \left(\overline{\nu}_{Rs} \gamma_\mu \nu_{Rt}\right), 
& \mathcal{Q}_{\underset{prst}{\nu e}}^{(6)} &\equiv \left(\overline{\nu}_{Rp} \gamma_\mu \nu_{Rr}\right) \left(\overline{E}_{Rs} \gamma_\mu E_{Rt}\right), \\[0.8mm]
\mathcal{Q}_{\underset{prst}{\nu u}}^{(6)} &\equiv \left(\overline{\nu}_{Rp} \gamma_\mu \nu_{Rr}\right) \left(\overline{U}_{Rs} \gamma_\mu U_{Rt}\right), 
& \mathcal{Q}_{\underset{prst}{\nu d}}^{(6)} &\equiv \left(\overline{\nu}_{Rp} \gamma_\mu \nu_{Rr}\right) \left(\overline{D}_{Rs} \gamma_\mu D_{Rt}\right), \\[0.8mm]
\mathcal{Q}_{\underset{prst}{\nu edu}}^{(6)} &\equiv \left(\overline{\nu}_{Rp} \gamma_\mu E_{Rr}\right) \left(\overline{D}_{Rs} \gamma_\mu U_{Rt}\right),
\end{align}
\item $\left(\overline{L}L\right) \left(\overline{R}R\right)$ class:
\begin{align}
\mathcal{Q}_{\underset{prst}{\ell \nu}}^{(6)} &\equiv \left(\overline{L}_{Lp} \gamma_\mu L_{Lr}\right) \left(\overline{\nu}_{Rs} \gamma_\mu \nu_{Rt}\right), 
& \mathcal{Q}_{\underset{prst}{q\nu}}^{(6)} &\equiv \left(\overline{Q}_{Lp} \gamma_\mu Q_{Lr}\right) \left(\overline{\nu}_{Rs} \gamma_\mu \nu_{Rt}\right),
\end{align}
\item $\left(\overline{L}R\right) \left(\overline{R}L\right)$ class:
\begin{align}
\mathcal{Q}_{\underset{prst}{\ell \nu uq}}^{(6)} \equiv \left(\overline{L}_{Lp}^j \nu_{Rr}\right) \left(\overline{U}_{Rs} Q_{Lt}^j\right),
\end{align}
\item $\left(\overline{L}R\right) \left(\overline{L}R\right)$ class:
\begin{align}
\mathcal{Q}_{\underset{prst}{\ell\nu\ell e}}^{(6)} &\equiv \left(\overline{L}_{Lp}^j \nu_{Rr}\right) \epsilon_{jk} \left(\overline{L}_{Ls}^k E_{Rt}\right), 
& \mathcal{Q}_{\underset{prst}{\ell\nu qd}}^{(1),(6)} &\equiv \left(\overline{L}_{Lp}^j \nu_{Rr}\right) \epsilon_{jk} \left(\overline{Q}_{Ls}^k D_{Rt}\right), \\[0.8mm]
\mathcal{Q}_{\underset{prst}{\ell\nu qd}}^{(3),(6)} &\equiv \left(\overline{L}_{Lp}^j \sigma_{\mu\nu} \nu_{Rr}\right) \epsilon_{jk} \left(\overline{Q}_{Ls}^k \sigma_{\mu\nu} D_{Rt}\right),
\end{align}
\end{itemize}
as well as three dimension-six operators that violate baryon and/or lepton number:
\begin{align}
\mathcal{Q}_{\underset{prst}{qqd\nu}}^{(6)} &\equiv \epsilon_{\alpha\beta\gamma} \epsilon_{jk} \left(Q_{L\alpha p}^{jT} C Q_{L\beta r}^k\right) \left(D_{R\gamma s}^T C \nu_{Rt}\right), 
& \mathcal{Q}_{\underset{prst}{\nu\nu\nu\nu}}^{(6)} &\equiv \left(\nu_{Rp}^T C \nu_{Rr}\right) \left(\nu_{Rs}^T C \nu_{Rt}\right), \\[0.8mm]
\mathcal{Q}_{\underset{prst}{udd\nu}}^{(6)} &\equiv \epsilon_{\alpha\beta\gamma} \left(U_{R\alpha p}^{T} C D_{R\beta r}\right) \left(D_{R\gamma s}^T C \nu_{Rt}\right).
\end{align}
Here, $\epsilon_{jk}$ and $\epsilon_{\alpha\beta\gamma}$ denote the antisymmetric tensors, $C \equiv i\gamma^0\gamma^2$ is the charge-conjugation matrix, and
\begin{equation}
\widetilde{H}_1^\dagger \overleftrightarrow{D}_\mu H_1 \equiv \widetilde{H}_1^\dagger \left(D_\mu - \overleftarrow{D}_\mu\right) H_1/2.
\end{equation}
Moreover, $p,r,s,t$ label fermion flavor indices, while $j,k$ and $\alpha,\beta,\gamma$ are isospin and color indices, respectively.

The evanescent operators relevant for our analysis are introduced as
\begin{align}
E_{\underset{prst}{\ell\nu\nu\ell}}^{S,(6)} &\equiv \mathcal{O}_{\underset{prst}{\ell\nu\nu\ell}}^{S,(6)} + \frac{1}{2} \mathcal{Q}_{\underset{ptsr}{\ell\nu}}^{(6)},& 
E_{\underset{prst}{\ell ee\ell}}^{S,(6)} &\equiv \mathcal{O}_{\underset{prst}{\ell ee\ell}}^{S,(6)} + \frac{1}{2} \mathcal{Q}_{\underset{ptsr}{\ell e}}^{(6)}, \label{eq:evan_beg}\\[0.8mm]
E_{\underset{prst}{quuq}}^{S,(6)} &\equiv \mathcal{O}_{\underset{prst}{quuq}}^{S,(6)} + \frac{1}{6} \mathcal{Q}_{\underset{ptsr}{qu}}^{(1),(6)} + \mathcal{Q}_{\underset{ptsr}{qu}}^{(8),(6)},& 
E_{\underset{prst}{quuq}}^{(8),(6)} &\equiv \mathcal{O}_{\underset{prst}{quuq}}^{(8),(6)} + \frac{1}{2} \mathcal{Q}_{\underset{ptsr}{qu}}^{(1),(6)}, \\[0.8mm]
E_{\underset{prst}{qddq}}^{S,(6)} &\equiv \mathcal{O}_{\underset{prst}{qddq}}^{S,(6)} + \frac{1}{6} \mathcal{Q}_{\underset{ptsr}{qd}}^{(1),(6)} + \mathcal{Q}_{\underset{ptsr}{qd}}^{(8),(6)},& 
E_{\underset{prst}{qddq}}^{(8),(6)} &\equiv \mathcal{O}_{\underset{prst}{qddq}}^{(8),(6)} + \frac{1}{2} \mathcal{Q}_{\underset{ptsr}{qd}}^{(1),(6)}, \\[0.8mm]
E_{\underset{prst}{\ell\nu\nu\ell}}^{T,(6)} &\equiv \mathcal{O}_{\underset{prst}{\ell\nu\nu\ell}}^{T,(6)} - 4 \left(1+a_{\rm ev} \epsilon\right) \mathcal{O}_{\underset{prst}{\ell\nu\nu\ell}}^{S,(6)},& 
E_{\underset{prst}{\ell ee\ell}}^{T,(6)} &\equiv \mathcal{O}_{\underset{prst}{\ell ee\ell}}^{T,(6)} - 4 \left(1+a_{\rm ev} \epsilon\right) \mathcal{O}_{\underset{prst}{\ell ee\ell}}^{S,(6)}, \\[0.8mm]
E_{\underset{prst}{quuq}}^{T,(6)} &\equiv \mathcal{O}_{\underset{prst}{quuq}}^{T,(6)} - 4 \left(1+a_{\rm ev} \epsilon\right) \mathcal{O}_{\underset{prst}{quuq}}^{S,(6)},& 
E_{\underset{prst}{quuq}}^{T,(8),(6)} &\equiv \mathcal{O}_{\underset{prst}{quuq}}^{T,(8),(6)} - 4 \left(1+a_{\rm ev} \epsilon\right) \mathcal{O}_{\underset{prst}{quuq}}^{(8),(6)}, \\[0.8mm]
E_{\underset{prst}{qddq}}^{T,(6)} &\equiv \mathcal{O}_{\underset{prst}{qddq}}^{T,(6)} - 4 \left(1+a_{\rm ev} \epsilon\right) \mathcal{O}_{\underset{prst}{qddq}}^{S,(6)},& 
E_{\underset{prst}{qddq}}^{T,(8),(6)} &\equiv \mathcal{O}_{\underset{prst}{qddq}}^{T,(8),(6)} - 4 \left(1+a_{\rm ev} \epsilon\right) \mathcal{O}_{\underset{prst}{qddq}}^{(8),(6)}, \\[0.8mm]
E_{\underset{prst}{\ell\nu uq}}^{T,(6)} &\equiv \mathcal{O}_{\underset{prst}{\ell\nu uq}}^{T,(6)} - 4 \left(1+a_{\rm ev} \epsilon\right) \mathcal{Q}_{\underset{prst}{\ell\nu uq}}^{(6)},& 
E_{\underset{prst}{\ell edq}}^{T,(6)} &\equiv \mathcal{O}_{\underset{prst}{\ell edq}}^{T,(6)} - 4 \left(1+a_{\rm ev} \epsilon\right) \mathcal{Q}_{\underset{prst}{\ell edq}}^{(6)},
\end{align}
corresponding to the $\left(\overline{L}R\right)\left(\overline{R}L\right)$ structures,
\begin{align}
E_{\underset{prst}{\ell\ell}}^{(3),(6)} &\equiv \mathcal{O}_{\underset{prst}{\ell\ell}}^{(3),(6)} - \mathcal{Q}_{\underset{ptsr}{\ell\ell}}^{(6)}, \\[0.8mm]
E_{\underset{prst}{qq}}^{(8),(6)} &\equiv \mathcal{O}_{\underset{prst}{qq}}^{(8),(6)} - \frac{1}{2} \mathcal{Q}_{\underset{ptsr}{qq}}^{(1),(6)} - \frac{1}{2} \mathcal{Q}_{\underset{ptsr}{qq}}^{(3),(6)},
\end{align}
for operators of the $\left(\overline{L}L\right)\left(\overline{L}L\right)$ type, 
\begin{align}
E_{\underset{prst}{ud}}^{(6)} &\equiv \mathcal{O}_{\underset{prst}{ud}}^{(6)} - \frac{1}{3} \mathcal{Q}_{\underset{ptsr}{ud}}^{(1),(6)} - 2 \mathcal{Q}_{\underset{ptsr}{ud}}^{(8),(6)},& 
E_{\underset{prst}{\nu e}}^{(6)} &\equiv \mathcal{O}_{\underset{prst}{\nu e}}^{(6)} - \mathcal{Q}_{\underset{ptsr}{\nu e}}^{(6)}, \\[0.8mm]
E_{\underset{prst}{uu}}^{(8),(6)} &\equiv \mathcal{O}_{\underset{prst}{uu}}^{(8),(6)} - \mathcal{Q}_{\underset{ptsr}{uu}}^{(6)},& 
E_{\underset{prst}{dd}}^{(8),(6)} &\equiv \mathcal{O}_{\underset{prst}{dd}}^{(8),(6)} - \mathcal{Q}_{\underset{ptsr}{dd}}^{(6)},
\end{align}
which belong to the $\left(\overline{R}R\right)\left(\overline{R}R\right)$ class,
\begin{align}
E_{\underset{prst}{quqd}}^{(1),(6)} &\equiv \mathcal{O}_{\underset{prst}{quqd}}^{(1),(6)} + \frac{1}{2} \mathcal{O}_{\underset{srpt}{quqd}}^{(8),(6)} - d \mathcal{Q}_{\underset{prst}{quqd}}^{(1),(6)} + \left(2 - \frac{d}{6}\right) \mathcal{Q}_{\underset{srpt}{quqd}}^{(1),(6)} + \left(12 - d\right) \mathcal{Q}_{\underset{srpt}{quqd}}^{(8),(6)}, \\[0.8mm]
E_{\underset{prst}{quqd}}^{(8),(6)} &\equiv \mathcal{O}_{\underset{prst}{quqd}}^{(8),(6)} + \frac{1}{2} \mathcal{O}_{\underset{srpt}{quqd}}^{(1),(6)} - \frac{d}{3} \mathcal{Q}_{\underset{prst}{quqd}}^{(1),(6)} + \left(6 - \frac{d}{2}\right) \mathcal{Q}_{\underset{srpt}{quqd}}^{(1),(6)} - 2d \mathcal{Q}_{\underset{prst}{quqd}}^{(8),(6)}, \\[0.8mm]
E_{\underset{prst}{\ell\nu \ell e}}^{(6)} &\equiv \mathcal{O}_{\underset{prst}{\ell\nu \ell e}}^{(6)} + \frac{1}{2} \mathcal{O}_{\underset{srpt}{\ell\nu \ell e}}^{(6)} - d \mathcal{Q}_{\underset{prst}{\ell\nu \ell e}}^{(6)} + \left(6 - \frac{d}{2}\right) \mathcal{Q}_{\underset{srpt}{\ell\nu \ell e}}^{(6)}, \label{eq:evan_end}
\end{align}
corresponding to $\left(\overline{L}R\right)\left(\overline{L}R\right)$ structures. The scheme-dependent parameters $a_{\rm ev}$ are defined as in ref.~\cite{Dekens:2019ept}, and the Green's operators $\mathcal{O}_i^{(n)}$ are listed in appendix~\ref{appB:Green_basis}. Although different choices of evanescent schemes, corresponding to different $a_{\rm ev}$, modify intermediate expressions, they leave all physical observables unchanged. In practical computations, a common and convenient choice is $a_{\rm ev}=1$~\cite{Buras:1998raa}. Within the DR scheme, these evanescent operators are of $\mathcal{O}(\epsilon)$ and therefore vanish in the four-dimensional limit $d\to4$.

\section{Green's basis during the functional matching}
\label{appB:Green_basis}

In addition to the physical and evanescent operators listed in appendix~\ref{appA:vSMEFT_basis}, we also present here a number of redundant but IBP-independent operators in the Green's basis after applying isospin, color and Dirac algebraic identities for ease of reference. These operators inevitably arise at intermediate stages of the one-loop matching calculation and may, of course, be further reduced by employing the EOMs of the light d.o.f., and other pertinent relations. In our setup, beyond those given in eqs.~\eqref{eq:evan_beg}--\eqref{eq:evan_end}, we encounter the following IBP-independent operators (along with their possible h.c.), which comprise bosonic operators with\footnote{Here we have ignored operators of the $\left(\overline{f} \gamma_\mu i\overleftrightarrow{D}_\nu f\right) F_{\mu\nu}$ and $\left(\overline{f} \gamma_\mu \gamma_\nu \gamma_\rho f\right) \left(\overline{f} \gamma_\mu \gamma_\nu \gamma_\rho f\right)$ types, because they lead to null contributions when matching the $\nu$2HDM onto $\nu$SMEFT.}
\begin{itemize}
\item $X^2D^2$ class:
\begin{align}
\mathcal{O}_{BD}^{(6)} \equiv \left(\partial_\mu B_{\mu\nu}\right)^2, \qquad \mathcal{O}_{WD}^{(6)} \equiv \left(D_\mu W_{\mu\nu}^I\right)^2,
\end{align}
\item $H_1^4D^2$ class:
\begin{align}
\mathcal{O}_{H_1D}^{(6)} \equiv \left(H_1^\dagger H_1\right) \left(H_1^\dagger D^2 H_1\right),
\end{align}
\end{itemize}
two-fermion operators with
\begin{itemize}
\item $\psi^2H_1^2D$ class:
\begin{align}
\mathcal{O}_{\underset{pr}{H_1\ell}}^{(1),(6)} &\equiv \left(H_1^\dagger H_1\right)           \left(\overline{L}_{Lp} \slashed{D} L_{Lr}\right),& \mathcal{O}_{\underset{pr}{H_1q}}^{(1),(6)} &\equiv \left(H_1^\dagger H_1\right) \left(\overline{Q}_{Lp} \slashed{D}\, Q_{Lr}\right), \\[0.8mm]
\mathcal{O}_{\underset{pr}{H_1\ell}}^{(3),(6)} &\equiv \left(H_1^{j\ast} H_1^k\right) \left(\overline{L}_{Lp}^k \slashed{D} L_{Lr}^j\right),& \mathcal{O}_{\underset{pr}{H_1q}}^{(3),(6)} &\equiv \left(H_1^{j\ast} H_1^k\right) \left(\overline{Q}_{Lp}^k \slashed{D}\, Q_{Lr}^j\right), \\[0.8mm]
\mathcal{O}_{\underset{pr}{H_1\nu}}^{(6)} &\equiv \left(H_1^\dagger H_1\right) \left(\overline{\nu}_{Rp} \slashed{D}\, \nu_{Rr}\right),& \mathcal{O}_{\underset{pr}{H_1u}}^{(6)} &\equiv \left(H_1^\dagger H_1\right) \left(\overline{U}_{Rp} \slashed{D}\, U_{Rr}\right), \\[0.8mm]
\mathcal{O}_{\underset{pr}{H_1e}}^{(6)} &\equiv \left(H_1^\dagger H_1\right) \left(\overline{E}_{Rp} \slashed{D} E_{Rr}\right),& \mathcal{O}_{\underset{pr}{H_1d}}^{(6)} &\equiv \left(H_1^\dagger H_1\right) \left(\overline{D}_{Rp} \slashed{D}\, D_{Rr}\right),
\end{align}
\item $\psi^2 D^3$ class:
\begin{align}
\mathcal{O}_{\underset{pr}{Dq}}^{(6)} &\equiv \overline{Q}_{Lp} \left(D^2 \slashed{D} + \slashed{D} D^2\right)\, Q_{Lr},& \mathcal{O}_{\underset{pr}{Du}}^{(6)} &\equiv \overline{U}_{Rp} \left(D^2 \slashed{D} + \slashed{D} D^2\right)\, U_{Rr}, \\[0.8mm]
\mathcal{O}_{\underset{pr}{Dd}}^{(6)} &\equiv \overline{D}_{Rp} \left(D^2 \slashed{D} + \slashed{D} D^2\right)\, D_{Rr},& \mathcal{O}_{\underset{pr}{D\ell}}^{(6)} &\equiv \overline{L}_{Lp} \left(D^2 \slashed{D} + \slashed{D} D^2\right)\, L_{Lr}, \\[0.8mm]
\mathcal{O}_{\underset{pr}{D\nu}}^{(6)} &\equiv \overline{\nu}_{Rp} \left(D^2 \slashed{D} + \slashed{D} D^2\right)\, \nu_{Rr},& \mathcal{O}_{\underset{pr}{De}}^{(6)} &\equiv \overline{E}_{Rp} \left(D^2 \slashed{D} + \slashed{D} D^2\right)\, E_{Rr},
\end{align}
\item $\psi^2H_1D^2$ class:
\begin{align}
\mathcal{O}_{\underset{pr}{H_1uD}}^{S,(6)} &\equiv \left(\overline{Q}_{Lp}^j D_\mu U_{Rr}\right) \left(D_\mu \widetilde{H}_1^j\right),& \mathcal{O}_{\underset{pr}{H_1dD}}^{S,(6)} &\equiv \left(\overline{Q}_{Lp}^j D_\mu D_{Rr}\right) \left(D_\mu H_1^j\right), \\[0.8mm]
\mathcal{O}_{\underset{pr}{H_1uD}}^{T,(6)} &\equiv \left(\overline{Q}_{Lp}^j \gamma_\nu \slashed{D} U_{Rr}\right) \left(D_\nu \widetilde{H}_1^j\right),& \mathcal{O}_{\underset{pr}{H_1dD}}^{T,(6)} &\equiv \left(\overline{Q}_{Lp}^j \gamma_\nu \slashed{D} D_{Rr}\right) \left(D_\nu H_1^j\right), \\[0.8mm]
\mathcal{O}_{\underset{pr}{H_1\nu D}}^{S,(6)} &\equiv \left(\overline{L}_{Lp}^j D_\mu \nu_{Rr}\right) \left(D_\mu \widetilde{H}_1^j\right),& \mathcal{O}_{\underset{pr}{H_1eD}}^{S,(6)} &\equiv \left(\overline{L}_{Lp}^j D_\mu E_{Rr}\right) \left(D_\mu H_1^j\right), \\[0.8mm]
\mathcal{O}_{\underset{pr}{H_1\nu D}}^{T,(6)} &\equiv \left(\overline{L}_{Lp}^j \gamma_\nu \slashed{D} \nu_{Rr}\right) \left(D_\nu \widetilde{H}_1^j\right),& \mathcal{O}_{\underset{pr}{H_1eD}}^{T,(6)} &\equiv \left(\overline{L}_{Lp}^j \gamma_\nu \slashed{D} E_{Rr}\right) \left(D_\nu H_1^j\right), \\[0.8mm]
\mathcal{O}_{\underset{pr}{H_1Du}}^{S,(6)} &\equiv \left(\overline{Q}_{Lp}^j D^2 U_{Rr}\right) \widetilde{H}_1^j,& \mathcal{O}_{\underset{pr}{H_1Dd}}^{S,(6)} &\equiv \left(\overline{Q}_{Lp}^j D^2 D_{Rr}\right) H_1^j, \\[0.8mm]
\mathcal{O}_{\underset{pr}{H_1D\nu}}^{S,(6)} &\equiv \left(\overline{L}_{Lp}^j D^2 \nu_{Rr}\right) \widetilde{H}_1^j,& \mathcal{O}_{\underset{pr}{H_1De}}^{S,(6)} &\equiv \left(\overline{L}_{Lp}^j D^2 E_{Rr}\right) H_1^j, \\[0.8mm]
\mathcal{O}_{\underset{pr}{DH_1u}}^{S,(6)} &\equiv \left(\overline{Q}_{Lp}^j U_{Rr}\right) \left(D^2 \widetilde{H}_1^j\right),& \mathcal{O}_{\underset{pr}{DH_1d}}^{S,(6)} &\equiv \left(\overline{Q}_{Lp}^j D_{Rr}\right) \left(D^2 H_1^j\right), \\[0.8mm]
\mathcal{O}_{\underset{pr}{DH_1\nu}}^{S,(6)} &\equiv \left(\overline{L}_{Lp}^j \nu_{Rr}\right) \left(D^2 \widetilde{H}_1^j\right),& \mathcal{O}_{\underset{pr}{DH_1e}}^{S,(6)} &\equiv \left(\overline{L}_{Lp}^j E_{Rr}\right) \left(D^2 H_1^j\right),
\end{align}
\item $\psi^2 X D$ class:
\begin{align}
\mathcal{O}_{\underset{pr}{DBq}}^{(6)} &\equiv \left(\overline{Q}_{Lp} \gamma_\nu Q_{Lr}\right) \partial_\mu B_{\mu\nu},& \mathcal{O}_{\underset{pr}{DB\ell}}^{(6)} &\equiv \left(\overline{L}_{Lp} \gamma_\nu L_{Lr}\right) \partial_\mu B_{\mu\nu}, \\[0.8mm]
\mathcal{O}_{\underset{pr}{DWq}}^{(6)} &\equiv \left(\overline{Q}_{Lp} S^I \gamma_\nu Q_{Lr}\right) D_\mu W_{\mu\nu}^I,& \mathcal{O}_{\underset{pr}{DW\ell}}^{(6)} &\equiv \left(\overline{L}_{Lp} S^I \gamma_\nu L_{Lr}\right) D_\mu W_{\mu\nu}^I, \\[0.8mm]
\mathcal{O}_{\underset{pr}{DBu}}^{(6)} &\equiv \left(\overline{U}_{Rp} \gamma_\nu U_{Rr}\right) \partial_\mu B_{\mu\nu},& \mathcal{O}_{\underset{pr}{DBd}}^{(6)} &\equiv \left(\overline{D}_{Rp} \gamma_\nu D_{Rr}\right) \partial_\mu B_{\mu\nu}, \\[0.8mm]
\mathcal{O}_{\underset{pr}{DGu}}^{(6)} &\equiv \left(\overline{U}_{Rp} \gamma_\nu T^A U_{Rr}\right) D_\mu G_{\mu\nu}^A,& \mathcal{O}_{\underset{pr}{DGd}}^{(6)} &\equiv \left(\overline{D}_{Rp} \gamma_\nu T^A D_{Rr}\right) D_\mu G_{\mu\nu}^A, \\[0.8mm]
\mathcal{O}_{\underset{pr}{DB\nu}}^{(6)} &\equiv \left(\overline{\nu}_{Rp} \gamma_\nu \nu_{Rr}\right) \partial_\mu B_{\mu\nu},& \mathcal{O}_{\underset{pr}{DBe}}^{(6)} &\equiv \left(\overline{E}_{Rp} \gamma_\nu E_{Rr}\right) \partial_\mu B_{\mu\nu}, \\[0.8mm]
\mathcal{O}_{\underset{pr}{DGq}}^{(6)} &\equiv \left(\overline{Q}_{Lp} \gamma_\nu T^A Q_{Lr}\right) D_\mu G_{\mu\nu}^A, \\[0.8mm]
\mathcal{O}_{\underset{pr}{BDq}}^{T,(6)} &\equiv \left(\overline{Q}_{Lp} \gamma_\mu \gamma_\nu \slashed{D} Q_{Lr}\right) B_{\mu\nu},& \mathcal{O}_{\underset{pr}{BD\ell}}^{T,(6)} &\equiv \left(\overline{L}_{Lp} \gamma_\mu \gamma_\nu \slashed{D} L_{Lr}\right) B_{\mu\nu}, \\[0.8mm]
\mathcal{O}_{\underset{pr}{BDu}}^{T,(6)} &\equiv \left(\overline{U}_{Rp} \gamma_\mu \gamma_\nu \slashed{D} U_{Rr}\right) B_{\mu\nu},& \mathcal{O}_{\underset{pr}{BD\nu}}^{T,(6)} &\equiv \left(\overline{\nu}_{Rp} \gamma_\mu \gamma_\nu \slashed{D} \nu_{Rr}\right) B_{\mu\nu}, \\[0.8mm]
\mathcal{O}_{\underset{pr}{BDd}}^{T,(6)} &\equiv \left(\overline{D}_{Rp} \gamma_\mu \gamma_\nu \slashed{D} D_{Rr}\right) B_{\mu\nu},& \mathcal{O}_{\underset{pr}{BDe}}^{T,(6)} &\equiv \left(\overline{E}_{Rp} \gamma_\mu \gamma_\nu \slashed{D} E_{Rr}\right) B_{\mu\nu}, \\[0.8mm]
\mathcal{O}_{\underset{pr}{GDq}}^{T,(6)} &\equiv \left(\overline{Q}_{Lp} \gamma_\mu \gamma_\nu T^A \slashed{D} Q_{Lr}\right) G_{\mu\nu}^A,& \mathcal{O}_{\underset{pr}{GDu}}^{T,(6)} &\equiv \left(\overline{U}_{Rp} \gamma_\mu \gamma_\nu T^A \slashed{D} U_{Rr}\right) G_{\mu\nu}^A, \\[0.8mm]
\mathcal{O}_{\underset{pr}{GDd}}^{T,(6)} &\equiv \left(\overline{D}_{Rp} \gamma_\mu \gamma_\nu T^A \slashed{D} D_{Rr}\right) G_{\mu\nu}^A,
\end{align}
\end{itemize}
and four-fermion operators with
\begin{itemize}
\item $\left(\overline{L}R\right) \left(\overline{R}L\right)$ class:
\begin{align}
\mathcal{O}_{\underset{prst}{quuq}}^{T,(6)} &\equiv \left(\overline{Q}_{Lp}^j \gamma_\mu \gamma_\nu U_{Rr}\right) \left(\overline{U}_{Rs} \gamma_\mu \gamma_\nu Q_{Lt}^j\right),& \mathcal{O}_{\underset{prst}{quuq}}^{S,(6)} &\equiv \left(\overline{Q}_{Lp}^j U_{Rr}\right) \left(\overline{U}_{Rs} Q_{Lt}^j\right), \\[0.8mm]
\mathcal{O}_{\underset{prst}{qddq}}^{T,(6)} &\equiv \left(\overline{Q}_{Lp}^j \gamma_\mu \gamma_\nu D_{Rr}\right) \left(\overline{D}_{Rs} \gamma_\mu \gamma_\nu Q_{Lt}^j\right),& \mathcal{O}_{\underset{prst}{qddq}}^{S,(6)} &\equiv \left(\overline{Q}_{Lp}^j D_{Rr}\right) \left(\overline{D}_{Rs} Q_{Lt}^j\right), \\[0.8mm]
\mathcal{O}_{\underset{prst}{\ell\nu\nu\ell}}^{T,(6)} &\equiv \left(\overline{L}_{Lp}^j \gamma_\mu \gamma_\nu \nu_{Rr}\right) \left(\overline{\nu}_{Rs} \gamma_\mu \gamma_\nu L_{Lt}^j\right),& \mathcal{O}_{\underset{prst}{\ell\nu\nu\ell}}^{S,(6)} &\equiv \left(\overline{L}_{Lp}^j \nu_{Rr}\right) \left(\overline{\nu}_{Rs} L_{Lt}^j\right), \\[0.8mm]
\mathcal{O}_{\underset{prst}{\ell ee\ell}}^{T,(6)} &\equiv \left(\overline{L}_{Lp}^j \gamma_\mu \gamma_\nu E_{Rr}\right) \left(\overline{E}_{Rs} \gamma_\mu \gamma_\nu L_{Lt}^j\right),& \mathcal{O}_{\underset{prst}{\ell ee\ell}}^{S,(6)} &\equiv \left(\overline{L}_{Lp}^j E_{Rr}\right) \left(\overline{E}_{Rs} L_{Lt}^j\right), \\[0.8mm]
\mathcal{O}_{\underset{prst}{\ell\nu uq}}^{T,(6)} &\equiv \left(\overline{L}_{Lp}^j \gamma_\mu \gamma_\nu \nu_{Rr}\right) \left(\overline{U}_{Rs} \gamma_\mu \gamma_\nu Q_{Lt}^j\right), \\[0.8mm]
\mathcal{O}_{\underset{prst}{\ell edq}}^{T,(6)} &\equiv \left(\overline{L}_{Lp}^j \gamma_\mu \gamma_\nu E_{Rr}\right) \left(\overline{D}_{Rs} \gamma_\mu \gamma_\nu Q_{Lt}^j\right),
\end{align}
and color operators
\begin{align}
\mathcal{O}_{\underset{prst}{qddq}}^{(8),(6)} &\equiv \left(\overline{Q}_{Lp}^{j\alpha} D_{Rr}^\beta\right) \left(\overline{D}_{Rs}^\beta Q_{Lt}^{j\alpha}\right), \\[0.8mm]
\mathcal{O}_{\underset{prst}{quuq}}^{(8),(6)} &\equiv \left(\overline{Q}_{Lp}^{j\alpha} U_{Rr}^\beta\right) \left(\overline{U}_{Rs}^\beta Q_{Lt}^{j\alpha}\right), \\[0.8mm]
\mathcal{O}_{\underset{prst}{quuq}}^{T,(8),(6)} &\equiv \left(\overline{Q}_{Lp}^{j\alpha} \gamma_\mu \gamma_\nu U_{Rr}^\beta\right) \left(\overline{U}_{Rs}^\beta \gamma_\mu \gamma_\nu Q_{Lt}^{j\alpha}\right), \\[0.8mm]
\mathcal{O}_{\underset{prst}{qddq}}^{T,(8),(6)} &\equiv \left(\overline{Q}_{Lp}^{j\alpha} \gamma_\mu \gamma_\nu D_{Rr}^\beta\right) \left(\overline{D}_{Rs}^\beta \gamma_\mu \gamma_\nu Q_{Lt}^{j\alpha}\right),
\end{align}
\item $\left(\overline{L}L\right) \left(\overline{L}L\right)$ class:
\begin{align}
\mathcal{O}_{\underset{prst}{\ell\ell}}^{(3),(6)} &\equiv \left(\overline{L}_{Lp}^j \gamma_\mu L_{Lr}^k\right) \left(\overline{L}_{Ls}^k \gamma_\mu L_{Lt}^j\right), \\[0.8mm]
\mathcal{O}_{\underset{prst}{qq}}^{(8),(6)} &\equiv \left(\overline{Q}_{Lp}^\alpha \gamma_\mu Q_{Lr}^\beta\right) \left(\overline{Q}_{Ls}^\beta \gamma_\mu Q_{Lt}^\alpha\right),
\end{align}
\item $\left(\overline{R}R\right) \left(\overline{R}R\right)$ class:
\begin{align}
\mathcal{O}_{\underset{prst}{ud}}^{(6)} \equiv \left(\overline{U}_{Rp} \gamma_\mu D_{Rr}\right) \left(\overline{D}_{Rs} \gamma_\mu U_{Rt}\right), \qquad \mathcal{O}_{\underset{prst}{\nu e}}^{(6)} \equiv \left(\overline{\nu}_{Rp} \gamma_\mu E_{Rr}\right) \left(\overline{E}_{Rs} \gamma_\mu \nu_{Rt}\right),
\end{align}
and color operators
\begin{align}
\mathcal{O}_{\underset{prst}{uu}}^{(8),(6)} &\equiv \left(\overline{U}_{Rp}^\alpha \gamma_\mu U_{Rr}^\beta\right) \left(\overline{U}_{Rs}^\beta \gamma_\mu U_{Rt}^\alpha\right),&  \\[0.8mm]
\mathcal{O}_{\underset{prst}{dd}}^{(8),(6)} &\equiv \left(\overline{D}_{Rp}^\alpha \gamma_\mu D_{Rr}^\beta\right) \left(\overline{D}_{Rs}^\beta \gamma_\mu D_{Rt}^\alpha\right),
\end{align}
\item $\left(\overline{L}R\right) \left(\overline{L}R\right)$ class:
\begin{align}
\mathcal{O}_{\underset{prst}{\ell\nu\ell e}}^{(6)} &\equiv \left(\overline{L}_{Lp}^j \gamma_\mu \gamma_\nu \nu_{Rr}\right) \epsilon_{jk} \left(\overline{L}_{Ls}^k \gamma_\mu \gamma_\nu E_{Rt}\right), \\[0.8mm]
\mathcal{O}_{\underset{prst}{quqd}}^{(1),(6)} &\equiv \left(\overline{Q}_{Lp}^j \gamma_\mu \gamma_\nu U_{Rr}\right) \epsilon_{jk} \left(\overline{Q}_{Ls}^k \gamma_\mu \gamma_\nu D_{Rt}\right), \\[0.8mm]
\mathcal{O}_{\underset{prst}{quqd}}^{(8),(6)} &\equiv \left(\overline{Q}_{Lp}^{j\alpha} \gamma_\mu \gamma_\nu U_{Rr}^\beta\right) \epsilon_{jk} \left(\overline{Q}_{Ls}^{k\beta} \gamma_\mu \gamma_\nu D_{Rt}^\alpha\right).
\end{align}
\end{itemize}

Making use of the EOMs in eqs.~\eqref{eq:EOM_fermion_1}--\eqref{eq:EOM_boson_4}, we can connect these Green's operators with the non-redundant operators collected in appendix~\ref{appA:vSMEFT_basis} through
\begin{align}
\mathcal{O}_{BD}^{(6)} =&\; g_1^2 Y_H^2 \mathcal{Q}_{H_1\Box}^{(6)} + 2g_1^2 \delta_{pr} \delta_{st} \left(Y_{U_R} Y_{D_R} \mathcal{Q}_{\underset{prst}{ud}}^{(1),(6)} + Y_{E_R} Y_{U_R} \mathcal{Q}_{\underset{prst}{eu}}^{(6)} + Y_{E_R} Y_{D_R} \mathcal{Q}_{\underset{prst}{ed}}^{(6)}\right) \notag\\[0.8mm]
&+ 4g_1^2 Y_H^2 \mathcal{Q}_{H_1D}^{(6)} + 2g_1^2 Y_{L_L} \delta_{pr} \delta_{st} \left(Y_{Q_L} \mathcal{Q}_{\underset{prst}{\ell q}}^{(1),(6)} + Y_{U_R} \mathcal{Q}_{\underset{prst}{\ell u}}^{(6)} + Y_{D_R} \mathcal{Q}_{\underset{prst}{\ell d}}^{(6)} + Y_{E_R} \mathcal{Q}_{\underset{prst}{\ell e}}^{(6)}\right) \notag\\[0.8mm]
&+ 2g_1^2 Y_H \delta_{pr} \left(Y_{Q_L} \mathcal{Q}_{\underset{pr}{H_1q}}^{(1),(6)} + Y_{U_R} \mathcal{Q}_{\underset{pr}{H_1u}}^{(6)} + Y_{D_R} \mathcal{Q}_{\underset{pr}{H_1d}}^{(6)} + Y_{L_L} \mathcal{Q}_{\underset{pr}{H_1\ell}}^{(1),(6)} + Y_{E_R} \mathcal{Q}_{\underset{pr}{H_1e}}^{(6)}\right) \notag\\[0.8mm]
&+ g_1^2 \delta_{pr} \delta_{st} \left(Y_{Q_L}^2 \mathcal{Q}_{\underset{prst}{qq}}^{(1),(6)} + Y_{U_R}^2 \mathcal{Q}_{\underset{prst}{uu}}^{(6)} + Y_{D_R}^2 \mathcal{Q}_{\underset{prst}{dd}}^{(6)} + Y_{L_L}^2 \mathcal{Q}_{\underset{prst}{\ell\ell}}^{(6)} + Y_{E_R}^2 \mathcal{Q}_{\underset{prst}{ee}}^{(6)}\right) \notag\\[0.8mm]
&+ 2g_1^2 Y_{Q_L} \delta_{pr} \delta_{st} \left(Y_{U_R} \mathcal{Q}_{\underset{prst}{qu}}^{(1),(6)} + Y_{D_R} \mathcal{Q}_{\underset{prst}{qd}}^{(1),(6)} + Y_{E_R} \mathcal{Q}_{\underset{prst}{qe}}^{(6)}\right), \\[1.2mm]
\mathcal{O}_{WD}^{(6)} =& - \frac{1}{2} g_2^2 \left(\mathcal{O}_{H_1D}^{(6)} + h.c.\right) + \frac{3}{4} g_2^2 \mathcal{Q}_{H_1\Box}^{(6)} - \frac{1}{4} g_2^2 \delta_{pr} \delta_{st} \mathcal{Q}_{\underset{prst}{\ell\ell}}^{(6)} + \frac{1}{2} g_2^2 \delta_{pr} \delta_{st} \mathcal{O}_{\underset{prst}{\ell\ell}}^{(3),(6)} \notag\\[0.8mm]
& + \frac{1}{2} g_2^2 \delta_{pr} \mathcal{Q}_{\underset{pr}{H_1\ell}}^{(3),(6)} + \frac{1}{2} g_2^2 \delta_{pr} \mathcal{Q}_{\underset{pr}{H_1q}}^{(3),(6)} + \frac{1}{2} g_2^2 \delta_{pr} \delta_{st} \mathcal{Q}_{\underset{prst}{\ell q}}^{(3),(6)} + \frac{1}{4} g_2^2 \delta_{pr} \delta_{st} \mathcal{Q}_{\underset{prst}{qq}}^{(3),(6)}
\end{align}
for $X^2D^2$ type,
\begin{align}
\mathcal{O}_{H_1D}^{(6)} =& - m_{11}^2 \mathcal{Q}_{H_1}^{(4)} - \lambda_1 \mathcal{Q}_{H_1}^{(6)} - \left(Y_U^1\right)_{pr} \mathcal{Q}_{\underset{pr}{uH_1}}^{(6)} - \left(Y_\nu^1\right)_{pr} \mathcal{Q}_{\underset{pr}{\nu H_1}}^{(6)} - \left(Y_D^{1\dagger}\right)_{pr} \mathcal{Q}_{\underset{pr}{dH_1}}^{(6)\dagger} \notag\\[0.8mm]
&- \left(Y_E^{1\dagger}\right)_{pr} \mathcal{Q}_{\underset{pr}{eH_1}}^{(6)\dagger}
\end{align}
for $H^4D^2$ type,
\begin{align}
\mathcal{O}_{\underset{pr}{H_1\ell}}^{(1),(6)} &= -i \left(Y_E^1\right)_{rs} \mathcal{Q}_{\underset{ps}{eH_1}}^{(6)} -i \left(Y_\nu^1\right)_{rs} \mathcal{Q}_{\underset{ps}{\nu H_1}}^{(6)},& \mathcal{O}_{\underset{pr}{H_1\nu}}^{(6)} &= -i \left(Y_\nu^{1\dagger}\right)_{rs} \mathcal{Q}_{\underset{ps}{\nu H_1}}^{(6)\dagger}, \\[0.2mm]
\mathcal{O}_{\underset{pr}{H_1\ell}}^{(3),(6)} &= -i \left(Y_E^1\right)_{rs} \mathcal{Q}_{\underset{ps}{eH_1}}^{(6)},& \mathcal{O}_{\underset{pr}{H_1e}}^{(6)} &= -i \left(Y_E^{1\dagger}\right)_{rs} \mathcal{Q}_{\underset{ps}{e H_1}}^{(6)\dagger}, \\[0.2mm]
\mathcal{O}_{\underset{pr}{H_1q}}^{(1),(6)} &= -i \left(Y_D^1\right)_{rs} \mathcal{Q}_{\underset{ps}{dH_1}}^{(6)} -i \left(Y_U^1\right)_{rs} \mathcal{Q}_{\underset{ps}{u H_1}}^{(6)},& \mathcal{O}_{\underset{pr}{H_1u}}^{(6)} &= -i \left(Y_U^{1\dagger}\right)_{rs} \mathcal{Q}_{\underset{ps}{u H_1}}^{(6)\dagger}, \\[0.2mm]
\mathcal{O}_{\underset{pr}{H_1q}}^{(3),(6)} &= - i \left(Y_D^1\right)_{rs} \mathcal{Q}_{\underset{ps}{dH_1}}^{(6)},& \mathcal{O}_{\underset{pr}{H_1d}}^{(6)} &= -i \left(Y_D^{1\dagger}\right)_{rs} \mathcal{Q}_{\underset{ps}{d H_1}}^{(6)\dagger}
\end{align}
for $\psi^2H^2D$ type,
\begin{align}
\mathcal{O}_{\underset{pr}{Dq}}^{(6)} =& -i \left(Y_D^1\right)_{rs} \left(\mathcal{O}_{\underset{ps}{DH_1d}}^{S,(6)} + \mathcal{O}_{\underset{ps}{H_1Dd}}^{S,(6)}\right) - i \left(Y_U^1\right)_{rs} \left(\mathcal{O}_{\underset{ps}{DH_1u}}^{S,(6)} + \mathcal{O}_{\underset{ps}{H_1Du}}^{S,(6)}\right) - 2i \left(Y_D^1\right)_{rs} \mathcal{O}_{\underset{ps}{H_1dD}}^{S,(6)} \notag\\[0.8mm]
&- 2i \left(Y_U^1\right)_{rs} \mathcal{O}_{\underset{ps}{H_1uD}}^{S,(6)} -i \left(Y_D^{1\dagger}\right)_{sp} \left(\mathcal{O}_{\underset{sr}{DH_1d}}^{S,(6)\dagger} + \mathcal{O}_{\underset{sr}{H_1Dd}}^{S,(6)\dagger}\right) - i\left(Y_U^{1\dagger}\right)_{sp} \left(\mathcal{O}_{\underset{sr}{DH_1u}}^{S,(6)\dagger} + \mathcal{O}_{\underset{sr}{H_1Du}}^{S,(6)\dagger}\right) \notag\\[0.8mm]
&- 2i \left(Y_D^{1\dagger}\right)_{sp} \mathcal{O}_{\underset{sr}{H_1dD}}^{S,(6)\dagger} - 2i \left(Y_U^{1\dagger}\right)_{sp} \mathcal{O}_{\underset{sr}{H_1uD}}^{S,(6)\dagger}, \\[1.2mm]
\mathcal{O}_{\underset{pr}{D\ell}}^{(6)} =& -i \left(Y_E^1\right)_{rs} \left(\mathcal{O}_{\underset{ps}{DH_1e}}^{S,(6)} + \mathcal{O}_{\underset{ps}{H_1De}}^{S,(6)}\right) - i \left(Y_\nu^1\right)_{rs} \left(\mathcal{O}_{\underset{ps}{DH_1\nu}}^{S,(6)} + \mathcal{O}_{\underset{ps}{H_1D\nu}}^{S,(6)}\right) - 2i \left(Y_E^1\right)_{rs} \mathcal{O}_{\underset{ps}{H_1eD}}^{S,(6)} \notag\\[0.8mm]
&- 2i \left(Y_\nu^1\right)_{rs} \mathcal{O}_{\underset{ps}{H_1\nu D}}^{S,(6)} -i \left(Y_E^{1\dagger}\right)_{sp} \left(\mathcal{O}_{\underset{sr}{DH_1e}}^{S,(6)\dagger} + \mathcal{O}_{\underset{sr}{H_1De}}^{S,(6)\dagger}\right) - i \left(Y_\nu^{1\dagger}\right)_{sp} \left(\mathcal{O}_{\underset{sr}{DH_1\nu}}^{S,(6)\dagger} + \mathcal{O}_{\underset{sr}{H_1D\nu}}^{S,(6)\dagger}\right) \notag\\[0.8mm]
&- 2i \left(Y_E^{1\dagger}\right)_{sp} \mathcal{O}_{\underset{sr}{H_1eD}}^{S,(6)\dagger} - 2i \left(Y_\nu^{1\dagger}\right)_{sp} \mathcal{O}_{\underset{sr}{H_1\nu D}}^{S,(6)\dagger}, \\[1.2mm]
\mathcal{O}_{\underset{pr}{Du}}^{(6)} =& - i \left(Y_U^1\right)_{sp} \mathcal{O}_{\underset{sr}{H_1Du}}^{S,(6)} - i \left(Y_U^{1\dagger}\right)_{rs} \mathcal{O}_{\underset{ps}{H_1Du}}^{S,(6)\dagger}, \\[1.2mm]
\mathcal{O}_{\underset{pr}{Dd}}^{(6)} =& - i \left(Y_D^1\right)_{sp} \mathcal{O}_{\underset{sr}{H_1Dd}}^{S,(6)} - i \left(Y_D^{1\dagger}\right)_{rs} \mathcal{O}_{\underset{ps}{H_1Dd}}^{S,(6)\dagger}, \\[1.2mm]
\mathcal{O}_{\underset{pr}{D\nu}}^{(6)} =& - i \left(Y_\nu^1\right)_{sp} \mathcal{O}_{\underset{sr}{H_1D\nu}}^{S,(6)} - i \left(Y_\nu^{1\dagger}\right)_{rs} \mathcal{O}_{\underset{ps}{H_1D\nu}}^{S,(6)\dagger}, \\[1.2mm]
\mathcal{O}_{\underset{pr}{De}}^{(6)} =& - i \left(Y_E^1\right)_{sp} \mathcal{O}_{\underset{sr}{H_1De}}^{S,(6)} - i \left(Y_E^{1\dagger}\right)_{rs} \mathcal{O}_{\underset{ps}{H_1De}}^{S,(6)\dagger}
\end{align}
for $\psi^2D^3$ type,
\begin{align}
\mathcal{O}_{\underset{pr}{H_1uD}}^{S,(6)} &= \frac{i}{4} \left(Y_U^{1\dagger}\right)_{sp} \mathcal{O}_{\underset{sr}{H_1u}}^{(6)} + \frac{i}{4} \left(Y_U^{1\dagger}\right)_{sp} \mathcal{O}_{\underset{sr}{H_1u}}^{(6)\dagger} + \frac{i}{4} \left(Y_U^{1\dagger}\right)_{rs} \mathcal{O}_{\underset{ps}{H_1q}}^{(1),(6)} + \frac{i}{4} \left(Y_U^{1\dagger}\right)_{rs} \mathcal{O}_{\underset{ps}{H_1q}}^{(1),(6)\dagger} \notag\\[0.8mm]
&- \frac{i}{4} \left(Y_U^{1\dagger}\right)_{rs} \mathcal{O}_{\underset{ps}{H_1q}}^{(3),(6)} - \frac{i}{4} \left(Y_U^{1\dagger}\right)_{rs} \mathcal{O}_{\underset{ps}{H_1q}}^{(3),(6)\dagger} + \frac{1}{4} \left(Y_U^{1\dagger}\right)_{sp} \mathcal{Q}_{\underset{sr}{H_1u}}^{(6)} + \frac{1}{8} \left(Y_U^{1\dagger}\right)_{rs} \mathcal{Q}_{\underset{ps}{H_1q}}^{(1),(6)} \notag\\[0.8mm]
&- \frac{1}{8} \left(Y_U^{1\dagger}\right)_{rs} \mathcal{Q}_{\underset{ps}{H_1q}}^{(3),(6)} - \frac{g_1}{4} Y_H \mathcal{Q}_{\underset{pr}{uB}}^{(6)} + \frac{g_2}{8} \mathcal{Q}_{\underset{pr}{uW}}^{(6)} - \frac{1}{2} \mathcal{O}_{\underset{pr}{DH_1u}}^{S,(6)} - \frac{1}{2} \left(Y_D^{1\dagger}\right)_{sp} \mathcal{Q}_{\underset{sr}{H_1ud}}^{(6)\dagger}, \\[0.8mm]
\mathcal{O}_{\underset{pr}{H_1uD}}^{T,(6)} &= \frac{i}{2} \left(Y_U^{1\dagger}\right)_{rs} \mathcal{O}_{\underset{ps}{H_1q}}^{(1),(6)} + \frac{i}{2} \left(Y_U^{1\dagger}\right)_{rs} \mathcal{O}_{\underset{ps}{H_1q}}^{(1),(6)\dagger} - \frac{i}{2} \left(Y_U^{1\dagger}\right)_{rs} \mathcal{O}_{\underset{ps}{H_1q}}^{(3),(6)} - \frac{i}{2} \left(Y_U^{1\dagger}\right)_{rs} \mathcal{O}_{\underset{ps}{H_1q}}^{(3),(6)\dagger} \notag\\[0.8mm]
&+ \frac{1}{4} \left(Y_U^{1\dagger}\right)_{rs} \mathcal{Q}_{\underset{ps}{H_1q}}^{(1),(6)} - \frac{1}{4} \left(Y_U^{1\dagger}\right)_{rs} \mathcal{Q}_{\underset{ps}{H_1q}}^{(3),(6)}, \\[0.8mm]
\mathcal{O}_{\underset{pr}{H_1dD}}^{S,(6)} &= \frac{i}{4} \left(Y_D^{1\dagger}\right)_{sp} \mathcal{O}_{\underset{sr}{H_1d}}^{(6)} + \frac{i}{4} \left(Y_D^{1\dagger}\right)_{sp} \mathcal{O}_{\underset{sr}{H_1d}}^{(6)\dagger} + \frac{i}{4} \left(Y_D^{1\dagger}\right)_{rs} \mathcal{O}_{\underset{ps}{H_1q}}^{(3),(6)} + \frac{i}{4} \left(Y_D^{1\dagger}\right)_{rs} \mathcal{O}_{\underset{ps}{H_1q}}^{(3),(6)\dagger} \notag\\[0.8mm]
&- \frac{1}{4} \left(Y_D^{1\dagger}\right)_{sp} \mathcal{Q}_{\underset{sr}{H_1d}}^{(6)} - \frac{1}{8} \left(Y_D^{1\dagger}\right)_{rs} \mathcal{Q}_{\underset{ps}{H_1q}}^{(1),(6)} - \frac{1}{8} \left(Y_D^{1\dagger}\right)_{rs} \mathcal{Q}_{\underset{ps}{H_1q}}^{(3),(6)} + \frac{g_1}{4} Y_H \mathcal{Q}_{\underset{pr}{dB}}^{(6)} \notag\\[0.8mm]
&+ \frac{g_2}{8} \mathcal{Q}_{\underset{pr}{dW}}^{(6)} - \frac{1}{2} \mathcal{O}_{\underset{pr}{DH_1d}}^{S,(6)} - \frac{1}{2} \left(Y_U^{1\dagger}\right)_{sp} \mathcal{Q}_{\underset{sr}{H_1ud}}^{(6)}, \\[0.8mm]
\mathcal{O}_{\underset{pr}{H_1dD}}^{T,(6)} &= \frac{i}{2} \left(Y_D^{1\dagger}\right)_{rs} \left(\mathcal{O}_{\underset{ps}{H_1q}}^{(3),(6)} + \mathcal{O}_{\underset{ps}{H_1q}}^{(3),(6)\dagger}\right) - \frac{1}{4} \left(Y_D^{1\dagger}\right)_{rs} \left(\mathcal{Q}_{\underset{ps}{H_1q}}^{(1),(6)} + \mathcal{Q}_{\underset{ps}{H_1q}}^{(3),(6)}\right), \\[0.8mm]
\mathcal{O}_{\underset{pr}{H_1\nu D}}^{S,(6)} &= \frac{i}{4} \left(Y_\nu^{1\dagger}\right)_{sp} \mathcal{O}_{\underset{sr}{H_1\nu}}^{(6)} + \frac{i}{4} \left(Y_\nu^{1\dagger}\right)_{sp} \mathcal{O}_{\underset{sr}{H_1\nu}}^{(6)\dagger} + \frac{i}{4} \left(Y_\nu^{1\dagger}\right)_{rs} \mathcal{O}_{\underset{ps}{H_1\ell}}^{(1),(6)} + \frac{i}{4} \left(Y_\nu^{1\dagger}\right)_{rs} \mathcal{O}_{\underset{ps}{H_1\ell}}^{(1),(6)\dagger} \notag\\[0.8mm]
&- \frac{i}{4} \left(Y_\nu^{1\dagger}\right)_{rs} \mathcal{O}_{\underset{ps}{H_1\ell}}^{(3),(6)} - \frac{i}{4} \left(Y_\nu^{1\dagger}\right)_{rs} \mathcal{O}_{\underset{ps}{H_1\ell}}^{(3),(6)\dagger} + \frac{1}{4} \left(Y_\nu^{1\dagger}\right)_{sp} \mathcal{Q}_{\underset{sr}{H_1\nu}}^{(6)} + \frac{1}{8} \left(Y_\nu^{1\dagger}\right)_{rs} \mathcal{Q}_{\underset{ps}{H_1\ell}}^{(1),(6)} \notag\\[0.8mm]
&- \frac{1}{8} \left(Y_\nu^{1\dagger}\right)_{rs} \mathcal{Q}_{\underset{ps}{H_1\ell}}^{(3),(6)} - \frac{g_1}{4} Y_H \mathcal{Q}_{\underset{pr}{\nu B}}^{(6)} + \frac{g_2}{8} \mathcal{Q}_{\underset{pr}{\nu W}}^{(6)} - \frac{1}{2} \mathcal{O}_{\underset{pr}{DH_1\nu}}^{S,(6)} - \frac{1}{2} \left(Y_E^{1\dagger}\right)_{sp} \mathcal{Q}_{\underset{sr}{H_1\nu e}}^{(6)\dagger}, \\[0.8mm]
\mathcal{O}_{\underset{pr}{H_1\nu D}}^{T,(6)} &= \frac{i}{2} \left(Y_\nu^{1\dagger}\right)_{rs} \mathcal{O}_{\underset{ps}{H_1\ell}}^{(1),(6)} + \frac{i}{2} \left(Y_\nu^{1\dagger}\right)_{rs} \mathcal{O}_{\underset{ps}{H_1\ell}}^{(1),(6)\dagger} - \frac{i}{2} \left(Y_\nu^{1\dagger}\right)_{rs} \mathcal{O}_{\underset{ps}{H_1\ell}}^{(3),(6)} - \frac{i}{2} \left(Y_\nu^{1\dagger}\right)_{rs} \mathcal{O}_{\underset{ps}{H_1\ell}}^{(3),(6)\dagger} \notag\\[0.8mm]
&+ \frac{1}{4} \left(Y_\nu^{1\dagger}\right)_{rs} \mathcal{Q}_{\underset{ps}{H_1\ell}}^{(1),(6)} - \frac{1}{4} \left(Y_\nu^{1\dagger}\right)_{rs} \mathcal{Q}_{\underset{ps}{H_1\ell}}^{(3),(6)}, \\[0.8mm]
\mathcal{O}_{\underset{pr}{H_1eD}}^{S,(6)} &= \frac{i}{4} \left(Y_E^{1\dagger}\right)_{sp} \mathcal{O}_{\underset{sr}{H_1e}}^{(6)} + \frac{i}{4} \left(Y_E^{1\dagger}\right)_{sp} \mathcal{O}_{\underset{sr}{H_1e}}^{(6)\dagger} + \frac{i}{4} \left(Y_E^{1\dagger}\right)_{rs} \mathcal{O}_{\underset{ps}{H_1\ell}}^{(3),(6)} + \frac{i}{4} \left(Y_E^{1\dagger}\right)_{rs} \mathcal{O}_{\underset{ps}{H_1\ell}}^{(3),(6)\dagger} \notag\\[0.8mm]
&- \frac{1}{4} \left(Y_E^{1\dagger}\right)_{sp} \mathcal{Q}_{\underset{sr}{H_1e}}^{(6)} - \frac{1}{8} \left(Y_E^{1\dagger}\right)_{rs} \mathcal{Q}_{\underset{ps}{H_1\ell}}^{(1),(6)} - \frac{1}{8} \left(Y_E^{1\dagger}\right)_{rs} \mathcal{Q}_{\underset{ps}{H_1\ell}}^{(3),(6)} + \frac{g_1}{4} Y_H \mathcal{Q}_{\underset{pr}{eB}}^{(6)} \notag\\[0.8mm]
&+ \frac{g_2}{8} \mathcal{Q}_{\underset{pr}{eW}}^{(6)} - \frac{1}{2} \mathcal{O}_{\underset{pr}{DH_1e}}^{S,(6)} - \frac{1}{2} \left(Y_\nu^{1\dagger}\right)_{sp} \mathcal{Q}_{\underset{sr}{H_1\nu e}}^{(6)}, \\[0.8mm]
\mathcal{O}_{\underset{pr}{H_1eD}}^{T,(6)} &= \frac{i}{2} \left(Y_E^{1\dagger}\right)_{rs} \left(\mathcal{O}_{\underset{ps}{H_1\ell}}^{(3),(6)} + \mathcal{O}_{\underset{ps}{H_1\ell}}^{(3),(6)\dagger}\right) - \frac{1}{4} \left(Y_E^{1\dagger}\right)_{rs} \left(\mathcal{Q}_{\underset{ps}{H_1\ell}}^{(1),(6)} + \mathcal{Q}_{\underset{ps}{H_1\ell}}^{(3),(6)}\right), \\[0.8mm]
\mathcal{O}_{\underset{pr}{H_1Du}}^{S,(6)} &= - \frac{i}{2} \left(Y_U^{1\dagger}\right)_{rs} \mathcal{O}_{\underset{ps}{H_1q}}^{(1),(6)} + \frac{i}{2} \left(Y_U^{1\dagger}\right)_{rs} \mathcal{O}_{\underset{ps}{H_1q}}^{(1),(6)\dagger} + \frac{i}{2} \left(Y_U^{1\dagger}\right)_{rs} \mathcal{O}_{\underset{ps}{H_1q}}^{(3),(6)} - \frac{i}{2} \left(Y_U^{1\dagger}\right)_{rs} \mathcal{O}_{\underset{ps}{H_1q}}^{(3),(6)\dagger} \notag\\[0.8mm]
&- \frac{1}{4} \left(Y_U^{1\dagger}\right)_{rs} \mathcal{Q}_{\underset{ps}{H_1q}}^{(1),(6)} + \frac{1}{4} \left(Y_U^{1\dagger}\right)_{rs} \mathcal{Q}_{\underset{ps}{H_1q}}^{(3),(6)} + \frac{g_1}{2} Y_{U_R} \mathcal{Q}_{\underset{pr}{uB}}^{(6)} + \frac{g_s}{2} \mathcal{Q}_{\underset{pr}{uG}}^{(6)}, \\[0.8mm]
\mathcal{O}_{\underset{pr}{H_1Dd}}^{S,(6)} &= - \frac{i}{2} \left(Y_D^{1\dagger}\right)_{rs} \mathcal{O}_{\underset{ps}{H_1q}}^{(3),(6)} + \frac{i}{2} \left(Y_D^{1\dagger}\right)_{rs} \mathcal{O}_{\underset{ps}{H_1q}}^{(3),(6)\dagger} + \frac{1}{4} \left(Y_D^{1\dagger}\right)_{rs} \mathcal{Q}_{\underset{ps}{H_1q}}^{(1),(6)} + \frac{1}{4} \left(Y_D^{1\dagger}\right)_{rs} \mathcal{Q}_{\underset{ps}{H_1q}}^{(3),(6)} \notag\\[0.8mm]
&+ \frac{g_1}{2} Y_{D_R} \mathcal{Q}_{\underset{pr}{dB}}^{(6)} + \frac{g_s}{2} \mathcal{Q}_{\underset{pr}{dG}}^{(6)}, \\[0.8mm]
\mathcal{O}_{\underset{pr}{H_1D\nu}}^{S,(6)} &= - \frac{i}{2} \left(Y_\nu^{1\dagger}\right)_{rs} \mathcal{O}_{\underset{ps}{H_1\ell}}^{(1),(6)} + \frac{i}{2} \left(Y_\nu^{1\dagger}\right)_{rs} \mathcal{O}_{\underset{ps}{H_1\ell}}^{(1),(6)\dagger} + \frac{i}{2} \left(Y_\nu^{1\dagger}\right)_{rs} \mathcal{O}_{\underset{ps}{H_1\ell}}^{(3),(6)} - \frac{i}{2} \left(Y_\nu^{1\dagger}\right)_{rs} \mathcal{O}_{\underset{ps}{H_1\ell}}^{(3),(6)\dagger} \notag\\[0.8mm]
&- \frac{1}{4} \left(Y_\nu^{1\dagger}\right)_{rs} \mathcal{Q}_{\underset{ps}{H_1\ell}}^{(1),(6)} + \frac{1}{4} \left(Y_\nu^{1\dagger}\right)_{rs} \mathcal{Q}_{\underset{ps}{H_1\ell}}^{(3),(6)} + \frac{g_1}{2} Y_{\nu_R} \mathcal{Q}_{\underset{pr}{\nu B}}^{(6)}, \\[0.8mm]
\mathcal{O}_{\underset{pr}{H_1De}}^{S,(6)} &= - \frac{i}{2} \left(Y_E^{1\dagger}\right)_{rs} \mathcal{O}_{\underset{ps}{H_1\ell}}^{(3),(6)} + \frac{i}{2} \left(Y_E^{1\dagger}\right)_{rs} \mathcal{O}_{\underset{ps}{H_1\ell}}^{(3),(6)\dagger} + \frac{1}{4} \left(Y_E^{1\dagger}\right)_{rs} \mathcal{Q}_{\underset{ps}{H_1\ell}}^{(1),(6)} + \frac{1}{4} \left(Y_E^{1\dagger}\right)_{rs} \mathcal{Q}_{\underset{ps}{H_1\ell}}^{(3),(6)} \notag\\[0.8mm]
&+ \frac{g_1}{2} Y_{E_R} \mathcal{Q}_{\underset{pr}{eB}}^{(6)}, \\[0.8mm]
\mathcal{O}_{\underset{pr}{DH_1u}}^{S,(6)} &= - m_{11}^2 \mathcal{Q}_{\underset{pr}{qu}}^{(4)} - \lambda_1 \mathcal{Q}_{\underset{pr}{uH_1}}^{(6)} - \left(Y_U^{1\dagger}\right)_{st} \mathcal{O}_{\underset{prst}{quuq}}^{S,(6)} + \left(Y_E^1\right)_{st} \mathcal{Q}_{\underset{stpr}{\ell equ}}^{(1),(6)} - \left(Y_D^1\right)_{st} \mathcal{Q}_{\underset{prst}{quqd}}^{(1),(6)} \notag\\[0.8mm]
&- \left(Y_\nu^{1\dagger}\right)_{st} \mathcal{Q}_{\underset{stpr}{\ell\nu uq}}^{(6)\dagger}, \\[0.8mm]
\mathcal{O}_{\underset{pr}{DH_1d}}^{S,(6)} &= - m_{11}^2 \mathcal{Q}_{\underset{pr}{qd}}^{(4)} - \lambda_1 \mathcal{Q}_{\underset{pr}{dH_1}}^{(6)} - \left(Y_D^{1\dagger}\right)_{st} \mathcal{O}_{\underset{prst}{qddq}}^{S,(6)} - \left(Y_\nu^1\right)_{st} \mathcal{Q}_{\underset{stpr}{\ell\nu qd}}^{(1),(6)} - \left(Y_U^1\right)_{st} \mathcal{Q}_{\underset{stpr}{quqd}}^{(1),(6)} \notag\\[0.8mm]
&- \left(Y_E^{1\dagger}\right)_{st} \mathcal{Q}_{\underset{stpr}{\ell edq}}^{(6)\dagger}, \\[0.8mm]
\mathcal{O}_{\underset{pr}{DH_1\nu}}^{S,(6)} &= - m_{11}^2 \mathcal{Q}_{\underset{pr}{\ell\nu}}^{(4)} - \lambda_1 \mathcal{Q}_{\underset{pr}{\nu H_1}}^{(6)} - \left(Y_\nu^{1\dagger}\right)_{st} \mathcal{O}_{\underset{prst}{\ell\nu\nu\ell}}^{S,(6)} - \left(Y_D^1\right)_{st} \mathcal{Q}_{\underset{prst}{\ell\nu qd}}^{(1),(6)} - \left(Y_E^1\right)_{st} \mathcal{Q}_{\underset{prst}{\ell\nu\ell e}}^{(6)} \notag\\[0.8mm]
&- \left(Y_U^{1\dagger}\right)_{st} \mathcal{Q}_{\underset{prst}{\ell\nu uq}}^{(6)}, \\[0.8mm]
\mathcal{O}_{\underset{pr}{DH_1e}}^{S,(6)} &= - m_{11}^2 \mathcal{Q}_{\underset{pr}{\ell e}}^{(4)} - \lambda_1 \mathcal{Q}_{\underset{pr}{eH_1}}^{(6)} - \left(Y_E^{1\dagger}\right)_{st} \mathcal{O}_{\underset{prst}{\ell ee\ell}}^{S,(6)} + \left(Y_U^1\right)_{st} \mathcal{Q}_{\underset{prst}{\ell equ}}^{(1),(6)} - \left(Y_\nu^1\right)_{st} \mathcal{Q}_{\underset{stpr}{\ell\nu\ell e}}^{(6)} \notag\\[0.8mm]
&- \left(Y_D^{1\dagger}\right)_{st} \mathcal{Q}_{\underset{prst}{\ell edq}}^{(6)}
\end{align}
for $\psi^2HD^2$ type,
\begin{align}
\mathcal{O}_{\underset{pr}{DGq}}^{(6)} &= - \frac{g_s}{2} \delta_{st} \mathcal{O}_{\underset{prst}{qq}}^{(8),(6)} + \frac{g_s}{6} \delta_{st} \mathcal{Q}_{\underset{prst}{qq}}^{(1),(6)} - g_s \delta_{st} \mathcal{Q}_{\underset{prst}{qd}}^{(8),(6)} - g_s \delta_{st} \mathcal{Q}_{\underset{prst}{qu}}^{(8),(6)}, \\[0.8mm]
\mathcal{O}_{\underset{pr}{DWq}}^{(6)} &= - \frac{g_2}{4} \mathcal{Q}_{\underset{pr}{H_1q}}^{(3),(6)} - \frac{g_2}{4} \delta_{st} \mathcal{Q}_{\underset{stpr}{\ell q}}^{(3),(6)} - \frac{g_2}{4} \delta_{st} \mathcal{Q}_{\underset{stpr}{qq}}^{(3),(6)}, \\[0.8mm]
\mathcal{O}_{\underset{pr}{DBq}}^{(6)} &= - g_1 Y_{E_R} \delta_{st} \mathcal{Q}_{\underset{prst}{qe}}^{(6)} - g_1 Y_H \mathcal{Q}_{\underset{pr}{H_1q}}^{(1),(6)} - g_1 Y_{L_L} \delta_{st} \mathcal{Q}_{\underset{stpr}{\ell q}}^{(1),(6)} - g_1 Y_{D_R} \delta_{st} \mathcal{Q}_{\underset{prst}{qd}}^{(1),(6)} \notag\\[0.8mm]
&- g_1 Y_{Q_L} \delta_{st} \mathcal{Q}_{\underset{prst}{qq}}^{(1),(6)} - g_1 Y_{U_R} \delta_{st} \mathcal{Q}_{\underset{prst}{qu}}^{(1),(6)}, \\[0.8mm]
\mathcal{O}_{\underset{pr}{DGu}}^{(6)} &= - \frac{g_s}{2} \delta_{st} \mathcal{O}_{\underset{prst}{uu}}^{(8),(6)} + \frac{g_s}{6} \delta_{st} \mathcal{Q}_{\underset{prst}{uu}}^{(6)} - g_s \delta_{st} \mathcal{Q}_{\underset{stpr}{qu}}^{(8),(6)} - g_s \delta_{st} \mathcal{Q}_{\underset{prst}{ud}}^{(8),(6)}, \\[0.8mm]
\mathcal{O}_{\underset{pr}{DBu}}^{(6)} &= - g_1 Y_{E_R} \delta_{st} \mathcal{Q}_{\underset{stpr}{eu}}^{(6)} - g_1 Y_H \mathcal{Q}_{\underset{pr}{H_1u}}^{(6)} - g_1 Y_{L_L} \delta_{st} \mathcal{Q}_{\underset{stpr}{\ell u}}^{(6)} - g_1 Y_{U_R} \delta_{st} \mathcal{Q}_{\underset{prst}{uu}}^{(6)} \notag\\[0.8mm]
&- g_1 Y_{Q_L} \delta_{st} \mathcal{Q}_{\underset{stpr}{qu}}^{(1),(6)} - g_1 Y_{D_R} \delta_{st} \mathcal{Q}_{\underset{prst}{ud}}^{(1),(6)}, \\[0.8mm]
\mathcal{O}_{\underset{pr}{DGd}}^{(6)} &= - \frac{g_s}{2} \delta_{st} \mathcal{O}_{\underset{prst}{dd}}^{(8),(6)} + \frac{g_s}{6} \delta_{st} \mathcal{Q}_{\underset{prst}{dd}}^{(6)} - g_s \delta_{st} \mathcal{Q}_{\underset{stpr}{qd}}^{(8),(6)} - g_s \delta_{st} \mathcal{Q}_{\underset{stpr}{ud}}^{(8),(6)}, \\[0.8mm]
\mathcal{O}_{\underset{pr}{DBd}}^{(6)} &= - g_1 Y_{E_R} \delta_{st} \mathcal{Q}_{\underset{stpr}{ed}}^{(6)} - g_1 Y_H \mathcal{Q}_{\underset{pr}{H_1d}}^{(6)} - g_1 Y_{L_L} \delta_{st} \mathcal{Q}_{\underset{stpr}{\ell d}}^{(6)} - g_1 Y_{D_R} \delta_{st} \mathcal{Q}_{\underset{prst}{dd}}^{(6)} \notag\\[0.8mm]
&- g_1 Y_{Q_L} \delta_{st} \mathcal{Q}_{\underset{stpr}{qd}}^{(1),(6)} - g_1 Y_{U_R} \delta_{st} \mathcal{Q}_{\underset{stpr}{ud}}^{(1),(6)}, \\[0.8mm]
\mathcal{O}_{\underset{pr}{DW\ell}}^{(6)} &= - \frac{g_2}{2} \delta_{st} \mathcal{O}_{\underset{stpr}{\ell\ell}}^{(3),(6)} + \frac{g_2}{4} \delta_{st} \mathcal{Q}_{\underset{prst}{\ell\ell}}^{(6)} - \frac{g_2}{4} \mathcal{Q}_{\underset{pr}{H_1\ell}}^{(3),(6)} - \frac{g_2}{4} \delta_{st} \mathcal{Q}_{\underset{prst}{\ell q}}^{(3),(6)}, \\[0.8mm]
\mathcal{O}_{\underset{pr}{DB\ell}}^{(6)} &= - g_1 Y_{D_R} \delta_{st} \mathcal{Q}_{\underset{prst}{\ell d}}^{(6)} - g_1 Y_H \mathcal{Q}_{\underset{pr}{H_1\ell}}^{(1),(6)} - g_1 Y_{Q_L} \delta_{st} \mathcal{Q}_{\underset{prst}{\ell q}}^{(1),(6)} - g_1 Y_{E_R} \delta_{st} \mathcal{Q}_{\underset{prst}{\ell e}}^{(6)} \notag\\[0.8mm]
&- g_1 Y_{L_L} \delta_{st} \mathcal{Q}_{\underset{prst}{\ell\ell}}^{(6)} - g_1 Y_{U_R} \delta_{st} \mathcal{Q}_{\underset{prst}{\ell u}}^{(6)}, \\[0.8mm]
\mathcal{O}_{\underset{pr}{DB\nu}}^{(6)} &= - g_1 Y_{D_R} \delta_{st} \mathcal{Q}_{\underset{prst}{\nu d}}^{(6)} - g_1 Y_H \mathcal{Q}_{\underset{pr}{H_1\nu}}^{(6)} - g_1 Y_{L_L} \delta_{st} \mathcal{Q}_{\underset{stpr}{\ell\nu}}^{(6)} - g_1 Y_{U_R} \delta_{st} \mathcal{Q}_{\underset{prst}{\nu u}}^{(6)} \notag\\[0.8mm]
&- g_1 Y_{Q_L} \delta_{st} \mathcal{Q}_{\underset{stpr}{q\nu}}^{(6)} - g_1 Y_{E_R} \delta_{st} \mathcal{Q}_{\underset{prst}{\nu e}}^{(6)}, \\[0.8mm]
\mathcal{O}_{\underset{pr}{DBe}}^{(6)} &= - g_1 Y_{E_R} \delta_{st} \mathcal{Q}_{\underset{prst}{ee}}^{(6)} - g_1 Y_H \mathcal{Q}_{\underset{pr}{H_1e}}^{(6)} - g_1 Y_{L_L} \delta_{st} \mathcal{Q}_{\underset{stpr}{\ell e}}^{(6)} - g_1 Y_{D_R} \delta_{st} \mathcal{Q}_{\underset{prst}{ed}}^{(6)} \notag\\[0.8mm]
&- g_1 Y_{Q_L} \delta_{st} \mathcal{Q}_{\underset{stpr}{qe}}^{(6)} - g_1 Y_{U_R} \delta_{st} \mathcal{Q}_{\underset{prst}{eu}}^{(6)}
\end{align}
and
\begin{align}
\mathcal{O}_{\underset{pr}{BDq}}^{T,(6)} &= - \left(Y_D^1\right)_{rs} \mathcal{Q}_{\underset{ps}{dB}}^{(6)} - \left(Y_U^1\right)_{rs} \mathcal{Q}_{\underset{ps}{uB}}^{(6)},& \mathcal{O}_{\underset{pr}{BDu}}^{T,(6)} &= - \left(Y_U^{1\dagger}\right)_{rs} \mathcal{Q}_{\underset{ps}{uB}}^{(6)\dagger}, \\[0.8mm]
\mathcal{O}_{\underset{pr}{BD\ell}}^{T,(6)} &= - \left(Y_E^1\right)_{rs} \mathcal{Q}_{\underset{ps}{eB}}^{(6)} - \left(Y_\nu^1\right)_{rs} \mathcal{Q}_{\underset{ps}{\nu B}}^{(6)},& \mathcal{O}_{\underset{pr}{BD\nu}}^{T,(6)} &= - \left(Y_\nu^{1\dagger}\right)_{rs} \mathcal{Q}_{\underset{ps}{\nu B}}^{(6)\dagger}, \\[0.8mm]
\mathcal{O}_{\underset{pr}{BDd}}^{T,(6)} &= - \left(Y_D^{1\dagger}\right)_{rs} \mathcal{Q}_{\underset{ps}{dB}}^{(6)\dagger},& \mathcal{O}_{\underset{pr}{BDe}}^{T,(6)} &= - \left(Y_E^{1\dagger}\right)_{rs} \mathcal{Q}_{\underset{ps}{eB}}^{(6)\dagger}, \\[0.8mm]
\mathcal{O}_{\underset{pr}{GDq}}^{T,(6)} &= - \left(Y_D^1\right)_{rs} \mathcal{Q}_{\underset{ps}{dG}}^{(6)} - \left(Y_U^1\right)_{rs} \mathcal{Q}_{\underset{ps}{uG}}^{(6)},& \mathcal{O}_{\underset{pr}{GDu}}^{T,(6)} &= - \left(Y_U^{1\dagger}\right)_{rs} \mathcal{Q}_{\underset{ps}{uG}}^{(6)\dagger}, \\[0.8mm]
\mathcal{O}_{\underset{pr}{GDd}}^{T,(6)} &= - \left(Y_D^{1\dagger}\right)_{rs} \mathcal{Q}_{\underset{ps}{dG}}^{(6)\dagger}& &
\end{align}
for $\psi^2XD$ type.

\section{Interaction matrices of the \texorpdfstring{$\boldsymbol{\nu}$2HDM}{v2HDM}}
\label{appC:Inter_Matrix}

From the complete $\nu$2HDM Lagrangian in eq.~\eqref{eq:v2HDM}, one can derive the interaction matrices $X$ introduced in eq.~\eqref{eq:Lnu2HDM_compact}~\cite{Cohen:2020fcu,Fuentes-Martin:2024agf}. In particular, we obtain
\begin{itemize}
\item scalar-scalar type:
\begin{align}
&\boxed{X_{H_1H_1}^{\cyan{\alpha,\beta}}} = \lambda_1 \begin{pmatrix} H_1^{\cyan{\alpha}} H_1^{\cyan{\beta}\ast} & H_1^{\cyan{\alpha}} H_1^{\cyan{\beta}} \\ H_1^{\cyan{\alpha}\ast} H_1^{\cyan{\beta}\ast} & H_1^{\cyan{\alpha}\ast} H_1^{\cyan{\beta}}
\end{pmatrix} + \left[\lambda_1 \left|H_1\right|^2 + \lambda_3 \left|H_2\right|^2 + \lambda_6 H_1^\dagger H_2 + \lambda_6^\ast H_2^\dagger H_1 \right] \cyan{\mathcal{I}_\mathbbm{1}^{\alpha\beta}} \notag\\[0.8mm]
&+ \begin{pmatrix} \lambda_4 H_2^{\cyan{\alpha}} H_2^{\cyan{\beta}\ast} & \lambda_5 H_2^{\cyan{\alpha}} H_2^{\cyan{\beta}} \\ \lambda_5^\ast H_2^{\cyan{\alpha}\ast} H_2^{\cyan{\beta}\ast} & \lambda_4 H_2^{\cyan{\alpha}\ast} H_2^{\cyan{\beta}} \end{pmatrix} + \begin{pmatrix} \lambda_6^\ast H_1^{\cyan{\alpha}} H_2^{\cyan{\beta}\ast} + \lambda_6 H_2^{\cyan{\alpha}} H_1^{\cyan{\beta}\ast} & \lambda_6 H_1^{\cyan{\alpha}} H_2^{\cyan{\beta}} + \lambda_6 H_2^{\cyan{\alpha}} H_1^{\cyan{\beta}} \\ \lambda_6^\ast H_1^{\cyan{\alpha}\ast} H_2^{\cyan{\beta}\ast} + \lambda_6^\ast H_2^{\cyan{\alpha}\ast} H_1^{\cyan{\beta}\ast} & \lambda_6 H_1^{\cyan{\alpha}\ast} H_2^{\cyan{\beta}} + \lambda_6^\ast H_2^{\cyan{\alpha}\ast} H_1^{\cyan{\beta}} \end{pmatrix}, \\[0.8mm]
&\boxed{X_{H_2H_2}^{\cyan{\alpha,\beta}}} = \lambda_2 \begin{pmatrix} H_2^{\cyan{\alpha}} H_2^{\cyan{\beta}\ast} & H_2^{\cyan{\alpha}} H_2^{\cyan{\beta}} \\ H_2^{\cyan{\alpha}\ast} H_2^{\cyan{\beta}\ast} & H_2^{\cyan{\alpha}\ast} H_2^{\cyan{\beta}} \end{pmatrix} + \left[\lambda_2 \left|H_2\right|^2 + \lambda_3 \left|H_1\right|^2 + \lambda_7 H_1^\dagger H_2 + \lambda_7^\ast H_2^\dagger H_1 \right] \cyan{\mathcal{I}_{\mathbbm{1}}^{\alpha\beta}} \notag\\[0.8mm]
&+ \begin{pmatrix} \lambda_4 H_1^{\cyan{\alpha}} H_1^{\cyan{\beta}\ast} & \lambda_5^\ast H_1^{\cyan{\alpha}} H_1^{\cyan{\beta}} \\ \lambda_5 H_1^{\cyan{\alpha}\ast} H_1^{\cyan{\beta}\ast} & \lambda_4 H_1^{\cyan{\alpha}\ast} H_1^{\cyan{\beta}} \end{pmatrix} + \begin{pmatrix} \lambda_7^\ast H_1^{\cyan{\alpha}} H_2^{\cyan{\beta}\ast} + \lambda_7 H_2^{\cyan{\alpha}} H_1^{\cyan{\beta}\ast} & \lambda_7^\ast H_1^{\cyan{\alpha}} H_2^{\cyan{\beta}} + \lambda_7^\ast H_2^{\cyan{\alpha}} H_1^{\cyan{\beta}} \\ \lambda_7 H_1^{\cyan{\alpha}\ast} H_2^{\cyan{\beta}\ast} + \lambda_7 H_2^{\cyan{\alpha}\ast} H_1^{\cyan{\beta}\ast} & \lambda_7 H_1^{\cyan{\alpha}\ast} H_2^{\cyan{\beta}} + \lambda_7^\ast H_2^{\cyan{\alpha}\ast} H_1^{\cyan{\beta}} \end{pmatrix}, \\[0.8mm]
&\boxed{X_{H_1H_2}^{\cyan{\alpha,\beta}}} = \boxed{X_{H_2H_1}^{\cyan{\beta,\alpha}\dagger}} = \lambda_3 \begin{pmatrix} H_1^{\cyan{\alpha}} H_2^{\cyan{\beta}\ast} & H_1^{\cyan{\alpha}} H_2^{\cyan{\beta}} \\ H_1^{\cyan{\alpha}\ast} H_2^{\cyan{\beta}\ast} & H_1^{\cyan{\alpha}\ast} H_2^{\cyan{\beta}} \end{pmatrix} + \begin{pmatrix} \lambda_5 H_2^{\cyan{\alpha}} H_1^{\cyan{\beta}\ast} & \lambda_4 H_2^{\cyan{\alpha}} H_1^{\cyan{\beta}} \\ \lambda_4 H_2^{\cyan{\alpha}\ast} H_1^{\cyan{\beta}\ast} & \lambda_5^\ast H_2^{\cyan{\alpha}\ast} H_1^{\cyan{\beta}} \end{pmatrix} \notag\\[0.8mm]
&+ \begin{pmatrix} \cyan{\mathbbm{1}^{\alpha\beta}} \mathcal{H}_{12} & 0 \\ 0 & \cyan{\mathbbm{1}^{\alpha\beta}} \mathcal{H}_{12}^\dagger \end{pmatrix} + \begin{pmatrix} \lambda_6 H_1^{\cyan{\alpha}} H_1^{\cyan{\beta}\ast} & \lambda_6^\ast H_1^{\cyan{\alpha}} H_1^{\cyan{\beta}} \\ \lambda_6 H_1^{\cyan{\alpha}\ast} H_1^{\cyan{\beta}\ast} & \lambda_6^\ast H_1^{\cyan{\alpha}\ast} H_1^{\cyan{\beta}} \end{pmatrix} + \begin{pmatrix} \lambda_7 H_2^{\cyan{\alpha}} H_2^{\cyan{\beta}\ast} & \lambda_7 H_2^{\cyan{\alpha}} H_2^{\cyan{\beta}} \\ \lambda_7^\ast H_2^{\cyan{\alpha}\ast} H_2^{\cyan{\beta}\ast} & \lambda_7^\ast H_2^{\cyan{\alpha}\ast} H_2^{\cyan{\beta}} \end{pmatrix},
\end{align}
with \cyan{$\mathbbm{1}$} being the $2\times 2$ unit matrix in the isospin space, \cyan{$\mathcal{I}_\mathbbm{1}^{\alpha\beta}$} defined by
\begin{equation}
\cyan{\mathcal{I}_\mathbbm{1}^{\alpha\beta}} = \begin{pmatrix} \cyan{\mathbbm{1}^{\alpha\beta}} & 0 \\ 0 & \cyan{\mathbbm{1}^{\alpha\beta}} \end{pmatrix},
\end{equation}
and $\mathcal{H}_{12} \equiv -m_{12}^2 + \lambda_4 H_2^\dagger H_1 + \lambda_5 H_1^\dagger H_2 + \lambda_6 \left|H_1\right|^2 + \lambda_7 \left|H_2\right|^2$.
\item scalar-fermion type:
\begin{align}
&\boxed{X_{Q_LH_i}^{\cyan{\alpha,\beta}}} = \begin{pmatrix} \cyan{\mathbbm{1}^{\alpha\beta}} Y_D^i D_R & \left(i\tau^2\right)^\cyan{\alpha\beta} Y_U^i U_R \\ \left(i\tau^2\right)^\cyan{\alpha\beta} Y_U^{i\ast} U_R^c & \cyan{\mathbbm{1}^{\alpha\beta}} Y_D^{i\ast} D_R^c \end{pmatrix}, \\[0.8mm]
&\boxed{X_{H_iU_R}^{\cyan{\alpha}}} = \begin{pmatrix} \left(\overline{Q}_L i\tau^2\right)^\cyan{\alpha} Y_U^i & 0 \\ 0 & \left(\overline{Q_L^c} i\tau^2\right)^\cyan{\alpha}Y_U^{i\ast} \end{pmatrix}, \\[0.8mm]
&\boxed{X_{L_LH_i}^{\cyan{\alpha,\beta}}} = \begin{pmatrix} \cyan{\mathbbm{1}^{\alpha\beta}} Y_E^i E_R & \left(i\tau^2\right)^\cyan{\alpha\beta} Y_\nu^i \nu_R \\ \left(i\tau^2\right)^\cyan{\alpha\beta} Y_\nu^{i\ast} \nu_R^c & \cyan{\mathbbm{1}^{\alpha\beta}} Y_E^{i\ast} E_R^c \end{pmatrix},\\[0.8mm]
&\boxed{X_{H_i\nu_R}^{\cyan{\alpha}}} = \begin{pmatrix} \left(\overline{L}_L i\tau^2\right)^\cyan{\alpha} Y_\nu^i & 0 \\ 0 & \left(\overline{L_L^c} i\tau^2\right)^\cyan{\alpha}Y_\nu^{i\ast} \end{pmatrix}, 
\end{align}
and
\begin{align}
&\boxed{X_{H_iQ_L}^{\cyan{\alpha,\beta}}} = \begin{pmatrix} \cyan{\mathbbm{1}^{\alpha\beta}} \overline{D}_R Y_D^{i\dagger} & - \left(i\tau^2\right)^{\cyan{\alpha\beta}} \overline{U^c_R} Y_U^{i\sf T} \\ - \left( i\tau^2\right)^{\cyan{\alpha\beta}} \overline{U}_R Y_U^{i\dagger} & \cyan{\mathbbm{1}^{\alpha\beta}} \overline{D^c_R} Y_D^{i\sf T} \end{pmatrix},\\[0.8mm]
&\boxed{X_{H_iD_R}^{\cyan{\alpha}}} = \begin{pmatrix} 0 & \left(\overline{Q_L^c}\right)^\cyan{\alpha} Y_D^{i\ast} \\ \overline{Q}_L^{\cyan{\alpha}} Y_D^i & 0 \end{pmatrix}, \\[0.8mm]
&\boxed{X_{U_RH_i}^{\cyan{\alpha}}} = - \begin{pmatrix} Y_U^{i\dagger} \left(i\tau^2 Q_L\right)^\cyan{\alpha} & 0 \\ 0 & Y_U^{i\sf T} \left(i\tau^2 Q_L^c\right)^\cyan{\alpha} \end{pmatrix},\\[0.8mm]
&\boxed{X_{D_RH_i}^{\cyan{\alpha}}} = \begin{pmatrix} 0 & Y_D^{i\dagger} Q_L^{\cyan{\alpha}} \\ Y_D^{i\sf T} \left(Q_L^c\right)^\cyan{\alpha} & 0 \end{pmatrix}, \\[0.8mm]
&\boxed{X_{H_iL_L}^{\cyan{\alpha,\beta}}} = \begin{pmatrix} \cyan{\mathbbm{1}^{\alpha\beta}} \overline{E}_R Y_E^{i\dagger} & - \left(i\tau^2\right)^{\cyan{\alpha\beta}} \overline{\nu^c_R} Y_\nu^{i\sf T} \\ - \left( i\tau^2\right)^{\cyan{\alpha\beta}} \overline{\nu}_R Y_\nu^{i\dagger} & \cyan{\mathbbm{1}^{\alpha\beta}} \overline{E^c_R} Y_E^{i\sf T} \end{pmatrix},\\[0.8mm]
&\boxed{X_{H_iE_R}^{\cyan{\alpha}}} = \begin{pmatrix} 0 & \left(\overline{L_L^c}\right)^\cyan{\alpha} Y_E^{i\ast} \\ \overline{L}_L^{\cyan{\alpha}} Y_E^i & 0 \end{pmatrix}, \\[0.8mm]
&\boxed{X_{\nu_RH_i}^{\cyan{\alpha}}} = - \begin{pmatrix} Y_\nu^{i\dagger} \left(i\tau^2 L_L\right)^\cyan{\alpha} & 0 \\ 0 & Y_\nu^{i\sf T} \left(i\tau^2 L_L^c\right)^\cyan{\alpha} \end{pmatrix}, \\[0.8mm]
&\boxed{X_{E_RH_i}^{\cyan{\alpha}}} = \begin{pmatrix} 0 & Y_E^{i\dagger} L_L^{\cyan{\alpha}} \\ Y_E^{i\sf T} \left(L_L^c\right)^\cyan{\alpha} & 0 \end{pmatrix}, 
\end{align}
where $i=1,2$, and $\tau^2$ is the second Pauli matrix.
\item scalar-vector type:
\begin{align}
&\boxed{X_{H_iB}^{\cyan{\alpha},\mu}} \equiv U_{H_iB}^{\cyan{\alpha},\mu} + i\widehat{D}^\mu Z_{H_iB}^{\cyan{\alpha}},&& \boxed{X_{H_iW}^{\cyan{\alpha},\mu\cyan{I}}} \equiv U_{H_iW}^{\cyan{\alpha},\mu\cyan{I}} + i\widehat{D}^\mu Z_{H_iW}^{\cyan{\alpha},\cyan{I}} , \\[0.8mm]
&\boxed{X_{BH_i}^{\mu,\cyan{\alpha}}} \equiv U_{BH_i}^{\mu,\cyan{\alpha}} + i\overline{Z}_{BH_i}^{\cyan{\alpha}}\widehat{D}^\mu,&& \boxed{X_{WH_i}^{\mu\cyan{I},\cyan{\alpha}}} \equiv U_{WH_i}^{\mu\cyan{I},\cyan{\alpha}} + i\overline{Z}_{WH_i}^{\cyan{I},\cyan{\alpha}} \widehat{D}^\mu, 
\end{align}
with the open covariant derivative $\widehat{D}^\mu$ acting on all the fields to its right, and 
\begin{align}
&U_{H_iB}^{\cyan{\alpha},\mu} \equiv ig_1Y_H \begin{pmatrix} - \left(D^\mu H_i\right)^{\cyan{\alpha}} \\ \left(D^{\mu} H_i\right)^{\ast\cyan{\alpha}} \end{pmatrix},&& U_{WH_i}^{\mu\cyan{I},\cyan{\alpha}} \equiv ig_2 \begin{pmatrix} \left(S^{\cyan{I}} D^{\mu} H_i\right)^{\ast\cyan{\alpha}} & - \left(S^{\cyan{I}} D^{\mu} H_i\right)^{\cyan{\alpha}} \end{pmatrix}, \\[0.8mm]
&Z_{H_iB}^{\cyan{\alpha}} \equiv g_1 Y_H \begin{pmatrix} -H_i^{\cyan{\alpha}} \\ H_i^{\ast{\cyan{\alpha}}} \end{pmatrix},&& \overline{Z}_{WH_i}^{\cyan{I},\cyan{\alpha}} \equiv g_2 \begin{pmatrix} - \left(H_i^{\dagger} S^{\cyan{I}}\right)^{\cyan{\alpha}} & \left(H_i^{\sf T} S^{\cyan{I}\ast}\right)^{\cyan{\alpha}} \end{pmatrix}, \\[0.8mm]
&U_{H_iW}^{\cyan{\alpha},\mu\cyan{I}} \equiv ig_2 \begin{pmatrix} - \left(S^{\cyan{I}} D^\mu H_i\right)^{\cyan{\alpha}} \\ \left(S^{\cyan{I}} D^{\mu} H_i\right)^{\ast\cyan{\alpha}} \end{pmatrix},&& U_{BH_i}^{\mu,\cyan{\alpha}} \equiv ig_1Y_H \begin{pmatrix} \left(D^{\mu} H_i\right)^{\ast\cyan{\alpha}} & - \left(D^{\mu} H_i\right)^{\cyan{\alpha}} \end{pmatrix}, \\[0.8mm]
&Z_{H_iW}^{\cyan{\alpha},\cyan{I}} \equiv g_2 \begin{pmatrix} - \left(S^{\cyan{I}} H_i\right)^{\cyan{\alpha}} \\ \left(S^{\cyan{I}} H_i\right)^{\ast{\cyan{\alpha}}} \end{pmatrix},&& \overline{Z}_{BH_i}^{\cyan{\alpha}} \equiv g_1Y_H \begin{pmatrix} - H_i^{\ast\cyan{\alpha}} & H_i^{\cyan{\alpha}} \end{pmatrix}. 
\end{align}
\item fermion-fermion type:
\begin{align}
&\boxed{X_{Q_LU_R}^{\cyan{\alpha}}} = \sum_{i=1}^2\begin{pmatrix}{\color{red}\mathbbm{1}} Y_U^i \widetilde{H}_i^{\cyan{\alpha}} & 0 \\ 0 & {\color{red}\mathbbm{1}} Y_U^{i\ast} \widetilde{H}_i^{\cyan{\alpha}\ast} \end{pmatrix},&& \boxed{X_{Q_LD_R}^{\cyan{\alpha}}} = \sum_{i=1}^2\begin{pmatrix} {\color{red}\mathbbm{1}} Y_D^i H_i^{\cyan{\alpha}} & 0 \\ 0 & {\color{red}\mathbbm{1}} Y_D^{i\ast} H_i^{\cyan{\alpha}\ast} \end{pmatrix}, \\[0.8mm]
&\boxed{X_{U_RQ_L}^{\cyan{\alpha}}} = \sum_{i=1}^2\begin{pmatrix} {\color{red}\mathbbm{1}} Y_U^{i\dagger} \widetilde{H}_i^{\cyan{\alpha}\ast} & 0 \\ 0 & {\color{red}\mathbbm{1}} Y_U^{i\sf T} \widetilde{H}_i^{\cyan{\alpha}} \end{pmatrix},&& \boxed{X_{D_RQ_L}^{\cyan{\alpha}}} = \sum_{i=1}^2\begin{pmatrix} {\color{red}\mathbbm{1}} Y_D^{i\dagger} H_i^{\cyan{\alpha}\ast} & 0 \\ 0 & {\color{red}\mathbbm{1}} Y_D^{i\sf T} H_i^{\cyan{\alpha}} \end{pmatrix}, \\[0.8mm]
&\boxed{X_{L_L\nu_R}^{\cyan{\alpha}}} = \sum_{i=1}^2\begin{pmatrix} Y_\nu^i \widetilde{H}_i^{\cyan{\alpha}} & 0 \\ 0 & Y_\nu^{i\ast} \widetilde{H}_i^{\cyan{\alpha}\ast} \end{pmatrix},&& \boxed{X_{L_LE_R}^{\cyan{\alpha}}} = \sum_{i=1}^2\begin{pmatrix} Y_E^i H_i^{\cyan{\alpha}} & 0 \\ 0 & Y_E^{i\ast} H_i^{\cyan{\alpha}\ast} \end{pmatrix}, \\[0.8mm]
&\boxed{X_{\nu_RL_L}^{\cyan{\alpha}}} = \sum_{i=1}^2\begin{pmatrix} Y_\nu^{i\dagger} \widetilde{H}_i^{\cyan{\alpha}\ast} & 0 \\ 0 & Y_\nu^{i\sf T} \widetilde{H}_i^{\cyan{\alpha}} \end{pmatrix},&& \boxed{X_{E_RL_L}^{\cyan{\alpha}}} = \sum_{i=1}^2\begin{pmatrix} Y_E^{i\dagger} H_i^{\cyan{\alpha}\ast} & 0 \\ 0 & Y_E^{i\sf T} H_i^{\cyan{\alpha}} \end{pmatrix}, 
\end{align}
where {\color{red}$\mathbbm{1}$} represents the $3\times3$ unit matrix in the color space.
\item fermion-vector type:
\begin{align}
&\boxed{X_{\mathcal{F}_1B}^{\cyan{\alpha},\mu}} = g_1 Y_{\mathcal{F}_1} \begin{pmatrix} -\gamma^\mu \mathcal{F}_1^{\cyan{\alpha}} \\ \gamma^\mu \left(\mathcal{F}_1^c\right)^\cyan{\alpha} \end{pmatrix},&& \boxed{X_{B\mathcal{F}_1}^{\mu,\cyan{\alpha}}} = g_1 Y_{\mathcal{F}_1} \begin{pmatrix} - \left(\overline{\mathcal{F}_1}\right)^\cyan{\alpha} \gamma^\mu & \left(\overline{\mathcal{F}_1^c}\right)^\cyan{\alpha} \gamma^\mu \end{pmatrix}, \\[0.8mm]
&\boxed{X_{\mathcal{F}_2W}^{\cyan{\alpha},\mu\cyan{I}}} = g_2 \begin{pmatrix} -\gamma^\mu \left(S^\cyan{I} \mathcal{F}_2\right)^{\cyan{\alpha}} \\ \gamma^\mu \left(S^{\cyan{I}\ast} \mathcal{F}_2^c\right)^\cyan{\alpha} \end{pmatrix},&& \boxed{X_{W\mathcal{F}_2}^{\mu\cyan{I},\cyan{\alpha}}} = g_2 \begin{pmatrix} - \left(\overline{\mathcal{F}_2} S^\cyan{I}\right)^\cyan{\alpha} \gamma^\mu  & \left(\overline{\mathcal{F}_2^c} S^{\cyan{I}\ast}\right)^\cyan{\alpha} \gamma^\mu  \end{pmatrix}, \\[0.8mm]
&\boxed{X_{\mathcal{F}_3G}^{\cyan{\alpha},\mu{\color{red}A}}} = g_s \begin{pmatrix} -\gamma^\mu T^{\color{red}A} \mathcal{F}_3^{\cyan{\alpha}} \\ \gamma^\mu T^{{\color{red}A}\ast} \left(\mathcal{F}_3^c\right)^\cyan{\alpha} \end{pmatrix},&& \boxed{X_{G\mathcal{F}_3}^{\mu{\color{red}A},\cyan{\alpha}}} = g_s \begin{pmatrix} - \left(\overline{\mathcal{F}_3}\right)^\cyan{\alpha} T^{\color{red}A} \gamma^\mu & \left(\overline{\mathcal{F}_3^c}\right)^\cyan{\alpha} T^{{\color{red}A}\ast} \gamma^\mu  \end{pmatrix}, 
\end{align}
where $\mathcal{F}_1 \in \{Q_L, U_R, D_R, L_L, E_R, \nu_R\}$, $\mathcal{F}_2 \in \{Q_L, L_L\}$, and $\mathcal{F}_3 \in \{Q_L, U_R, D_R\}$.
\item vector-vector type:
\begin{align}
&\boxed{X_{BB}^{\mu,\nu}} = - 2g_1^2Y_H^2 g^{\mu\nu} \left(\left|H_1\right|^2 + \left|H_2\right|^2\right), \\[0.8mm]
&\boxed{X_{WW}^{\mu\cyan{I},\nu\cyan{J}}} = 2g_2\epsilon^\cyan{IJK} W^{\cyan{K}\mu\nu} - 2g_2^2g^{\mu\nu} H_i^\dagger S^\cyan{I} S^\cyan{J} H_i, \\[0.8mm]
&\boxed{X_{BW}^{\mu,\nu\cyan{I}}} = \boxed{X_{WB}^{\mu\cyan{I},\nu}} = -2g_1g_2Y_H g^{\mu\nu} \left(H_1^\dagger S^\cyan{I} H_1 + H_2^\dagger S^\cyan{I} H_2\right), \\[0.8mm]
&\boxed{X_{GG}^{\mu{\color{red}A},\nu{\color{red}B}}} = 2g_sf^{\color{red}ABC} G^{{\color{red}C}\mu\nu}. 
\end{align}
\end{itemize}
In these expressions, the cyan indices \cyan{$\alpha,\beta,I,J,K$} and the red indices {\color{red}$A,B,C$} refer to isospin and color spaces, respectively. In addition, we have omitted all ghost-related interactions, as they do not contribute to the one-loop matching calculation considered here.

\section{One-loop Fierz transformations}
\label{appD:1-loop_Fierz}

The impact of evanescent operators can be accounted for by removing the evanescent components from the EFT Lagrangian and compensating them through finite shifts of the physical WCs~\cite{Fuentes-Martin:2022vvu}. In this work, we generalize the analysis of ref.~\cite{Fuentes-Martin:2022vvu} by incorporating three right-handed Dirac neutrinos. Employing eqs.~\eqref{eq:evan_beg}--\eqref{eq:evan_end} and the Greek reduction identity~\cite{Buras:2020xsm}
\begin{align}
\gamma_\mu \gamma_\nu \gamma_\lambda P_A \otimes \gamma^\mu \gamma^\nu \gamma^\lambda P_B &= 4\left(1+c_{\rm ev}\epsilon\right) \gamma_\mu P_A \otimes \gamma^\mu P_B,
\end{align}
with $c_{\rm ev} = 1$ and $P_{A,B} \in \{P_L,P_R\}$, $P_A \ne P_B$, we evaluate the six categories of covariant diagrams shown in ref.~\cite{Fuentes-Martin:2022vvu} in the NDR scheme and derive the following one-loop substitutions for the four scalar evanescent operators:
\begin{align}
E_{\underset{prst}{\ell\nu\nu\ell}}^{S,(6)} \Rightarrow & - \frac{1}{4} \left(Y_\nu^{1\dagger}\right)_{rv} \left(Y_\nu^1\right)_{us} \mathcal{Q}_{\underset{ptuv}{\ell\ell}}^{(6)} - \frac{1}{4} \left(Y_\nu^{1\dagger}\right)_{ru} \left(Y_\nu^1\right)_{tv} \mathcal{Q}_{\underset{pusv}{\ell\nu}}^{(6)} \notag\\[0.8mm]
&+ \frac{1}{4} \left(Y_\nu^{1\dagger}\right)_{uv} \left(Y_\nu^1\right)_{ts} \mathcal{Q}_{\underset{pvur}{\ell\nu}}^{(6)} + \frac{1}{4} \left(Y_\nu^{1\dagger}\right)_{rp} \left(Y_\nu^1\right)_{uv} \mathcal{Q}_{\underset{utsv}{\ell\nu}}^{(6)} - \frac{1}{4} \left(Y_\nu^{1\dagger}\right)_{vp} \left(Y_\nu^1\right)_{us} \mathcal{Q}_{\underset{utvr}{\ell\nu}}^{(6)} \notag\\[0.8mm]
&- \frac{1}{2} \left[\left(Y_E^1\right)_{uv} \left(Y_\nu^1\right)_{ts} + \left(Y_E^1\right)_{tv} \left(Y_\nu^1\right)_{us}\right] \mathcal{Q}_{\underset{pruv}{\ell\nu\ell e}}^{(6)} + \frac{g_1}{8} \left(Y_\nu^1\right)_{ts} \mathcal{Q}_{\underset{pr}{\nu B}}^{(6)} - \frac{g_2}{8} \left(Y_\nu^1\right)_{ts} \mathcal{Q}_{\underset{pr}{\nu W}}^{(6)} \notag\\[0.8mm]
&+ \frac{1}{4} \left(Y_E^1\right)_{tv} \left(Y_\nu^1\right)_{us} \mathcal{Q}_{\underset{urpv}{\ell\nu\ell e}}^{(6)} - \frac{1}{2} \left(Y_U^{1\dagger}\right)_{uv} \left(Y_\nu^1\right)_{ts} \mathcal{Q}_{\underset{pruv}{\ell\nu uq}}^{(6)} - \frac{1}{2} \left(Y_E^{1\dagger}\right)_{up} \left(Y_E^1\right)_{tv} \mathcal{Q}_{\underset{sruv}{\nu e}}^{(6)} \notag\\[0.8mm]
&- \frac{1}{2} \left(Y_D^1\right)_{uv} \left(Y_\nu^1\right)_{ts} \mathcal{Q}_{\underset{pruv}{\ell\nu qd}}^{(1),(6)} - \frac{1}{2} \left(Y_\nu^{1\dagger}\right)_{up} \left(Y_\nu^1\right)_{tv} \mathcal{Q}_{\underset{sruv}{\nu\nu}}^{(6)} - \frac{m_{11}^2}{2} \left(Y_\nu^1\right)_{ts} \mathcal{Q}_{\underset{pr}{\ell\nu}}^{(4)} \notag\\[0.8mm]
&+ \left[\left(Y_\nu^1 Y_\nu^{1\dagger} Y_\nu^1\right)_{ts} - \frac{\lambda_1}{2} \left(Y_\nu^1\right)_{ts}\right] \mathcal{Q}_{\underset{pr}{\nu H_1}}^{(6)} + \frac{g_1}{8} \left(Y_\nu^{1\dagger}\right)_{rp} \mathcal{Q}_{\underset{st}{\nu B}}^{(6)\dagger} - \frac{g_2}{8} \left(Y_\nu^{1\dagger}\right)_{rp} \mathcal{Q}_{\underset{st}{\nu W}}^{(6)\dagger} \notag\\[0.8mm]
&- \frac{1}{2} \left[\left(Y_E^{1\dagger}\right)_{uv} \left(Y_\nu^{1\dagger}\right)_{rp} + \left(Y_E^{1\dagger}\right)_{up} \left(Y_\nu^{1\dagger}\right)_{rv}\right] \mathcal{Q}_{\underset{stuv}{\ell\nu\ell e}}^{(6)\dagger} - \frac{m_{11}^2}{2} \left(Y_\nu^{1\dagger}\right)_{rp} \mathcal{Q}_{\underset{st}{\ell\nu}}^{(4)\dagger} \notag\\[0.8mm]
&+ \left[\left(Y_\nu^{1\dagger} Y_\nu^1 Y_\nu^{1\dagger}\right)_{rp} - \frac{\lambda_1}{2} \left(Y_\nu^{1\dagger}\right)_{rp}\right] \mathcal{Q}_{\underset{st}{\nu H_1}}^{(6)\dagger} + \frac{1}{4} \left(Y_E^{1\dagger}\right)_{up} \left(Y_\nu^{1\dagger}\right)_{rv} \mathcal{Q}_{\underset{svut}{\ell\nu\ell e}}^{(6)\dagger} \notag\\[0.8mm]
&- \frac{1}{2} \left(Y_\nu^{1\dagger}\right)_{rp} \left(Y_U^1\right)_{uv} \mathcal{Q}_{\underset{stuv}{\ell\nu uq}}^{(6)\dagger} - \frac{1}{2} \left(Y_D^{1\dagger}\right)_{uv} \left(Y_\nu^{1\dagger}\right)_{rp} \mathcal{Q}_{\underset{stuv}{\ell\nu qd}}^{(1),(6)\dagger}, \\[1.5mm]
E_{\underset{prst}{\ell ee\ell}}^{S,(6)} \Rightarrow & - \frac{1}{2} \left(Y_E^{1\dagger}\right)_{up} \left(Y_E^1\right)_{tv} \mathcal{Q}_{\underset{sruv}{ee}}^{(6)} + \frac{1}{4} \left(Y_E^1\right)_{us} \left(Y_\nu^1\right)_{tv} \mathcal{Q}_{\underset{pvur}{\ell\nu\ell e}}^{(6)} \notag\\[0.8mm]
&+ \left[\left(Y_E^1 Y_E^{1\dagger} Y_E^1\right)_{ts} - \frac{\lambda_1}{2} \left(Y_E^1\right)_{ts}\right] \mathcal{Q}_{\underset{pr}{eH_1}}^{(6)} - \frac{g_2}{8} \left(Y_E^1\right)_{ts} \mathcal{Q}_{\underset{pr}{eW}}^{(6)} - \frac{1}{2} \left(Y_\nu^{1\dagger}\right)_{up} \left(Y_\nu^1\right)_{tv} \mathcal{Q}_{\underset{uvsr}{\nu e}}^{(6)} \notag\\[0.8mm]
&- \frac{1}{4} \left(Y_E^{1\dagger}\right)_{ru} \left(Y_E^1\right)_{tv} \mathcal{Q}_{\underset{pusv}{\ell e}}^{(6)} + \frac{1}{4} \left(Y_E^{1\dagger}\right)_{uv} \left(Y_E^1\right)_{ts} \mathcal{Q}_{\underset{pvur}{\ell e}}^{(6)} + \frac{1}{4} \left(Y_E^{1\dagger}\right)_{rp} \left(Y_E^1\right)_{uv} \mathcal{Q}_{\underset{utsv}{\ell e}}^{(6)} \notag\\[0.8mm]
&- \frac{1}{4} \left(Y_E^{1\dagger}\right)_{vp} \left(Y_E^1\right)_{us} \mathcal{Q}_{\underset{utvr}{\ell e}}^{(6)} - \frac{1}{2} \left(Y_D^{1\dagger}\right)_{uv} \left(Y_E^1\right)_{ts} \mathcal{Q}_{\underset{pruv}{\ell edq}}^{(6)} - \frac{1}{4} \left(Y_E^{1\dagger}\right)_{rv} \left(Y_E^1\right)_{us} \mathcal{Q}_{\underset{ptuv}{\ell\ell}}^{(6)} \notag\\[0.8mm]
&- \frac{1}{2} \left[\left(Y_E^1\right)_{us} \left(Y_\nu^1\right)_{tv} + \left(Y_E^1\right)_{ts} \left(Y_\nu^1\right)_{uv}\right] \mathcal{Q}_{\underset{uvpr}{\ell\nu\ell e}}^{(6)} + \frac{1}{2} \left(Y_E^1\right)_{ts} \left(Y_U^1\right)_{uv} \mathcal{Q}_{\underset{pruv}{\ell equ}}^{(1),(6)} \notag\\[0.8mm]
&+ \frac{3g_1}{8} \left(Y_E^1\right)_{ts} \mathcal{Q}_{\underset{pr}{eB}}^{(6)} - \frac{m_{11}^2}{2} \left(Y_E^1\right)_{ts} \mathcal{Q}_{\underset{pr}{\ell e}}^{(4)} + \left[\left(Y_E^{1\dagger} Y_E^1 Y_E^{1\dagger}\right)_{rp} - \frac{\lambda_1}{2} \left(Y_E^{1\dagger}\right)_{rp}\right] \mathcal{Q}_{\underset{st}{eH_1}}^{(6)\dagger} \notag\\[0.8mm]
&- \frac{1}{2} \left(Y_E^{1\dagger}\right)_{rp} \left(Y_D^1\right)_{uv} \mathcal{Q}_{\underset{stuv}{\ell edq}}^{(6)\dagger} + \frac{1}{4} \left(Y_E^{1\dagger}\right)_{rv} \left(Y_\nu^{1\dagger}\right)_{up} \mathcal{Q}_{\underset{utsv}{\ell\nu\ell e}}^{(6)\dagger} \notag\\[0.8mm]
&+ \frac{1}{2} \left(Y_E^{1\dagger}\right)_{rp} \left(Y_U^{1\dagger}\right)_{uv} \mathcal{Q}_{\underset{stuv}{\ell equ}}^{(1),(6)\dagger} - \frac{1}{2} \left[\left(Y_E^{1\dagger}\right)_{rv} \left(Y_\nu^{1\dagger}\right)_{up} + \left(Y_E^{1\dagger}\right)_{rp} \left(Y_\nu^{1\dagger}\right)_{uv}\right] \mathcal{Q}_{\underset{uvst}{\ell\nu\ell e}}^{(6)\dagger} \notag\\[0.8mm]
&- \frac{m_{11}^2}{2} \left(Y_E^{1\dagger}\right)_{rp} \mathcal{Q}_{\underset{st}{\ell e}}^{(4)\dagger} + \frac{3g_1}{8} \left(Y_E^{1\dagger}\right)_{rp} \mathcal{Q}_{\underset{st}{eB}}^{(6)\dagger} - \frac{g_2}{8} \left(Y_E^{1\dagger}\right)_{rp} \mathcal{Q}_{\underset{st}{eW}}^{(6)\dagger}, \\[1.5mm]
E_{\underset{prst}{quuq}}^{S,(6)} \Rightarrow & - \frac{3m_{11}^2}{2} \left(Y_U^1\right)_{ts} \mathcal{Q}_{\underset{pr}{qu}}^{(4)} - \frac{3m_{11}^2}{2} \left(Y_U^{1\dagger}\right)_{rp} \mathcal{Q}_{\underset{st}{qu}}^{(4)\dagger} - \frac{5g_1}{8} \left(Y_U^1\right)_{ts} \mathcal{Q}_{\underset{pr}{uB}}^{(6)} \notag\\[0.8mm]
&+ \left[3\left(Y_U^1 Y_U^{1\dagger} Y_U^1\right)_{ts} - \frac{3\lambda_1}{2} \left(Y_U^1\right)_{ts}\right] \mathcal{Q}_{\underset{pr}{uH_1}}^{(6)} - \frac{1}{2} \left(Y_U^{1\dagger}\right)_{up} \left(Y_U^1\right)_{tv} \mathcal{Q}_{\underset{svur}{uu}}^{(6)} \notag\\[0.8mm]
&+ \frac{3}{2} \left(Y_U^{1\dagger}\right)_{uv} \left(Y_U^1\right)_{ts} \mathcal{Q}_{\underset{pvur}{qu}}^{(8),(6)} + \frac{3}{2} \left(Y_U^{1\dagger}\right)_{rp} \left(Y_U^1\right)_{uv} \mathcal{Q}_{\underset{utsv}{qu}}^{(8),(6)} \notag\\[0.8mm]
&- \frac{3}{2} \left(Y_U^{1\dagger}\right)_{rp} \left(Y_\nu^1\right)_{uv} \mathcal{Q}_{\underset{uvst}{\ell\nu uq}}^{(6)} + \frac{1}{4} \left(Y_U^{1\dagger}\right)_{uv} \left(Y_U^1\right)_{ts} \mathcal{Q}_{\underset{pvur}{qu}}^{(1),(6)} \notag\\[0.8mm]
&+ \frac{1}{4} \left(Y_U^{1\dagger}\right)_{rp} \left(Y_U^1\right)_{uv} \mathcal{Q}_{\underset{utsv}{qu}}^{(1),(6)} - \frac{1}{4} \left(Y_U^{1\dagger}\right)_{vp} \left(Y_U^1\right)_{us} \mathcal{Q}_{\underset{utvr}{qu}}^{(1),(6)} \notag\\[0.8mm]
&- \frac{1}{8} \left(Y_U^{1\dagger}\right)_{rv} \left(Y_U^1\right)_{us} \mathcal{Q}_{\underset{pvut}{qq}}^{(3),(6)} + \frac{3}{2} \left(Y_E^1\right)_{uv} \left(Y_U^1\right)_{ts} \mathcal{Q}_{\underset{uvpr}{\ell equ}}^{(1),(6)} \notag\\[0.8mm]
&- \frac{1}{8} \left(Y_U^{1\dagger}\right)_{rv} \left(Y_U^1\right)_{us} \mathcal{Q}_{\underset{pvut}{qq}}^{(1),(6)} + \frac{1}{12} \left(Y_D^1\right)_{tv} \left(Y_U^1\right)_{us} \mathcal{Q}_{\underset{urpv}{quqd}}^{(1),(6)} \notag\\[0.8mm]
&- \frac{1}{4} \left(Y_U^{1\dagger}\right)_{ru} \left(Y_U^1\right)_{tv} \mathcal{Q}_{\underset{pusv}{qu}}^{(1),(6)} - \frac{3}{2} \left(Y_\nu^{1\dagger}\right)_{uv} \left(Y_U^1\right)_{ts} \mathcal{Q}_{\underset{uvpr}{\ell\nu uq}}^{(6)\dagger} \notag\\[0.8mm]
&+ \frac{1}{2} \left(Y_D^1\right)_{tv} \left(Y_U^1\right)_{us} \mathcal{Q}_{\underset{urpv}{quqd}}^{(8),(6)} - \frac{1}{6} \left(Y_D^{1\dagger}\right)_{up} \left(Y_D^1\right)_{tv} \mathcal{Q}_{\underset{sruv}{ud}}^{(1),(6)} \notag\\[0.8mm]
&- \left(Y_D^{1\dagger}\right)_{up} \left(Y_D^1\right)_{tv} \mathcal{Q}_{\underset{sruv}{ud}}^{(8),(6)} - \frac{3g_2}{8} \left(Y_U^1\right)_{ts} \mathcal{Q}_{\underset{pr}{uW}}^{(6)} \notag\\[0.8mm]
&+ \left[3\left(Y_U^{1\dagger} Y_U^1 Y_U^{1\dagger}\right)_{rp} - \frac{3\lambda_1}{2} \left(Y_U^{1\dagger}\right)_{rp}\right] \mathcal{Q}_{\underset{st}{uH_1}}^{(6)\dagger} + \frac{3}{2} \left(Y_E^{1\dagger}\right)_{uv} \left(Y_U^{1\dagger}\right)_{rp} \mathcal{Q}_{\underset{uvst}{\ell equ}}^{(1),(6)\dagger} \notag\\[0.8mm]
&- \frac{3g_2}{8} \left(Y_U^{1\dagger}\right)_{rp} \mathcal{Q}_{\underset{st}{uW}}^{(6)\dagger} - \frac{1}{2} \left[3\left(Y_D^{1\dagger}\right)_{uv} \left(Y_U^{1\dagger}\right)_{rp} + \left(Y_D^{1\dagger}\right)_{up} \left(Y_U^{1\dagger}\right)_{rv}\right] \mathcal{Q}_{\underset{stuv}{quqd}}^{(1),(6)\dagger} \notag\\[0.8mm]
&+ \frac{1}{2} \left(Y_D^{1\dagger}\right)_{up} \left(Y_U^{1\dagger}\right)_{rv} \mathcal{Q}_{\underset{svut}{quqd}}^{(8),(6)\dagger} - \frac{1}{2} \left[3\left(Y_D^1\right)_{uv} \left(Y_U^1\right)_{ts} + \left(Y_D^1\right)_{tv} \left(Y_U^1\right)_{us}\right] \mathcal{Q}_{\underset{pruv}{quqd}}^{(1),(6)} \notag\\[0.8mm]
&- \frac{5g_1}{8} \left(Y_U^{1\dagger}\right)_{rp} \mathcal{Q}_{\underset{st}{uB}}^{(6)\dagger} + \frac{1}{12} \left(Y_D^{1\dagger}\right)_{up} \left(Y_U^{1\dagger}\right)_{rv} \mathcal{Q}_{\underset{svut}{quqd}}^{(1),(6)\dagger}, \\[1.5mm]
E_{\underset{prst}{qddq}}^{S,(6)} \Rightarrow & - \frac{3m_{11}^2}{2} \left(Y_D^1\right)_{ts} \mathcal{Q}_{\underset{pr}{qd}}^{(4)} - \frac{3m_{11}^2}{2} \left(Y_D^{1\dagger}\right)_{rp} \mathcal{Q}_{\underset{st}{qd}}^{(4)\dagger} - \frac{1}{2} \left(Y_D^{1\dagger}\right)_{up} \left(Y_D^1\right)_{tv} \mathcal{Q}_{\underset{svur}{dd}}^{(6)} \notag\\[0.8mm]
&+ \left[3\left(Y_D^1 Y_D^{1\dagger} Y_D^1\right)_{ts} - \frac{3\lambda_1}{2} \left(Y_D^1\right)_{ts}\right] \mathcal{Q}_{\underset{pr}{dH_1}}^{(6)} - \frac{3g_2}{8} \left(Y_D^1\right)_{ts} \mathcal{Q}_{\underset{pr}{dW}}^{(6)} \notag\\[0.8mm]
&- \frac{1}{4} \left(Y_D^{1\dagger}\right)_{ru} \left(Y_D^1\right)_{tv} \mathcal{Q}_{\underset{pusv}{qd}}^{(1),(6)} + \frac{1}{4} \left(Y_D^{1\dagger}\right)_{uv} \left(Y_D^1\right)_{ts} \mathcal{Q}_{\underset{pvur}{qd}}^{(1),(6)} \notag\\[0.8mm]
&+ \frac{1}{4} \left(Y_D^{1\dagger}\right)_{rp} \left(Y_D^1\right)_{uv} \mathcal{Q}_{\underset{utsv}{qd}}^{(1),(6)} - \frac{1}{4} \left(Y_D^{1\dagger}\right)_{vp} \left(Y_D^1\right)_{us} \mathcal{Q}_{\underset{utvr}{qd}}^{(1),(6)} \notag\\[0.8mm]
&+ \frac{3}{2} \left(Y_D^{1\dagger}\right)_{uv} \left(Y_D^1\right)_{ts} \mathcal{Q}_{\underset{pvur}{qd}}^{(8),(6)} - \frac{3}{2} \left(Y_D^{1\dagger}\right)_{rp} \left(Y_E^1\right)_{uv} \mathcal{Q}_{\underset{uvst}{\ell edq}}^{(6)} \notag\\[0.8mm]
&+ \frac{3}{2} \left(Y_D^{1\dagger}\right)_{rp} \left(Y_D^1\right)_{uv} \mathcal{Q}_{\underset{utsv}{qd}}^{(8),(6)} - \frac{1}{8} \left(Y_D^{1\dagger}\right)_{rv} \left(Y_D^1\right)_{us} \mathcal{Q}_{\underset{pvut}{qq}}^{(1),(6)} \notag\\[0.8mm]
&- \frac{1}{8} \left(Y_D^{1\dagger}\right)_{rv} \left(Y_D^1\right)_{us} \mathcal{Q}_{\underset{pvut}{qq}}^{(3),(6)} + \frac{1}{12} \left(Y_D^1\right)_{us} \left(Y_U^1\right)_{tv} \mathcal{Q}_{\underset{pvur}{quqd}}^{(1),(6)} \notag\\[0.8mm]
&+ \frac{1}{2} \left(Y_D^1\right)_{us} \left(Y_U^1\right)_{tv} \mathcal{Q}_{\underset{pvur}{quqd}}^{(8),(6)} - \frac{1}{6} \left(Y_U^{1\dagger}\right)_{up} \left(Y_U^1\right)_{tv} \mathcal{Q}_{\underset{uvsr}{ud}}^{(1),(6)} + \frac{g_1}{8} \left(Y_D^1\right)_{ts} \mathcal{Q}_{\underset{pr}{dB}}^{(6)} \notag\\[0.8mm]
&+ \frac{g_1}{8} \left(Y_D^{1\dagger}\right)_{rp} \mathcal{Q}_{\underset{st}{dB}}^{(6)\dagger} + \left[3\left(Y_D^{1\dagger} Y_D^1 Y_D^{1\dagger}\right)_{rp} - \frac{3\lambda_1}{2} \left(Y_D^{1\dagger}\right)_{rp}\right] \mathcal{Q}_{\underset{st}{dH_1}}^{(6)\dagger} \notag\\[0.8mm]
&- \frac{3g_2}{8} \left(Y_D^{1\dagger}\right)_{rp} \mathcal{Q}_{\underset{st}{dW}}^{(6)\dagger} - \frac{3}{2} \left(Y_E^{1\dagger}\right)_{uv} \left(Y_D^1\right)_{ts} \mathcal{Q}_{\underset{uvpr}{\ell edq}}^{(6)\dagger} - \frac{3}{2} \left(Y_D^{1\dagger}\right)_{rp} \left(Y_\nu^{1\dagger}\right)_{uv} \mathcal{Q}_{\underset{uvst}{\ell\nu qd}}^{(1),(6)\dagger} \notag\\[0.8mm]
&+ \frac{1}{12} \left(Y_D^{1\dagger}\right)_{rv} \left(Y_U^{1\dagger}\right)_{up} \mathcal{Q}_{\underset{utsv}{quqd}}^{(1),(6)\dagger} + \frac{1}{2} \left(Y_D^{1\dagger}\right)_{rv} \left(Y_U^{1\dagger}\right)_{up} \mathcal{Q}_{\underset{utsv}{quqd}}^{(8),(6)\dagger} \notag\\[0.8mm]
&- \frac{1}{2} \left[\left(Y_D^{1\dagger}\right)_{rv} \left(Y_U^{1\dagger}\right)_{up} + 3\left(Y_D^{1\dagger}\right)_{rp} \left(Y_U^{1\dagger}\right)_{uv}\right] \mathcal{Q}_{\underset{uvst}{quqd}}^{(1),(6)\dagger} - \left(Y_U^{1\dagger}\right)_{up} \left(Y_U^1\right)_{tv} \mathcal{Q}_{\underset{uvsr}{ud}}^{(8),(6)} \notag\\[0.8mm]
&- \frac{3}{2} \left(Y_D^1\right)_{ts} \left(Y_\nu^1\right)_{uv} \mathcal{Q}_{\underset{uvpr}{\ell\nu qd}}^{(1),(6)} - \frac{1}{2} \left[\left(Y_D^1\right)_{us} \left(Y_U^1\right)_{tv} + 3\left(Y_D^1\right)_{ts} \left(Y_U^1\right)_{uv}\right] \mathcal{Q}_{\underset{uvpr}{quqd}}^{(1),(6)}.
\end{align}
The one-loop Fierz transformations of the scalar operators $\mathcal{O}^{S,(6)}_i$ are then obtained by substituting the evanescent operators in the above relations with their definitions provided in appendix~\ref{appA:vSMEFT_basis}.

\bibliographystyle{JHEP}
\bibliography{ref}
\end{document}